\pdfoutput=1
\documentclass[a4paper,12pt]{article}
\usepackage{amsfonts}
\usepackage{mathrsfs}
\usepackage{amsmath}
\usepackage{amssymb}
\usepackage{framed}
\usepackage[medium]{titlesec}
\usepackage{bm}
\usepackage{cite}
\usepackage[normalem]{ulem}
\usepackage{extarrows}
\usepackage{slashed}
\usepackage{isodateo}
\usepackage{graphicx}
\usepackage{xcolor}
\usepackage[bookmarksnumbered=true,bookmarksopen=true]{hyperref}
 \hypersetup{colorlinks,%
             linkcolor=[rgb]{0,0.3,0.6}, %
             citecolor=[rgb]{0,0.3,0.6}, %
             urlcolor=[rgb]{0,0.3,0.6}}
\usepackage[hmargin=.7in,vmargin=1.1in]{geometry}
\usepackage{indentfirst}
\usepackage{booktabs}

\newcommand{\FR}[2]{\displaystyle\frac{\,{#1}\,}{#2}}
\newcommand{\fr}[2]{\mbox{$\frac{\,{#1}\,}{#2}$}}
\newcommand{\n}{\nonumber}

\graphicspath{{fig/}}

\def\bge{\begin{equation}}
\def\ede{\end{equation}}
\def\bga{\begin{aligned}}
\def\eda{\end{aligned}}
\def\bgp{\begin{pmatrix}}
\def\edp{\end{pmatrix}}
\def\bgm{\begin{matrix}}
\def\edm{\end{matrix}}
\def\bgs{\begin{subequations}}
\def\eds{\end{subequations}}
\newcommand{\order}[1]{\mathcal{O}({#1})}
\def\di{{\mathrm{d}}}
\def\D{{\mathrm{D}}}

\def\mb{\mathbf}

\def\pd{\partial}
\def\ld{{\mathscr{L}}}

\def\la{\langle}\def\ra{\rangle}

\def\to{\rightarrow}
\def\To{\Rightarrow}
\def\ii{\mathrm{i}}

\def\al{\alpha}

\def\de{\delta}
\def\ep{\epsilon}

\def\lam{\lambda}

\def\si{\sigma}

\def\Re{\mathrm{Re}\,}

\usepackage{mdframed} 
                   
\newmdenv[backgroundcolor=gray!15,%
          skipabove=0pt,%
          skipbelow=5pt,%
          leftmargin=0pt,%
          rightmargin=0pt,%
          innertopmargin=-5pt,%
          innerbottommargin=7pt,%
          innerleftmargin=2pt,%
          innerrightmargin=2pt,%
          splittopskip=0pt,%
          splitbottomskip=0pt,%
          linewidth=0pt]%
          {keyeqn}

\newcommand{\wt}[1]{\mkern 2mu \widetilde{\mkern -2mu #1 \mkern -2mu}\mkern 2mu}
\newcommand{\wh}[1]{\mkern 2mu \widehat{\mkern-2mu#1\mkern-2mu}\mkern 2mu}

\newcommand{\fnemail}[1]{\footnote{Email: \href{mailto:#1}{\nolinkurl{#1}}}}

\begin{document}


\title{\Large\textbf{Phase Information in Cosmological Collider Signals}}
\author{Zhehan Qin$^{1,\,}$\fnemail{qzh21@mails.tsinghua.edu.cn}~~~~~ and ~~~~~Zhong-Zhi Xianyu$^{1,2,\,}$\fnemail{zxianyu@tsinghua.edu.cn}\\[5mm]
\normalsize{${}^{1}\,$\emph{Department of Physics, Tsinghua University, Beijing 100084, China}}\\
\normalsize{${}^{2}\,$\emph{Collaborative Innovation Center of Quantum Matter, Beijing 100084, China}}}

\date{}
\maketitle

\vspace{20mm}

\begin{abstract}
  Massive particles produced during the cosmic inflation can imprint in the primordial non-Gaussianities as characteristic oscillating functions of various momentum ratios, known as cosmological collider signals. We initiate a study of the phase of the oscillating signals which can be unambiguously defined and measured. The phase can provide useful new information about the spin and the couplings of the intermediate heavy particles that cannot be obtained from the signal frequency and angular dependences alone. We also present new analytical results for full nonlocal signals from two typical 1-loop processes, enabling precise determination of the signal phase away from the squeezed limit.
\end{abstract}

\newpage
\tableofcontents

\newpage
\section{Introduction}

The cosmological collider (CC) physics has emerged in recent years as a new tool to probe potential new physics at the inflation scale \cite{Arkani-Hamed:2015bza}, built on earlier works of the quasi-single-field inflation \cite{Chen:2009we,Chen:2009zp,Baumann:2011nk,Chen:2012ge,Pi:2012gf,Noumi:2012vr,Gong:2013sma}. The essential idea of the CC physics is the following: During the inflation, either the spacetime expansion or the inflaton rolling can trigger copious production of particles with masses comparable or even much higher than the inflation Hubble scale. After the production, such a massive particle quickly loses its momentum due to inflationary expansion and its mode function would oscillate with a fixed frequency given by its mass. By coupling to the curvature perturbation, the oscillating mode function gives rise to characteristic shapes of the $n$-point ($n\geq 3$) correlation functions of the curvature perturbation, in the form of oscillatory dependences in various momentum ratios, known as CC signals. It was shown that measuring the frequency and the angular dependence of these oscillating signals can reveal the mass and the spin of the massive particles. Given that the inflation scale is probably much higher than the reach of any terrestrial experiments, the CC physics provides us a unique window to fundamental particle physics at very high energies. 

Since the advent of this idea, progresses have been made in various directions related to the CC physics. On the observation side, new ideas of probing CC signals have been put forward, and the detectability of CC signals has been forecast, using cosmological data from large-scale structures \cite{MoradinezhadDizgah:2018ssw}, galaxy imaging \cite{Kogai:2020vzz}, to 21-cm tomography \cite{Meerburg:2016zdz}. At the theory frontier, particle models of CC signals have been investigated in various directions, both within and beyond the particle Standard Model \cite{Chen:2015lza,Chen:2016nrs,Chen:2016uwp,Chen:2016hrz,Lee:2016vti,An:2017hlx,Kumar:2017ecc,Chen:2017ryl,Tong:2018tqf,Chen:2018sce,Chen:2018xck,Chen:2018cgg,Chua:2018dqh,Domenech:2018bnf,Wu:2018lmx,Li:2019ves,Lu:2019tjj,Liu:2019fag,Hook:2019zxa,Hook:2019vcn,Kumar:2019ebj,Alexander:2019vtb,Wang:2019gbi,Wang:2019gok,Wang:2020uic,Li:2020xwr,Wang:2020ioa,Fan:2020xgh,Aoki:2020zbj,Bodas:2020yho,Maru:2021ezc,Lu:2021gso,Sou:2021juh,Lu:2021wxu,Pinol:2021aun,Cui:2021iie,Tong:2022cdz,Reece:2022soh}. Meanwhile, theoretical understanding of inflationary correlation functions has been advanced with multiple approaches, including the cosmological bootstrap program \cite{Arkani-Hamed:2018kmz,Baumann:2019oyu,Baumann:2020dch,Pajer:2020wnj,Pajer:2020wxk,Cabass:2021fnw,Pimentel:2022fsc},  the AdS-inspired Mellin space approach \cite{Sleight:2019mgd,Sleight:2019hfp,Sleight:2020obc,Sleight:2021plv}, and a direct numerical approach \cite{Wang:2021qez}. Results have also been obtained on the analytical structure of cosmic correlators \cite{Goodhew:2020hob,Melville:2021lst,Goodhew:2021oqg,DiPietro:2021sjt,Tong:2021wai}. See also \cite{Meerburg:2019qqi,Achucarro:2022qrl,Baumann:2022jpr}. 

As mentioned, the CC signal manifests itself in the cosmic correlators as the oscillatory dependences in the momentum ratios, in the limit where the intermediate particle's momentum approaches zero. In this paper, we shall focus on the 4-point correlators $\mathcal{T}=\la\varphi_{\mb k_1}\varphi_{\mb k_2}\varphi_{\mb k_3}\varphi_{\mb k_4}\ra'$  where $\varphi_{\mb k_i}$ represent Fourier modes of inflaton fluctuations and are proportional to the curvature fluctuations in typical inflation scenarios. As will be detailed in Sec.\;\ref{sec_dissect}, for a massive particle exchange in the $s$-channel at either the  tree level or the 1-loop level, the nonlocal CC signal, denoted by $\mathcal{T}_\text{NL}$, takes the following schematic form:
\begin{align}
\label{eq_TNLscheme}
  \lim_{k_s\to 0}\mathcal{T}_\text{NL} \sim \mathcal{A}\Big(\FR{k_s^2}{k_{12}k_{34}}\Big)^\al\cos\bigg[\omega\log\FR{k_s^2}{k_{12}k_{34}}+\vartheta\Big(\FR{k_s}{k_{12}},\FR{k_s}{k_{34}}\Big)\bigg]\mathcal{K}(\theta_{1s},\theta_{3s},\theta_{13}),
\end{align}
where $\mb k_s\equiv \mb k_1+\mb k_2$ is the momentum of the intermediate particle in the $s$-channel, and $k_{ij}=|\mb k_i|+|\mb k_j|$ $(i=1,2,3,4)$. We see that the oscillation appears in logarithms of the momentum ratio $k_s^2/k_{12}k_{34}$. The parameters $\mathcal{A}$, $\al$, $\omega$, $\vartheta$ all characterize the shape of the signal, and the factor $\mathcal{K}$ characterizes the dependence on various angles $\theta_{ij}$ formed by the momenta $\mb k_i$ and $\mb k_j$ (including $\mb k_s$). We shall elaborate on these parameters and functions in Sec.\;\ref{sec_dissect}.

It has been emphasized that the oscillation frequency $\omega$ is related to the mass of the particle while the angular dependence $\mathcal{K}$ tells the spin \cite{Arkani-Hamed:2015bza}. For instance, a particle of mass $m$ and spin $s\neq 0$ can mediate a signal at the tree level with frequency $\omega=\sqrt{(m/H)^2-(s-1/2)^2}$ and the angular dependence $\mathcal{K}\propto \text{P}_s(\theta_{13})+\cdots$ where $H$ is the Hubble scale of the inflation, and $\text{P}_s(z)$ is the Legendre polynomial.\footnote{The omitted terms in the angular dependence contains other angles such as $\theta_{1s}$ or $\theta_{3s}$. The omitted terms disappear for signals in 3-point function, in which case there is only one independent angle $\theta_{13}$ in the squeezed limit $k_3\to 0$.} Below, we will often take $H=1$ with a few exceptions.

However, the relationship between the observables and model parameters are not always as simple as in the above example. Typically, a particle receives mass corrections from various sources during the inflation \cite{Chen:2016hrz}, and the oscillation frequency only tells the corrected mass rather than the intrinsic mass. Also, the spin of the intermediate particle is usually obscured by the form of the inflaton coupling, and cannot be read from the angular dependence of the signal. 

The problem of spin measurement is particularly notable for 1-loop processes, where the angular dependence only tells the \emph{total angular momentum} of the two loop particles rather than the \emph{spin} of each individual particle. As a result, if the signal frequency and the angular dependence are all what we can measure, it would be very difficult to learn the particle species and the interaction form. Given that many particle models predict CC signals chiefly at 1-loop level \cite{Chen:2016uwp,Chen:2016hrz,Chen:2018xck,Lu:2019tjj,Liu:2019fag,Hook:2019zxa,Hook:2019vcn,Wang:2019gbi,Wang:2020ioa,Cui:2021iie}, this potential degeneracy in the signal would be a significant limitation of CC physics as a probe of new physics.  

Fortunately, as is clear from (\ref{eq_TNLscheme}), there are more parameters in the nonlocal signals besides the frequency and the angular dependence. These additional parameters can be used to extract more information from the observed signal. For instance, it was known that the scaling power $\al$ can be used to distinguish between tree and 1-loop signals \cite{Wang:2021qez}, or to tell loop corrected masses from tree-level masses \cite{Lu:2021wxu}.

\begin{figure}[t]
\centering
\includegraphics[width=0.63
\textwidth]{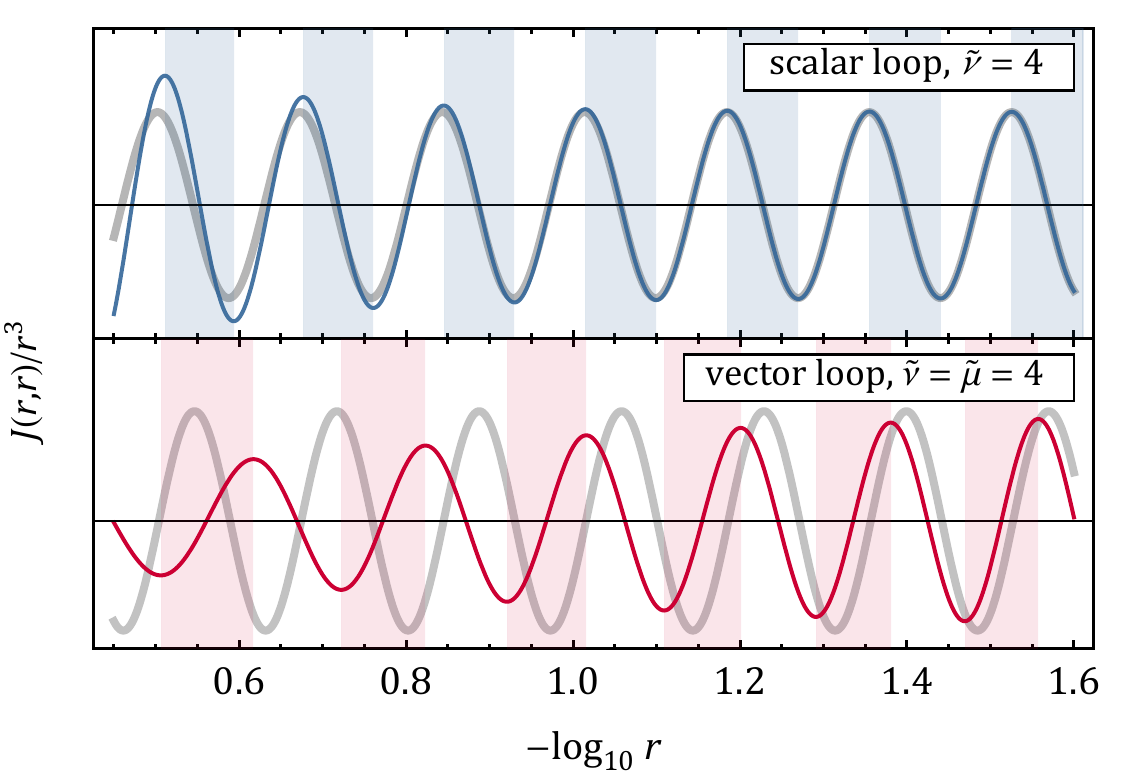}
\caption{Two samples of nonlocal CC signals in the 4-point function, from a scalar loop (upper) and a vector loop with chemical potential enhancement (lower). The colored curves show the complete results for nonlocal signals, while the light gray curves show the squeezed limit results. The colored bands in the background highlight the phase information in the signal. }
\label{fig_signal}
\end{figure}

There is one more parameter remained to be explored, namely the phase $\vartheta$ of the oscillation. This parameter was somehow considered unimportant and has been largely ignored in previous studies.\footnote{The phase was originally mentioned in \cite{Arkani-Hamed:2015bza} and was briefly discussed recently in \cite{Pimentel:2022fsc}.} However, as we shall show, the information in the phase can actually be very useful for better measuring and better theoretical understanding of CC signals. In particular, the phase can provide independent new information at 1-loop level which alleviates the aforementioned degeneracy of CC signals.

An immediate remark we would like to make is that the oscillating CC signal (\ref{eq_TNLscheme}) is in the logarithm of the momentum ratio. As a result, there is a well-defined reference point, $k_s^2/k_{12}k_{34}=1$, so that we have no freedom to arbitrarily shift the phase. In addition, we can define the amplitude factor $\mathcal{A}$ in (\ref{eq_TNLscheme}) to be real and positive, with a possible overall minus sign absorbed into the phase. Furthermore, with nontrivial angular dependences, there can be complex phases and overall sign flips in the angular dependent factor $\mathcal{K}$. In such cases, it turns out that we have freedom to independently vary the momentum ratio $k_{s}^2/k_{12}k_{34}$ and $\mathcal{K}$. Therefore, we can factor $\mathcal{K}$ out when measuring the signal phase. With all these points taken into consideration, we see that the phase of the oscillatory signal can be unambiguously defined and measured. 

In this work, we shall demonstrate the use of phase information with several examples of nonlocal signals (to be explained below) at both the tree level and the 1-loop level, mediated by either massive scalars or massive vector bosons. For signals from spin-1 gauge bosons, we pay special attention to a possible parity-violating chemical potential, which is a known source of large signals \cite{Chen:2018xck,Liu:2019fag,Wang:2019gbi,Wang:2020ioa,Sou:2021juh,Tong:2022cdz}. 

The determination of the phase requires a more delicate calculation of the signal, instead of just taking the squeezed limit (\ref{eq_TNLscheme}) as in most previous studies. In particular, the phase itself can evolve with the momentum ratio $k_s^2/k_{12}k_{34}$ when this ratio is not too small, shown in Fig.\ \ref{fig_signal}. Therefore, to reliably connect the measured phases with theory predictions, one should be able to compute this phase evolution. The phase evolution for the nonsqueezed configurations is itself a useful piece of information that can help to further break the degeneracy in the signals, as we shall show in the following sections. Furthermore, understanding the phase evolution for nonsqueezed configurations is crucial for 1-loop processes, since the loop signals decrease faster in the squeezed limit and thus they are to be measured only for not-too-squeezed configurations \cite{Wang:2021qez}.

To compute more precisely the phase evolution for nonsqueezed configurations, we push the calculation of the signal to all orders in the momentum ratio $k_s^2/k_{12}k_{34}$. In particular, for the two 1-loop processes we will be considering, we obtain for the first time the full analytical expressions of the nonlocal signals in form of power series.  This is made possible by using a partial Mellin-Barnes (MB) representation to be introduced in App.\ \ref{app_MB}. By a partial MB representation, we mean that we apply the inverse Mellin transformation only to intermediate massive modes, while the massless external modes are left untouched. This differs from the complete Mellin-space approach in \cite{Sleight:2019mgd,Sleight:2019hfp,Sleight:2020obc,Sleight:2021plv} where all modes are transformed into the Mellin space.

In Fig.\;\ref{fig_signal}, we show two representative signals that we obtained in this work, including the signal from a scalar loop (Fig.\;\ref{fig_scalar_loop}) and the signal from a vector loop with chemical potential enhancement (Fig.\;\ref{fig_vector_loop}). In both cases we choose the mass parameter $\wt\nu=4$, and, for the vector loop, we choose the chemical potential $\wt\mu=4$. Explanations of these parameters will be provided below. The two signals have the identical oscillation frequency $\omega=2\wt\nu=8$ in the squeezed limit, and have the identical angular dependence. Therefore, it is impossible to know the origin of the signal if we only measure the frequency and the angular dependence. However, it is clear from the figure that the two signals possess very different phases in the squeezed limit, and it is also clear that the two phases have significant dependence on the momentum ratio $r^2=k_s^2/k_{12}k_{34}$. Therefore, measuring the phases and their evolution in $r$ can be a useful way to extract more information about the underlying particle physics processes. In Fig.\;\ref{fig_signal} we also show the squeezed-limit results in light gray curves. Comparing the gray curves with the colored curves should make it clear that the deviations from squeezed-limit results are significant and must be included if we wish to fully utilize the phase information. Technical details about this figure will be given at the end of Sec.\ \ref{sec_loop}.

The rest of the paper is organized as follows. In Sec.\ \ref{sec_dissect}, we   review for non-experts the essential physics of 4-point correlation functions, the CC signals, and the signal parameters. We also take this chance to explain our notations and conventions. Then, in Sec.\ \ref{sec_tree}, we compute the full nonlocal signals from two tree-level processes, including a massive scalar exchange, and a massive vector boson exchange with nonzero chemical potential enhancement. The result in this section is known in the literature, and we reproduce them here to highlight the use of the signal phase. In Sec.\ \ref{sec_loop}, we present the main results of our paper, namely the full nonlocal signals from two 1-loop processes, including a massive scalar loop and a massive vector loop with nonzero chemical potential. These results are new, and we show that the phases in these signals provide useful information about the particle species and interaction types which cannot be revealed from signal frequency and the angular dependence. The analytical expressions for the nonlocal signals are presented as power series in the momentum ratios. We demonstrate the quick convergence of these power series which enables an efficient numerical implementation. More discussions are given in Sec.\ \ref{sec_discussions}. We summarize our conventions and notations in App.\ \ref{app_notation}. In App.\ \ref{app_MB} we introduce the Mellin-Barnes representation for intermediate propagators that allows us to analytically complete the Schwinger-Keldysh integral. In App.\ \ref{app_details} we provide further details of our calculations.

\section{Dissecting Cosmological Collider Signals}
\label{sec_dissect}

In principle, the cosmological collider signals can be looked for in all kinds of soft limits of arbitrary $n$-point $(n\geq 3)$ correlators of the inflaton fluctuation $\varphi$. In this paper we will focus on the 4-point function. 

\paragraph{4-point correlator.} In general, a four-point correlator $\la\varphi_{\mb k_1}\varphi_{\mb k_2}\varphi_{\mb k_3}\varphi_{\mb k_4}\ra'$ is a function of four external momenta $\mb k_i~(i=1,2,3,4)$. These objects are correlators of inflaton fluctuations $\varphi_{\mb k}$ in an inflationary universe with 3-dimensional flat spatial slices. They can be computed in a Feynman-diagram approach. See \cite{Chen:2017ryl} for an introduction. 

\begin{figure}[t]
\centering
\includegraphics[width=0.28\textwidth]{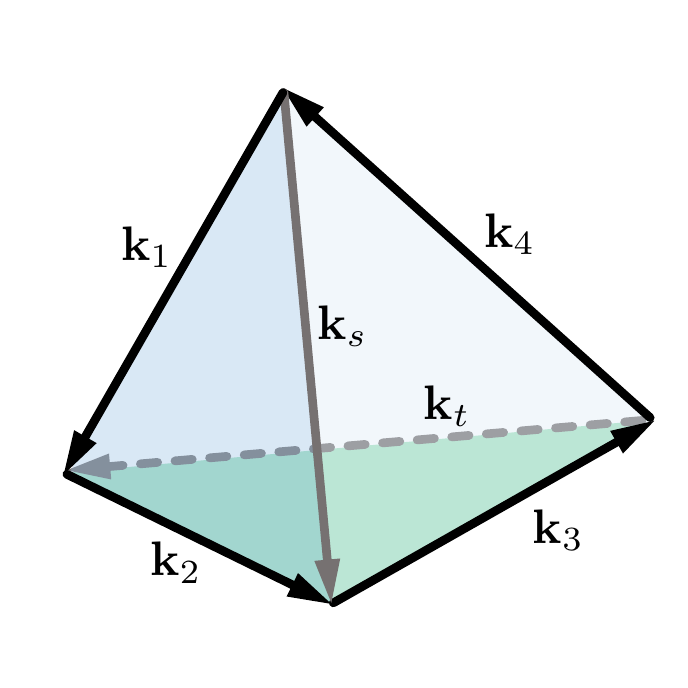}
\includegraphics[width=0.28\textwidth]{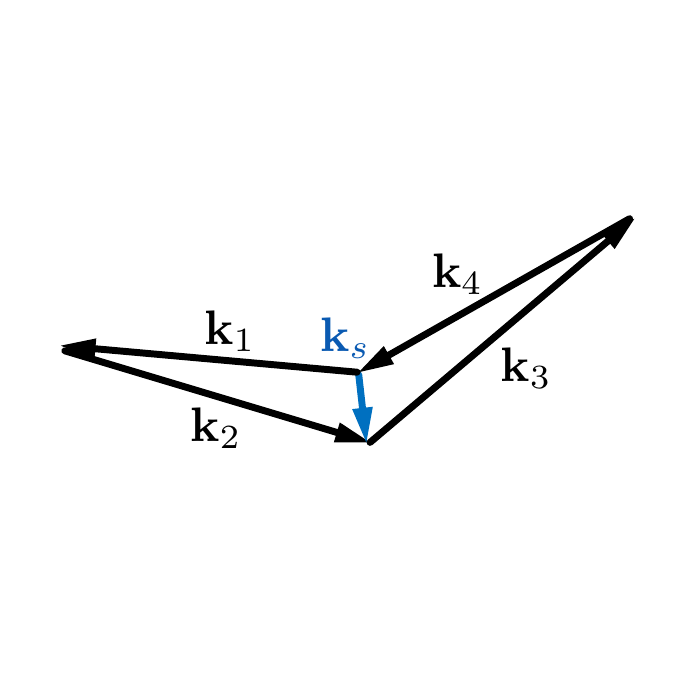}
\raisebox{-0.07\height}{\includegraphics[width=0.33\textwidth]{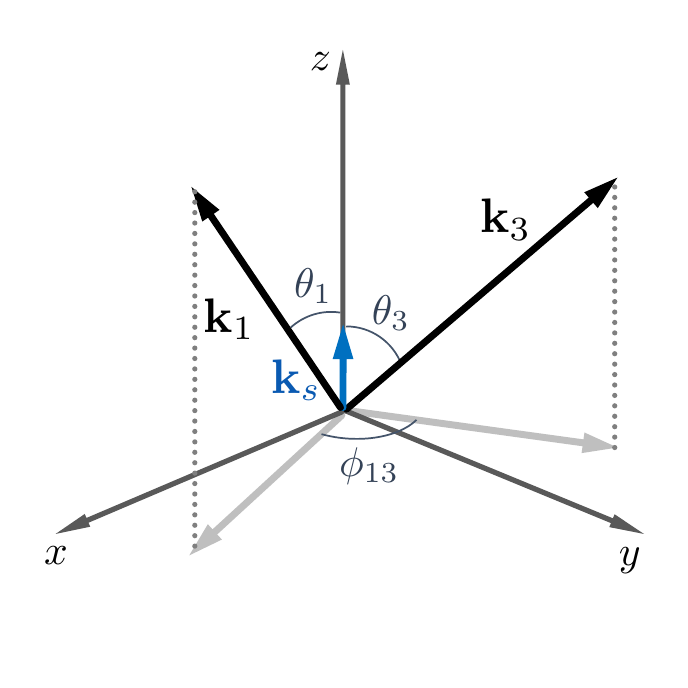}}
\caption{The momentum configuration of a 4-point function. Left: The six independent momentum variables; Middle: The squeezed limit; Right: The angles used in the text.}
\label{fig_4pt_config}
\end{figure}

The 3-dimensional space translation imposes the momentum conservation so that the correlator is proportional to a $\de$-function $\de(\sum \mb k_i)$. As a convention, this $\de$-function factor is always removed whenever we add a prime to the correlator $\la\cdots\ra'$. It is then clear that the four momenta form a closed tetragon in the 3-dimensional space, which can be identified as 4 edges of a tetrahedron. (See Fig.\ \ref{fig_4pt_config}.) The 3-dimensional rotation symmetry further imposes the condition that the correlator is invariant under a rigid 3d rotation of the tetrahedron. Therefore, the 4-point correlator is uniquely specified if we give the lengths of all 6 edges of the momentum tetrahedron, including the lengths of the four external momentum $k_i~(i=1,2,3,4)$, together with the lengths of the $s$-channel momentum $\mb k_s\equiv \mb k_1+\mb k_2$ and of the $t$-channel momentum $\mb k_t\equiv \mb k_1+\mb k_4$. That is, we can write
\begin{align}
\label{eq_varphi4vev}
  \la\varphi_{\mb k_1}\varphi_{\mb k_2}\varphi_{\mb k_3}\varphi_{\mb k_4}\ra'=\mathcal{T}(k_1,k_2,k_3,k_4,k_s,k_t).
\end{align}
We note in particular that knowing only the lengths of the four external momenta does not fix the correlator. 

In general, the 3d translations and rotations are the only unbroken spacetime symmetries in an inflationary universe. Although the spacetime metric is close to the dS, the other 4 dS isometries, namely the dilatation and the 3 dS boosts (or special conformal transformations), are broken by the rolling inflaton background ($\dot\phi\neq 0$). Nevertheless, the near scale invariance of the power spectrum tells us that the inflaton sector may respect approximately a shift symmetry $\phi\to \phi+\text{const.}$, and the rolling inflaton background breaks the dilatation and the shift symmetry down to their diagonal subgroup. That is, the whole system is invariant under the simultaneous transformations $\phi\to \phi+c$ and $t\to t-\dot\phi^{-1}c$. This can be viewed as a scale symmetry since the scale factor $a=e^{Ht}$ is transformed by a scaling $a\to e^{-Hc/\dot\phi}a$. The scale symmetry has the consequence that an $n$-point correlator scales as $1/K^{3n-3}$ under the simultaneous rigid scaling of all comoving quantities (such as the external momenta $\mb k_i$), where $K$ is any comoving momentum.\footnote{More explicitly, we have one $1/K^3$ from each of $n$ Fourier modes $\varphi_{\mb k}$, and another $1/K^{-3}$ from the removal of the $\de$-function of the 3-momentum conservation.} This implies, in particular, that the 4-point correlator can be written as a factor $\sim 1/K^9$, multiplied by a function $f$ that depends only on the \emph{shape} of the momentum tetrahedron. 

From the near scale invariance of the power spectrum, we expect that any interactions breaking the scale invariance would be slow-roll suppressed. Therefore, in this work, we only consider interactions that do not break the scale invariance. In particular, we will focus on interactions in which the inflaton fluctuation is \emph{derivatively} coupled. This type of derivative couplings automatically preserves the scale invariance. 

On the other hand, the 3 dS boosts, or the special conformal transformations, are spontaneously broken by the inflaton rolling background. We have no evidences for these three symmetries from observational data of primordial fluctuations. Therefore we should in general consider interactions that break the dS boosts. In particular, a recently studied class of models with chemical-potential-enhanced CC signals explicitly break the dS boosts \cite{Chen:2018xck,Liu:2019fag,Wang:2019gbi,Wang:2020ioa,Tong:2022cdz}.\footnote{ The breaking of dS boosts at the fluctuation level is explicit, but from the viewpoint of underlying Lorentz-invariant Lagrangian, the breaking is spontaneous, and is a consequence of rolling inflaton background.} The chemical potential originates from the dim-5 axion-type couplings between the inflaton field and a massive particle with nonzero spin \cite{Wang:2019gbi,Wang:2020ioa}. Such axion-type couplings are the leading order interactions between the inflaton and the spectator sector that preserve the shift symmetry. These couplings are also well motivated from particle physics considerations. The resulting signal features a much enhanced signal amplitude that is free from the usual Boltzmann suppression $e^{-\pi m/H}$, and thus are promising targets of future probes of CC signals. Therefore, we shall also consider these chemical-potential-enhanced but dS-boost-breaking signals in this paper.

With the above symmetry conditions plus simple dimensional analysis, we see that a 4-point correlator  has the following form: 
\begin{align}
  &\mathcal{T}(k_I)\equiv\la\varphi_{\mb k_1}\varphi_{\mb k_2}\varphi_{\mb k_3}\varphi_{\mb k_4}\ra'=\FR{M^4}{K^9}f(k_I/k_J), ~~~~~I,J\in\{1,2,3,4,s,t\},
\end{align}
where $M$ is a coefficient of \emph{physical} mass dimension 1, and $K^9$ represents any combination of $k_I$ with total power fixed at $9$ by the scale symmetry. The function $f$ is dimensionless, and depends only on momentum ratios.

\paragraph{Squeezed limit.} In a 4-point correlator, there are two distinct ways to take soft limits. One is to take an external momentum to zero, say, $k_4\to 0$, and is called the triangular limit. The other is to take one ``internal'' momentum to zero, e.g., $k_s \to 0$, and is called the squeezed limit.\footnote{The $k_s\to 0$ limit is also called the collapsed limit in the literature, while the term squeezed limit is usually given to the $k_3\to 0$ limit of a 3-point correlator. We choose to avoid the term ``collapsed limit,'' since it was also used to describe the colinear limit such as $k_s\to k_1+k_2$, which has different physical meaning from a soft limit. } In this paper we focus on the squeezed limit with $k_s\to 0$. CC signals appear in the squeezed limit due to $s$-channel exchange of massive particles, either at the tree level or at the loop level.  Below we use a well-understood tree level process \cite{Arkani-Hamed:2015bza,Arkani-Hamed:2018kmz} to illustrate the structure of the signal. 

In the squeezed limit $k_s\to 0$, it proves useful to use the following two variables:
\begin{align}
  &r_1\equiv\FR{k_s}{k_{12}},
  &&r_2\equiv\FR{k_s}{k_{34}},
\end{align}
where $k_{12}=k_1+k_2$ and $k_{34}=k_3+k_4$.
Now, consider a process with $s$-channel exchange of a massive particle $\si$ with mass $m$ and spin $s$. The oscillatory signal appears when the massive particle belongs to the principal-series representation of the dS's isometry group SO(4,1), that is, when $m>3H/2$ for $s=0$ and when $m >(s-\frac{1}{2})H$ for $s\geq 1/2$ \cite{Arkani-Hamed:2015bza}. 

When considering an $s$-channel exchange in the squeezed limit $k_s\to 0$, it is useful to reorganize the 4-point function in the following way:
\begin{align}
\label{eq_Tsqueezed}
  \mathcal{T}(k_I)=\mathcal{J}(k_1,k_3,k_s)\mathcal{K}(\theta_1,\theta_3,\phi_{13}).
\end{align}
Here $\mathcal{J}(k_1,k_3,k_s)$ is a dynamical piece that depends on the interactions and the dynamics of the intermediate particles, which we shall call the \emph{dynamical factor}. The oscillatory dependence in the momentum ratios is fully encoded in the dynamical factor. On the other hand, $\mathcal{K}(\theta_1,\theta_3,\phi_{13})$ is the \emph{kinematic factor} that depends only on various angles. The various angles in (\ref{eq_Tsqueezed}) are shown in the right panel of Fig.\;\ref{fig_4pt_config}, where we choose the direction of $\mb k_s$ as the $z$ axis. Then, $\theta_1$ and $\theta_3$ are the azimuthal angles of $\mb k_1$ and $\mb k_3$, respectively, and $\phi_{13}\equiv \phi_1-\phi_3$ is the longitude difference between $\mb k_1$ and $\mb k_3$. The angular dependence in the kinematic factor $\mathcal{K}$ contains the information about the spin of the intermediate particle, although the relation between the angular dependence and the spin of the intermediate particle could be ambiguous, at both the tree and the 1-loop levels.

The six-parameter set $(k_1,k_2,k_s,\theta_1,\theta_3,\phi_{13})$ also forms a complete set characterizing the shape of 4-momentum configurations. However, we note that the parameter dependence shown in (\ref{eq_Tsqueezed}) holds only in the squeezed limit. When we move away from the squeezed limit, $\mathcal{J}$ can also depends on $k_2$ and $k_4$, and $\mathcal{K}$ can also depend on $\theta_2$ and $\theta_4$. The separation of $\mathcal{T}$ into $\mathcal{J}$ and $\mathcal{K}$ is thus somewhat ambiguous.  

\paragraph{Signals in the squeezed limit.}
Now let us take a closer look at the dynamical factor $\mathcal{J}$, which is the main focus of this paper. For concreteness, let us consider a massive scalar $\si$ derivatively coupled to the inflaton fluctuation $\varphi$ via $\lam\varphi'^2\si$, where a prime denotes the derivative with respect to the conformal time $\tau=-1/(aH)$. The complete analytical result for this process has been obtained either by solving the conformal Ward identity \cite{Arkani-Hamed:2018kmz} or using the Mellin-Barnes representation \cite{Sleight:2019mgd}.\footnote{More precisely, \cite{Arkani-Hamed:2018kmz,Sleight:2019mgd} considered the fully dS-covariant interactions with the external fields being conformal scalars. The result with derivatively coupled external massless scalars can be obtained from the dS-covariant correlators by acting appropriate ``weight-shifting'' operators. See \cite{Arkani-Hamed:2018kmz,Baumann:2019oyu} for details.} The kinematic factor $\mathcal{K}=1$ in this case is trivial, and thus the whole 4-point function $\mathcal{T}=\mathcal{J}$. When taking the squeezed limit $k_s\to 0$, the 4-dynamical factor $\mathcal{J}$ breaks into three distinct pieces, defined by their analytical properties at $k_s=0$:
\begin{align}
\label{eq_TEFTLNL}
  &\mathcal{J}(k_1,k_2,k_3,k_4,k_s)=\mathcal{J}_\text{EFT} +\mathcal{J}_\text{L} +\mathcal{J}_\text{NL}.
\end{align}  
The first term $\mathcal{J}_\text{EFT}$ is fully analytic in both $r_1=k_s/k_{12}$ and $r_2=k_s/k_{12}$ when $k_s\to 0$, and possesses a Taylor expansion in terms of $r_1$ and $r_2$. We call this piece the \emph{EFT term}, since this term is preserved when we integrate out the intermediate particle $\si$. In the flat-space language, the various terms in the Taylor expansion in $r_1$ and $r_2$ in $\mathcal{J}_\text{EFT}$ simply correspond to various effective vertices from expanding the massive propagator $1/(\square-m^2)$ for large $m^2$. We also call $\mathcal{J}_\text{EFT}$ the \emph{background piece}, since it contains no oscillatory signals in $r_1$ or $r_2$, and thus constitutes the background for CC signals. 

The second and the third terms in (\ref{eq_TEFTLNL}) are \emph{signal} terms. They are both nonanalytic in $r_1$ and $r_2$ when sending $k_s$ to zero:
\begin{align}
  \label{eq_TL}
  &\lim_{k_s\to 0}\mathcal{J}_\text{L} =\FR{1}{(k_1k_3)^{9/2}}\mathcal{G}_\text{L}(r_1,r_2)\Big(\FR{r_1}{r_2}\Big)^{\ii\wt\nu}+\text{c.c.},\\
  \label{eq_TNL}
  &\lim_{k_s\to 0}\mathcal{J}_\text{NL} =\FR{1}{(k_1k_3)^{9/2}}\mathcal{G}_\text{NL}(r_1,r_2)\big(r_1r_2\big)^{\ii\wt\nu}+\text{c.c.}.
\end{align}
Here $\wt\nu$ is a dimensionless parameter which has one-to-one relationship with the mass $m$ for fixed spin $s$. When $s=0$, we have $\wt\nu\equiv\sqrt{m^2/H^2-9/4}$ and for $s\neq 0$, $\wt\nu\equiv\sqrt{m^2/H^2-(s-1/2)^2}$. For convenience, we call $\wt\nu$ the \emph{mass parameter} in this paper. The two functions $\mathcal{G}_\text{L}$ and $\mathcal{G}_\text{NL}$ are functions analytic in the $k_s\to 0$ limit, and allows a Taylor expansion in $r_1$ and $r_2$. For not-too-small $r_1$ and $r_2$, these functions can also depends on $k_2$ and $k_4$, which we omitted here but will spell out explicitly in specific examples in the next two sections.

As we can see, both $\mathcal{J}_L$ and $\mathcal{J}_\text{NL}$ contain nonanalytic dependences on various momentum ratios, in the form of oscillatory signals. However, the two pieces differ in their specific nonanalytic behaviors. The oscillatory signal in $\mathcal{J}_\text{L}$ appears when varying $k_{12}/k_{34}$. This part of signal is analytic in $k_s$. We call it the \emph{local signal} for the reason to be explained below. On the other hand, $\mathcal{J}_\text{NL}$ term is nonanalytic in $k_s$, and the signal in this term is from another momentum ratio $r_1r_2=k_s^2/k_{12}k_{34}$. We call it the \emph{nonlocal signal}.

The origin of the oscillatory signal is most easily seen by analyzing the propagator $D_>(k;\tau_1,\tau_2)=\la\si_{\mb k}(\tau_1)\si_{-\mb k}(\tau_2)\ra'$ of the intermediate massive particle $\si$. In an ordinary diagrammatic approach following the Schwinger-Keldysh path integral \cite{Chen:2017ryl}, the propagator $D_>(k;\tau_1,\tau_2)$ is a function of 3-momentum $k$ and the two times $\tau_1$ and $\tau_2$ of its two endpoints. For squeezed configurations, one can expand the propagator in the $k_s\tau_{1,2}\to 0$ limit. The result breaks into two parts:
\begin{align}
  &D_>(k;\tau_1,\tau_2)=D_\text{L}(k;\tau_1,\tau_2)+D_\text{NL}(k;\tau_1,\tau_2);\\
  &D_\text{L}(k;\tau_1,\tau_2)=\FR{H^2}{4\pi}(\tau_1\tau_2)^{3/2}\Gamma(-\ii\wt\nu)\Gamma(\ii\wt\nu)\bigg[e^{\pi\wt\nu}\Big(\FR{\tau_1}{\tau_2}\Big)^{\ii\wt\nu}+e^{-\pi\wt\nu}\Big(\FR{\tau_2}{\tau_1}\Big)^{-\ii\wt\nu}\bigg],\\
  &D_\text{NL}(k,\tau_1,\tau_2)= \FR{H^2}{4\pi}(\tau_1\tau_2)^{3/2}\bigg[\Gamma^2(-\ii\wt\nu)\Big(\FR{k^2\tau_1\tau_2}{4}\Big)^{\ii\wt\nu}+\Gamma^{2}(+\ii\wt\nu)\Big(\FR{k^2\tau_1\tau_2}{4}\Big)^{-\ii\wt\nu}\bigg].
\end{align}
When $k_s\to 0$, the \emph{local} part $D_\text{L}$ is analytic in $k_s$ while the \emph{nonlocal} part $D_\text{NL}$ is nonanalytic in $k_s$. The local part contains the EFT piece of the propagator. When $m\gg H$, the local part goes like $D_\text{L}\sim 1/m$, and reproduces the flat-space result.\footnote{Note that we are Fourier transforming the spatial coordinates to momentum space while leaving the time coordinate intact. In flat space, the propagator in this ``mixed'' representation is given by $e^{-\ii E(t_1-t_2)}/(2E)$ with $E=\sqrt{k^2+m^2}$, which goes like $1/m$ for large $m$. This differs from the 4-momentum representation where the propagator $-\ii/(k^2+m^2-\ii\ep)$ goes like $1/m^2$ in the large $m$ limit. See the appendix of \cite{Wang:2020ioa} for an example of computing flat-space correlators in this language. } On the other hand, the nonlocal part $D_\text{NL}$ goes like $e^{-\pi m/H}$ when $m\gg H$. This part cannot be captured by an EFT expansion, and represents the on-shell particle production in an inflationary spacetime. 

The three pieces of the 4-point function in (\ref{eq_TEFTLNL}) can be traced back to the separation of the propagator into $D_\text{L}$ and $D_\text{NL}$. In particular, the EFT part $\mathcal{J}_\text{EFT}$ is originated from the EFT part of $D_\text{L}$, namely the term in $D_\text{L}$ that goes like $1/m$ in the large $m$ limit. The local signal $\mathcal{J}_L$ is also originated from $D_\text{L}$. The analyticity of $D_\text{L}$ in the  limit $k_s\to 0$ translates to the same analyticity of the local signal $\mathcal{J}_\text{L}$ in the same limit. The nonanalytic behavior $(k_{12}/k_{34})^{\pm\ii\wt\nu}$ in the local signal is from the nonanalytic time dependence $(\tau_1/\tau_2)^{\pm\ii\wt\nu}$ in $D_\text{L}$. On the other hand, the nonlocal signal $\mathcal{J}_\text{NL}$ is totally from the nonlocal part of the propagator $D_\text{NL}$, and the both pieces share the same nonanalytic behavior in the squeezed limit $k_s\to 0$ . 

The discussion above is based on the squeezed limit $k_s\to 0$ of the propagator. However, the separation of the 4-point function into three parts (\ref{eq_TEFTLNL}) still applies when we move away from the squeezed limit. In the general situation, it is known that the signal is contributed by a suitably cut propagator, which is usually the part invariant under the exchange $\tau_1\leftrightarrow \tau_2$, although the symmetrization is a bit nontrivial in more general cases especially when there are boost-breaking chemical potentials \cite{Tong:2021wai}. Due to this cutting rule, the time ordering in the Schwinger-Keldysh diagrammatic calculation becomes irrelevant, and one can finish the time integral of each interaction vertex sepaprately. This essentially provides a cutting rule for computing the signal, similar to the cutting rule for on-shell propagators in flat-space QFT. The tree-level cutting rule for cosmological collider signals has been spelled out in \cite{Tong:2021wai}. See also \cite{Sleight:2021plv,Melville:2021lst,Goodhew:2021oqg} for related studies.  

In 4-point correlators, one can exploit the two independent variables $r_1$ and $r_2$ to extract both $\mathcal{J}_\text{L}$ and $\mathcal{J}_\text{NL}$ independently from the full result. Therefore, we can treat $\mathcal{J}_\text{L}$ and $\mathcal{J}_\text{NL}$ as two separated observables. However, in the 3-point function which is more often discussed, there is only one independent soft limit $k_3/k_1\to 0$, and one can no longer separate the local signal from the nonlocal signal. This is most easily seen by taking the triangular limit of a 4-point correlator: By sending $k_4\to 0$, we see that $k_s\simeq k_3$ and $r_2\simeq 1$. So, there is only one ratio $r_1\simeq k_3/k_1$ left. The local signal $\propto (r_1/r_2)^{\ii\wt\nu}$ and the nonlocal signal $\propto(r_1r_2)^{\ii\wt\nu}$ thus become inseparable.

The oscillatory signals also appear in loop diagrams. Many models produce observably large signals starting at 1-loop level, and many of them break the dS boost symmetry. However, much less is known about the analytical structure of 1-loop processes in inflationary background, especially the ones that break dS boost. Numerical study shows that a 1-loop 4-point correlator in the squeezed limit still possess a background piece and an oscillatory signal piece \cite{Wang:2021qez}. However, the spin information in the loop particle is normally obscured in the loop processes. The angular dependence in the kinematic factor $\mathcal{K}$ reveals the \emph{total angular momentum} of the two loop modes rather than the spin of each individual particle in the loop. Therefore, one cannot rely on the angular dependence to extract the spin of the intermediate particle. \emph{It is a main purpose of this paper to show that one can exploit the phase of the oscillatory signal in the loop process to learn more information about the intermediate particle including its spin and the interaction.} For this purpose, we need to compute the nonlocal signal at 1-loop level, which turns out to be achievable using the Mellin-Barnes representation.

\paragraph{Parameters of the nonlocal signal.} The main point of this paper is to show the use of phase information in the oscillatory signals. Throughout the paper, we focus on the nonlocal signal $\mathcal{J}_\text{NL}$ in (\ref{eq_TNL}) in the $s$-channel. Now, let us look more closely at the structure of the nonlocal signal. Quite generally, we can put $\mathcal{J}_\text{NL}$ into the following form:
\begin{align}
\label{eq_JNLpar}
&\mathcal{J}_\text{NL}(r_1,r_2)=\FR{\mathcal{A}}{K^9}\mathcal{G}_\text{NL}(r_1,r_2)\big(r_1r_2\big)^{\al+\ii\omega}+\text{c.c.}, 
\end{align}
where $K^9$ represents a product of momenta from $\{k_1,k_2,k_3,k_4,k_{12},k_{34}\}$, with total power 9 fixed by the dimension counting. We stress that our convention is that this prefactor is independent of the intermediate momentum $k_s$. We explain the rest functions and parameters in (\ref{eq_JNLpar}) as below.
\begin{enumerate}
  \item The amplitude $\mathcal{A}$ is a momentum-independent positive real number that accounts for the coupling dependence and also various numerical factors of $2$'s and $\pi$'s.  It is also our convention that this amplitude is chosen positive and real, and any possible complex phase is moved into the function $\mathcal{G}_\text{NL}$. 
  
  \item The function $\mathcal{G}_\text{NL}$ can depend on momentum ratios $r_1$ and $r_2$, and it is analytic in $r_1$ and $r_2$ in the squeezed limit $k_s\to 0$. This function arises from integrating over the time variables in the Schwinger-Keldysh integrals, and contains nontrivial information about the intermediate particle. This information can be extracted by measuring the signal phase, as we shall show. The analyticity of $\mathcal{G}_\text{NL}$ in $r_1$ and $r_2$ means that it goes to a constant $\mathcal{G}_\text{NL}(0,0)$ when $k_s\to 0$. Away from the squeezed limit, the function $\mathcal{G}_\text{NL}$ may also develop dependence on other momentum variables such as $k_2$ and $k_4$.

  \item The parameter $\al$ is a real number, and describes the scaling behavior of the signal in the squeezed limit. Physically, it is from the comoving dilution of the particle, although it can also receive corrections from other physical effects. For example, a tree-level exchange of scalar particle gives $\al=0$, which corresponds to the comoving dilution of the scalar mode $\si_{\mb k}\sim (-k\tau)^{3/2}$ as $\tau\to 0$.\footnote{The number density dilutes as $|\si_{\mb k}|^2\sim (-k\tau)^3\sim a^{-3}$ where $a$ is the scale factor. This is the reason that we call it the comoving dilution.} A 1-loop process mediated by a pair of scalars is doubly diluted, and thus $\al=3/2$. There are other possible physics entering $\al$. For instance, an $s$-channel exchange of a massive spin-1 particle would give $\al=1$ due to the cancelation of the leading $P$-wave contribution. (See the next section for details.) Furthermore, a loop-induced mass correction to the intermediate particle can introduce \emph{noninteger} correction to $\al$ \cite{Lu:2021wxu}, a fact sometimes interpreted as particle decay width \cite{Arkani-Hamed:2015bza}. 
  
  \item Last, the parameter $\omega$ is a positive real number and controls the (dimensionless) frequency of the oscillatory signal. Very often it is directly related to the mass of the intermediate signal. In tree level processes, we have $\omega=\wt\nu$, and in 1-loop processes, we have $\omega=2\wt\nu$. This parameter may as well receive other contributions. First, there are various possible sources of mass corrections during the inflation, and these effect correct $\omega$ directly \cite{Chen:2016uwp,Chen:2016hrz}. There are also examples beyond a mass correction. For instance, in a 1-loop process mediated by a fermion of mass $m$ and enhanced by a parity-odd chemical potential $\mu$, the oscillatory frequency is given by $\omega=2\sqrt{m^2+\mu^2}/H$ \cite{Chen:2018xck,Wang:2019gbi}.
  
\end{enumerate}

It is clear that the signal (\ref{eq_JNLpar}) can be rewritten as
\begin{keyeqn}
\begin{align}
\label{eq_JNLpar2}
&\mathcal{J}_\text{NL}(r_1,r_2)=\FR{2\mathcal{A}}{K^9}(r_1r_2)^\al \big|\mathcal{G}_\text{NL}(r_1,r_2)\big|\cos\Big[\omega\log(r_1r_2)+\vartheta(r_1,r_2)\Big], \\&\vartheta(r_1,r_2)\equiv\text{Arg}\,\mathcal{G}_\text{NL}(r_1,r_2). 
\end{align}
\end{keyeqn}
In the next two sections, we are going to compute the phase term, Arg\;$\mathcal{G}_\text{NL}(r_1,r_2)$, for several tree and 1-loop processes. The phase of a complex number Arg\;$z$ is defined up to addition of multiples of $2\pi$, and our convention is that Arg\;$z$ takes value in $(-\pi,\pi]$. 

We conclude the general discussion of this section by two remarks about the phase. First, we note again that the phase of $\mathcal{G}_\text{NL}$ is measured with respect to a well-defined original point $r_1=r_2=1$, and thus is unambiguous from both theoretical and observational points of view. It can therefore be a useful observable that connects theory with observations. Second, we note that there can be nontrivial $r_1$ and $r_2$ dependence in the phase, making the oscillations not strictly sinusoidal even in the dS limit. Therefore, calculating precisely the $r_1$ and $r_2$ dependence in the phase, on the one hand, is important for high-quality template making, and on the other hand reveals more information that cannot be seen in the deeply squeezed limit.

\section{Phase Information at the Tree Level}
\label{sec_tree}

Before tackling the main problem of this work, the loop processes, we first illustrate in this section the use of phase information for tree-level processes. For typical tree level processes, the mass and spin of a massive particle can be more directly read from the corresponding oscillating signal, and in typical cases the phase does not provide independent new information. Therefore, measuring the phase can serve as a useful consistency check. The tree-level cosmic correlators are relatively better understood than loop processes. In particular, the oscillatory signals in tree level processes usually have closed analytical expression in terms of special functions, as we shall show. 

\subsection{Massive Scalar}

\begin{figure}[tbph]
\centering
  \parbox{0.35\textwidth}{\includegraphics[width=0.31\textwidth]{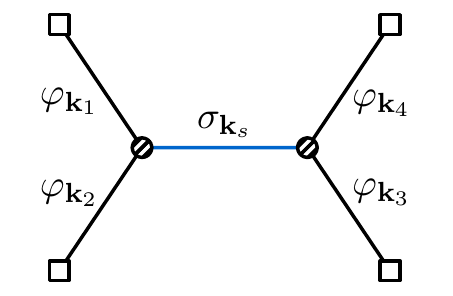}}
\caption{The 4-point function mediated by a massive scalar boson $\si$ at tree level.}
\label{fig_scalar_tree}
\end{figure}

To be concrete, we consider the 4-point function of the inflaton fluctuation $\varphi$ mediated by a scalar field $\si$ of mass $m\geq 3H/2$ in the $s$-channel, shown in Fig.\ \ref{fig_scalar_tree}. Our convention for drawing diagrams follows \cite{Chen:2017ryl}. As will be clear, the phase of the CC signal is also dependent on the form of the coupling between $\varphi$ and $\si$. As mentioned in the previous section, we require that the inflaton is derivatively coupled. Then, the lowest dimension couplings contributing to the process in Fig.\ \ref{fig_scalar_tree} are of dimension 5, and consist of the following two terms:
\begin{align}
\label{eq_lTlS}
  \Delta\ld=\FR{1}{2}\lam_T a^{2}\varphi'^2\si+\FR{1}{2}\lam_S a^2(\pd_i\varphi)^2\si.
\end{align}
Here $\lam_T$ and $\lam_S$ are couplings of dimension $-1$, and are independent parameters \emph{a priori}, at the level of inflaton perturbations. However, there are two important cases that are of main interest in CC model buildings:
\begin{enumerate}
  
  \item The purely time-derivate coupled case: $\lam_T\neq 0$ and $\lam_S=0$. There are multiple ways to get this coupling from an underlying model. An often appeared case is that the inflaton $\phi$ couples to a third scalar field $\chi$ via $\sqrt{-g}(\pd_\mu\phi)^2\chi$, and $\chi$ couples to $\si$ through a direct coupling $\chi^2\si$. After being evaluated with the inflaton background, the $\phi$-$\chi$ coupling produces a two-point mixing between $\varphi$ and $\chi$ in the form of $a^3\varphi'\chi$. If the field $\chi$ is heavy, we can integrate it out and this produces an effective coupling $\varphi'^2\si$. \footnote{As one can see in this example and more examples below, integrating out heavy fields is a convenient way to introduce effective inflaton couplings. Very often, the resulting effective couplings would produce additional EFT shapes, corresponding to $\mathcal{J}_\text{EFT}$ in (\ref{eq_TEFTLNL}). We note that, at least in principle, theses EFT shapes can always be separated from the oscillatory signals due to the distinct momentum dependence. Practically, it is possible to distill an oscillatory signal out of a much larger EFT ``background'' using various filtering techniques, as demonstrated in \cite{Wang:2021qez}.}

  \item The Lorentz covariant couplings $-\lam_T=\lam_S\equiv \lam$. This choice appears naturally from an underline coupling $ \lam\sqrt{-g}(\pd_\mu\phi)^2\si$ evaluated with the inflaton background $\phi=\phi_0+\varphi$.
  
\end{enumerate}

The nonlocal signal $\mathcal{T}_\text{NL}$ in the squeezed limit contributed by tree-level $s$-channel exchange of $\si$ via the couplings (\ref{eq_lTlS}) can be calculated in closed form, following the rules reviewed in \cite{Chen:2017ryl}. The calculation of this diagram is put in App.\ \ref{app_details}, and the result can be put into a factorized form:
\begin{align}
  \mathcal{T}_\text{NL}=\FR{\big(1+\ii\sinh\pi\wt\nu\big)\Gamma^2(-\ii\wt\nu)}{32\pi k_1k_2k_3k_4(k_{12}k_{34})^{5/2}}\mathcal{B}(k_1,k_2,k_s)\mathcal{B}(k_3,k_4,k_s)\Big(\FR{r_1r_2}{4}\Big)^{\ii\wt\nu}+\text{c.c.},
\end{align}
where
\begin{align}
\label{eq_Bfunction}
  &\mathcal{B}(k_1,k_2,k_s)= \lam_T \Gamma\Big(\FR{5}{2}+\ii\wt\nu\Big) {}_2\mathrm{F}_1\left[\bgm \fr{5}{4}+\fr{\ii\wt\nu}{2},\fr{7}{4}+\fr{\ii\wt\nu}{2} \\ 1+\ii\wt\nu \edm \middle|\, \FR{ k_s^2}{ k_{12}^2}\right]\n\\
  &-\lam_S\FR{ k_s^2- k_1^2- k_2^2}{2 k_1 k_2}\FR{ k_{12}^2}{ k_1 k_2}\Gamma\Big(\FR{1}{2}+\ii\wt\nu\Big)\bigg\{\bigg(1-\FR{ k_1 k_2}{ k_{12}^2- k_s^2}(\fr14+\wt\nu^2)\bigg)
{}_2\mathrm{F}_1\left[\bgm\fr{1}{4}+\fr{\ii\wt\nu}{2},\fr{3}{4}+\fr{\ii\wt\nu}{2}\\1+\ii\wt\nu\edm\middle|\,\FR{ k_s^2}{ k_{12}^2}\right]
\n\\
&~+\bigg(1+\FR{2 k_1 k_2}{ k_{12}^2- k_s^2}\bigg)\Big(\FR{1}{2}+\ii\wt\nu\Big){}_2\mathrm{F}_1\left[\bgm\fr{3}{4}+\fr{\ii\wt\nu}{2},\fr{5}{4}+\fr{\ii\wt\nu}{2}\\1+\ii\wt\nu\edm\middle|\,\FR{ k_s^2}{ k_{12}^2}\right]\bigg\}.
\end{align} 
Here we wrote $ k_{12}\equiv k_1+ k_2$ for brevity, and ${}_2\text{F}_1$ is the hypergeometric function. This somewhat complicated expression gets much simplified if we consider the squeezed limit $k_s\to 0$. In this limit we can make the approximations $ k_1\simeq k_2\simeq  k_{12}/2$, $ k_3\simeq k_4\simeq  k_{34}/2$, and $ k_s\simeq 0$. Then, we have
\begin{align}
  \lim_{k_s\to 0}\mathcal{T}_\text{NL}=\FR{\big(1+\ii\sinh\pi\wt\nu\big)\Gamma^2(-\ii\wt\nu)}{1024\pi( k_1 k_3)^{9/2} }\mathcal{B}( k_1, k_1,0)\mathcal{B}( k_3, k_3,0)\Big(\FR{r_1r_2}{4}\Big)^{\ii\wt\nu}+\text{c.c.},
\end{align}
where
\begin{align}
    &\mathcal{B}( k , k ,0)= \lam_T \Gamma\Big(\FR{5}{2}+\ii\wt\nu\Big) +\lam_S\Big(\FR{27}{4}+6\ii\wt\nu-\wt\nu^2  \Big) \Gamma\Big(\FR{1}{2}+\ii\wt\nu\Big) .
\end{align}
Then it is straightforward to read the phase of oscillatory signal from this expression. As explained above, we shall not consider arbitrary $\lam_T$ and $\lam_S$, but only focus on the two important cases. Below we present the phase terms for these two cases, respectively.

\paragraph{Case 1: Time-derivative coupling.} In this case we set $\lam_S=0$. Then we have
\begin{align}
  \lim_{k_s\to 0}\mathcal{T}_\text{NL}= \FR{\lam_T^2\big(1+\ii\sinh\pi\wt\nu\big)\Gamma^2(-\ii\wt\nu)}{1024\pi( k_1 k_3)^{9/2} }\Gamma^2\Big(\FR{5}{2}+\ii\wt\nu\Big)\Big(\FR{r_1r_2}{4}\Big)^{\ii\wt\nu}+\text{c.c.},
\end{align}
Comparing this result with (\ref{eq_JNLpar2}), we find, in the squeezed limit:
\begin{keyeqn}
\begin{align} 
 \lim_{k_s\to 0} \vartheta_{\varphi'^2\si} =\,\text{Arg}\,\bigg[2^{-2\ii\wt\nu}\big(1+\ii\sinh\pi\wt\nu\big)\Gamma^2(-\ii\wt\nu)\Gamma^2\Big(\FR{5}{2}+\ii\wt\nu\Big)\bigg]. 
\end{align}
\end{keyeqn}
The case of time-derivative coupling is actually simple enough that we can easily write down the full result of the phase without taking the squeezed limit:

\begin{align} 
\label{eq_thetaTr1r2}
  \vartheta_T(r_1,r_2) =\,\text{Arg}\,\bigg[2^{-2\ii\wt\nu}\big(1+\ii\sinh\pi\wt\nu\big)\Gamma^2(-\ii\wt\nu)\Gamma^2\Big(\FR{5}{2}+\ii\wt\nu\Big)\mathbf{F}_{\wt\nu}(r_1)\mathbf{F}_{\wt\nu}(r_2)\bigg],
\end{align}
where
\bge
\label{eq_Fbold}
   \mathbf{F}_{\wt\nu}(r)\equiv{}_2\mathrm{F}_1\left[\bgm \fr{5}{4}+\fr{\ii\wt\nu}{2},\fr{7}{4}+\fr{\ii\wt\nu}{2} \\ 1+\ii\wt\nu \edm \middle|\, r^2\right].
\ede

\begin{figure}[t]
\centering
\includegraphics[width=0.45\textwidth]{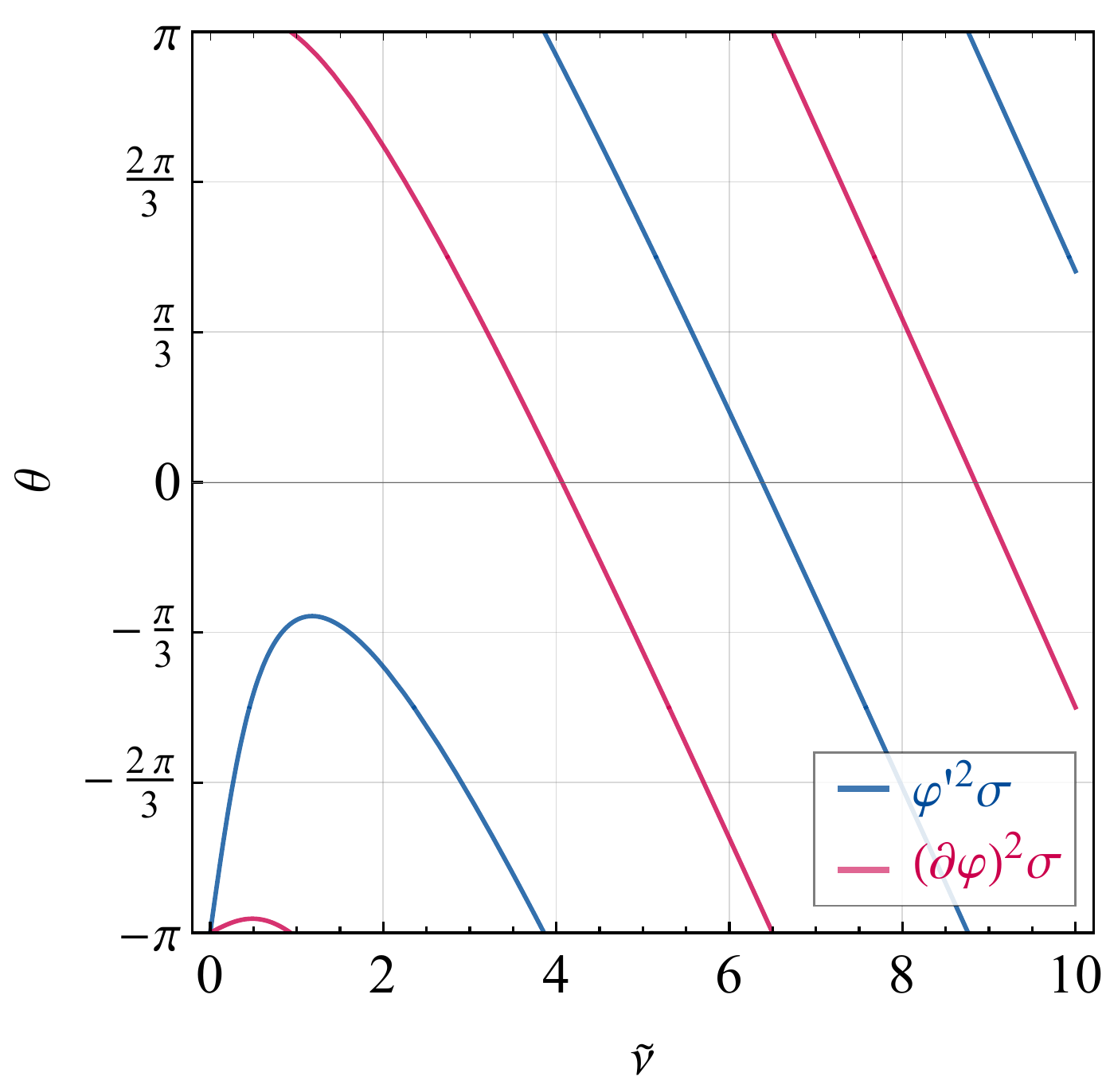} \hspace{5mm}
\includegraphics[width=0.45\textwidth]{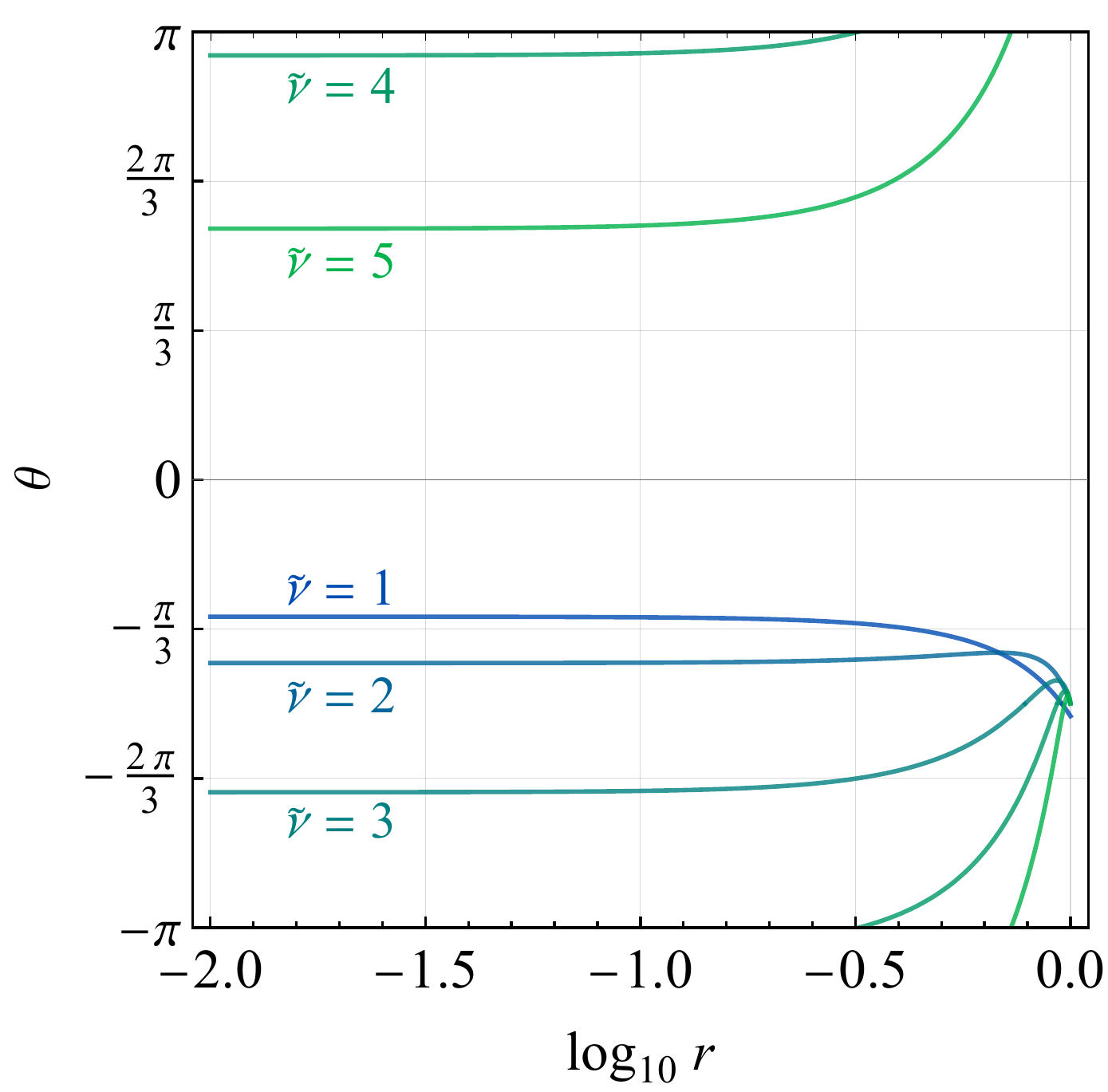} 
\caption{The phases $\theta_T(r,r)$ of the nonlocal signal from a tree-level exchange of a massive scalar. The left panel shows the phases in the squeezed limit $r\to 0$ as functions of scalar mass parameter $\wt\nu=\sqrt{m^2/H^2-9/4}$, for both time-derivative coupling (blue) and Lorentz covariant coupling (magenta). The right panel shows the phases from time-derivative coupling 
 as functions of momentum ratio $r$, for several values of $\wt\nu$.}
\label{fig_phasevsr}
\end{figure}

From the above result we see that the phase in the squeezed limit is completely determined by the parameter $\wt\nu$, namely, by the mass of the intermediate particle. See Fig.\ \ref{fig_phasevsr}. Since the parameter $\wt\nu$ can be independently measured from the frequency of the oscillatory signal, the information from the above phase can be used as a consistency check. When we look away from the very squeezed limit and consider small but non-negligible $r_1$ and $r_2$, the phase picks up dependences on $r_1$ and $r_2$ in the form of power series.  In the current case, the power series is summed into hypergeometric functions.  This correction would be important when extracting the phase information from not-so-squeezed configurations. 

It is useful to check how squeezed a configuration we need to see an $r$-independent phase. For this purpose we Taylor expand (\ref{eq_Fbold}) and get
\bge
  \mathbf{F}_{\wt\nu}(r)=1+\FR{(7+2\ii\wt\nu)(5+2\ii\wt\nu)}{16(1+\ii\wt\nu)}r^2+\order{r^4}.
\ede
An $r$-insensitive phase requires that the $r^2$-term is much smaller than the leading term 1. For CC signals, we typically expect that $\wt\nu$ is an $\order{1}$-ish number, since larger $\wt\nu$ would give a Boltzmann suppression factor $e^{-\pi\wt\nu}$ to the signal. Therefore, the coefficient of $r^2$-term is generally $\order{1}$, showing that we only need to consider mildly squeezed configurations to see constant phase. In Fig.\;\ref{fig_phasevsr}, we set $r_1=r_2=r$ and plot $\theta(r,r)$ as a function of $r$ for several choices of $\wt\nu$. As is clear from the plot, the phase becomes essentially a constant when $r<0.1$. The deviation from a constant phase becomes obvious when $r>0.1$, which can be important for constructing templates of CC signals.

\paragraph{Case 2: Lorentz covariant coupling.} In this case we take $\lam_T=-\lam_S=-\lam$. The $\mathcal{B}$ function defined in (\ref{eq_Bfunction}) now takes the following simple form:
\bge
  \mathcal{B}( k_1, k_2, k_s)= \lam(6+4\ii\wt\nu)\Gamma\Big(\FR{1}{2}+\ii\wt\nu\Big).
\ede
Hence the nonlocal signal in the squeezed limit is
\begin{align}
  \lim_{k_s\to 0}\mathcal{T}_\text{NL}= \FR{\lam^2\big(1+\ii\sinh\pi\wt\nu\big)(6+4\ii\wt\nu)^2\Gamma^2(-\ii\wt\nu)}{1024\pi( k_1 k_3)^{9/2} }\Gamma^2\Big(\FR{1}{2}+\ii\wt\nu\Big)\Big(\FR{r_1r_2}{4}\Big)^{\ii\wt\nu}+\text{c.c.}.
\end{align}
Therefore, the phase from the Lorentz covariant coupling in the squeezed limit is
\begin{keyeqn}
\begin{align} 
  \lim_{k_s\to 0}\vartheta_{(\pd\varphi)^2\si} =\,\text{Arg}\,\bigg[2^{-2\ii\wt\nu}\big(1+\ii\sinh\pi\wt\nu\big)(6+4\ii\wt\nu)^2\Gamma^2(-\ii\wt\nu)\Gamma^2\Big(\FR{1}{2}+\ii\wt\nu\Big)\bigg]. 
\end{align}
\end{keyeqn}
While this result also depends only on the mass parameter $\wt\nu$, the functional dependence is different from the case of time-derivative coupling, showing that the phase also encodes information of the interaction vertices. In the current case, both the time-derivative coupling and the space-derivative coupling contribute to the process. In the squeezed limit, both couplings lead to identical kinematic structure in the nonlocal signal, so that the whole result is a direct superposition of two couplings. However, once we move away from the squeezed limit, the space-derivative coupling can lead to much more complicated dependence on various momenta, as shown in (\ref{eq_Bfunction}). In principle, one can use this information to lift the degeneracy of two couplings in the squeezed limit, so that the signals from each of the two interactions can be measured separately.

\subsection{Massive gauge boson with nonzero chemical potential}

Now we consider a more interesting case where the tree-level 4-point function is mediated by a massive spin-1 gauge boson in the $s$-channel. We introduce a nonzero chemical potential to the gauge boson to enhance the signal strength. The parity-violating chemical potential from the axion-type coupling $\phi F\wt F$ has been studied in the context of CC physics \cite{Liu:2019fag,Wang:2019gbi,Wang:2020ioa,Wang:2021qez}.  

At the level of the perturbations, the quadratic Lagrangian for a massive gauge boson $A_\mu$ is:
\begin{align}
\label{eq_ALagrangian}
  \Delta\ld = -\FR{1}{4}F_{\mu\nu}F^{\mu\nu}-\FR{1}{2}a^2m^2A_\mu A^\mu-\FR{1}{2}\mu\ep^{ijk}A_i F_{jk}.
\end{align}
Here we have factored out all scalar factors $a$ and thus all indices are raised or lowered by Minkowski metric $\eta_{\mu\nu}$ or its spatial part $\de_{ij}$. The chemical potential term $-\mu\ep^{ijk}A_i F_{jk}/2$ can be derived from the Lorentz invariant dim-5 coupling $\phi F\wt F/(4\Lambda)$ with the rolling inflaton background $\phi=\dot\phi_0 t$, and the chemical potential $\mu=\dot\phi_0/\Lambda$. We refer the readers to \cite{Wang:2020ioa} for details about massive spin-1 boson with nonzero chemical potential.

\begin{figure}[t]
\centering  
  \parbox{0.31\textwidth}{\includegraphics[width=0.31\textwidth]{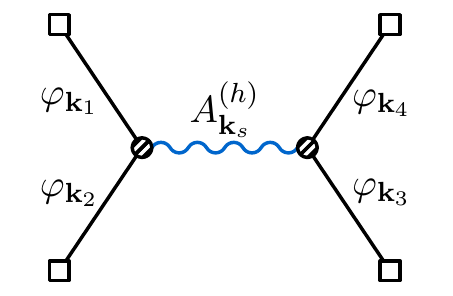}}
  \parbox{0.31\textwidth}{\includegraphics[width=0.31\textwidth]{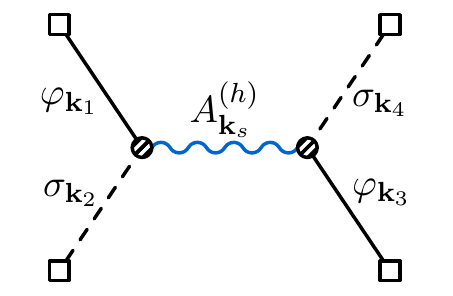}}
  \parbox{0.31\textwidth}{\includegraphics[width=0.31\textwidth]{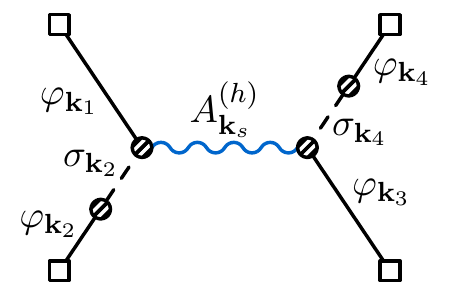}}
\caption{The 4-point function mediated by a massive spin-1 boson $A_\mu$ at tree level.}
\label{fig_vector_tree}
\end{figure}

To have a tree-level exchange of gauge boson, we also need an interaction term that is linear in the gauge field $A_\mu$. There are several possible choices. The simplest choice respecting the shift symmetry of the inflaton fluctuation may be the following one:
 \bge
 \label{eq_phiphiA}
  \Delta\ld = \FR{1}{2}\lam a\varphi'(\pd^i\varphi) A_i.
\ede
This coupling can generate a tree-level process shown in the left diagram in Fig.\;\ref{fig_vector_tree}. There is nothing wrong with this coupling, and we shall show below that it does generate a nonzero signal in the squeezed limit. The only concern is that this signal decreases faster than naively expected: Phrased in terms of the $\al$ parameter defined in (\ref{eq_JNLpar}), the signal from this vertex has $\al=1$ rather than $\al=0$ that one would naively expect for a tree-level process. The reason of this faster decay is the cancelation of identical external states when permuting the two external momenta on each side of the $s$-channel propagator, as required by the Feynman rule. As we shall see, the canceled leading order contribution has the familiar $P$-wave shape, $\mathcal{K}\propto \cos\theta_1\cos \theta_3$ or $\sin\theta_1\sin \theta_3$. After the cancelation, the remaining leading term would generate a $D$-wave shape, $\mathcal{K}\propto \cos 2\theta_1\cos 2\theta_3$ or $\sin 2\theta_1\sin 2\theta_3$. (See Fig.\;\ref{fig_4pt_config} for our definition of the angles.)

It is known that one has to use two real scalar degrees for the external states in order to see the $P$-wave contribution \cite{Arkani-Hamed:2018kmz,Liu:2019fag}. Therefore, we shall also consider the following coupling with a second scalar field $\si$ that could be either massive or massless:
\bge
\label{eq_phisiA}
  \Delta\ld = \lam a\varphi'(\pd^i\si) A_i.
\ede
This coupling can generate a two-scalar process $\la\varphi_{\mb k_1}\si_{\mb k_2}\varphi_{\mb k_3}\si_{\mb k_4}\ra'$, shown in the middle diagram of Fig.\;\ref{fig_vector_tree}. The momentum permutation is absent in this case, and the result has the familiar $P$-wave form, with $\al=0$. As we shall see, this process differs from the one-scalar amplitude (left diagram of Fig.\;\ref{fig_vector_tree}) only in the values of $\al$ and the kinematic factor $\mathcal{K}$. The phase of the two processes are identical, up to the difference from a possible overall sign flip.

In realistic CC phenomenology, only the inflaton fluctuation $\varphi$ is observed. In order to produce an observable signal from the above coupling, we need to introduce a two-point mixing in the form of $\varphi'\si$ which converts external $\si$ fields to $\varphi$. The corresponding diagram is the right diagram in Fig.\;\ref{fig_vector_tree}.
The phenomenology of this process has been investigated in \cite{Liu:2019fag}. Our result for the middle diagram in Fig.\;\ref{fig_vector_tree} can be applied to the right diagram in Fig.\;\ref{fig_vector_tree} by including an $r_{1,2}$-independent overall phase from the two-point mixing. Incidentally, we note that, we can integrate out $\si$ in the right diagram of Fig.\;\ref{fig_vector_tree} if $\si$ is heavier than the Hubble scale. This  will recover the left diagram of Fig.\;\ref{fig_vector_tree}, with the coupling given precisely by (\ref{eq_phiphiA}).

The chemical potential $\mu$ of the spin-1 boson affects the mode functions in a helicity dependent way \cite{Wang:2020ioa}. As a result, different helicity states contribute differently to the signal.\footnote{Note that the chemical potential term in (\ref{eq_ALagrangian}) breaks dS boosts. Consequently, no symmetry transformation can bring different helicity states into each other. So, the helicity is unambiguously defined even for massive particles.} Furthermore, the angular dependences, encoded in the kinematic factor $\mathcal{K}$, are also different for different helicity states. Therefore, one can in principle measure the signal from each helicity state separately. So, it is useful to present the result from each single polarization state with $h=(+,-,L)$, which we shall do next. We shall first give the complete results without assuming the squeezed limit, and then examine their behaviors in the squeezed limit.

\paragraph{Full results.} The details of computing the nonlocal signal of a massive spin-1 boson is given in App.\ \ref{app_details}. Here we present the results. First, we present the full result for the nonlocal signal in the two-scalar amplitude $\la\varphi_{\mb k_1}\si_{\mb k_2}\varphi_{\mb k_3}\si_{\mb k_4}\ra'$ with massless $\si$, denoted by $\mathcal{T}_\text{NL,2}^{(h)}$, and represented by the middle diagram of Fig.\ \ref{fig_vector_tree}. For the two transversely polarized states $h=\pm$, the result is:
\begin{align}
 \mathcal{T}_\text{NL,2}^{(\pm)}(k_I) =&  -\FR{\lam^2e^{\mp\pi\wt\mu}(\cosh 2\pi\wt\mu+\cosh2\pi\wt\nu)}{32\pi^2 k_1\cdots k_4(k_{12}k_{34})^{3/2}}\FR{\big(\wh{\mb k}_2 \cdot \mb e_\pm(\wh{\mb k}_s)\big)\big(\wh{\mb k}_4 \cdot \mb e_\pm^{*}(\wh{\mb k}_s)\big)}{k_2k_4}\n\\
&\times\Big[\mb{J}(u_2,u_4,r_1,r_2)\theta(r_2-r_1)+\mb{J}(u_4,u_2,r_2,r_1)\theta(r_1-r_2)\Big]
\end{align}
where $\mb e_\pm$ is the two transverse polarization vectors for the spin-1 field, $\wt\nu=\sqrt{m^2/H^2-1/2}$ is the mass parameter for the spin-1 boson of mass $m$, $\wt\mu=\mu/H$ is the dimensionless chemical potential, $u_2=k_2/k_{12}$, $u_4=k_4/k_{34}$, and the function $\mb{J}(u_2,u_4,r_1,r_2)$ is defined by
\begin{align}
  \mb{J}(u_2,u_4,r_1,r_2)=&~ \Big[\Big({\mathbf{G}}_{\pm \wt\mu,\wt\nu}(u_2,r_1){\mathbf{G}}_{\mp \wt\mu,\wt\nu}(u_4,r_2)-\ii {\mathbf{G}}_{\pm \wt\mu,\wt\nu}(u_2,e^{-\ii\pi}r_1){\mathbf{G}}_{\mp \wt\mu,\wt\nu}(u_4,r_2)
 \Big)
 \n\\
 &~+(\wt\nu\to-\wt\nu)\Big]+\text{c.c.}.
\end{align}
The function ${\mathbf{G}}_{\wt\mu,\wt\nu}(u,r)$ is defined by
\begin{align}
\label{eq_Gmunu}
  & {\mb G}_{\wt\mu,\wt\nu}(u,r) =  \FR{1}{(1+r)^{3/2}}\Big(\FR{2r}{1+r}\Big)^{\ii\wt\nu}\Gamma(-2\ii\wt\nu)\Gamma\Big(\FR32+\ii\wt\nu\Big)\Gamma\Big(\FR12-\ii\wt\mu+\ii\wt\nu\Big)\n\\
 &\times \bigg\{{}_2\mathrm{F}_1\left[\bgm\fr{3}{2}+\ii\wt\nu,\fr12-\ii\wt\mu+\ii\wt\nu\\1+2\ii\wt\nu\edm\middle|\,\FR{2r}{1+r}\right]
  +\FR{u(\fr32+\ii\wt\nu)}{1+r}{}_2\mathrm{F}_1\left[\bgm\fr{5}{2}+\ii\wt\nu,\fr12-\ii\wt\mu+\ii\wt\nu\\1+2\ii\wt\nu\edm\middle|\,\FR{2r}{1+r}\right]\bigg\}.
\end{align}
The appearance of Heaviside $\theta$-function is a result of the cutting rule for extracting the oscillatory signals \cite{Tong:2021wai}. We use the prescription that $\theta(z)=1$ for $z>0$, $\theta(z)=0$ for $z<0$, and $\theta(z)=1/2$ for $z=0$.

The nonlocal signal from the longitudinal polarization $(L)$ is:
\begin{align}
  \mathcal{T}_\text{NL,2}^{(L)}(k_I)  =& -\FR{\lam^2\big(1-\ii\sinh\pi\wt\nu\big)}{32\pi m^2 k_1\cdots k_4(k_{12}k_{34})^{3/2}}
   \FR{\big(\wh{\mb k}_2\cdot\mb{e}_L\big)\big(\wh{\mb k}_4\cdot\mb{e}_L^*\big)}{k_2k_4} {\mathbf G}^L_{\wt\nu}(u_2,r_1){\mathbf G}^L_{\wt\nu}(u_4,r_2)+\text{c.c.}  ,
\end{align}
where $\mb e_L$ is the longitudinal polarization vector, and the function $\mb G_{\wt\mu}^L$ is defined as
\begin{align}
  {\mathbf G}^L_{\wt\nu}(u,r)\equiv&\;                    
    \Gamma(-\ii\wt\nu)\Gamma\Big(\FR{3}{2}+\ii\wt\nu\Big)\bigg\{\Big(3+\FR{u}{1-r^2}(\fr{13}{4}+\wt\nu^2+3r^2)\Big)
  {}_2\mathrm{F}_1\left[\bgm\fr{3}{4}+\fr{\ii\wt\nu}{2},\fr{5}{4}+\fr{\ii\wt\nu}{2}\\ 1+\ii\wt\nu\edm \middle|\,r^2\right]\n\\
  &-\Big(1+\FR{u}{1-r^2}(1+3r^2)\Big)\Big(\FR52+\ii\wt\nu\Big)
  {}_2\mathrm{F}_1\left[\bgm\fr{3}{4}+\fr{\ii\wt\nu}{2},\fr{9}{4}+\fr{\ii\wt\nu}{2}\\ 1+\ii\wt\nu\edm \middle|\,r^2\right]\bigg\}\Big(\FR{r}{2}\Big)^{\ii\wt\nu}.
\end{align}
In the above expressions, the contractions of momentum with the polarization vector can be worked out to be:
\begin{align}
&\big(\wh{\mb k}_2\cdot\mb{e}_\pm\big)\big(\wh{\mb k}_4\cdot\mb{e}_\pm^*\big) =\FR{1}{2}\sin\theta_2\sin\theta_4e^{\pm\ii\phi_{24}},\\
&\big(\wh{\mb k}_2\cdot\mb{e}_L\big)\big(\wh{\mb k}_4\cdot\mb{e}_L^*\big)=\cos\theta_2\cos\theta_4,
\end{align}
where $\theta_i$ ($i=1,2,3,4$) is the angle between $\mb k_i$, and $\mb k_s$, and $\phi_{ij}=\phi_i-\phi_j$. Here $\phi_i$ is the longitude of $\mb k_i$ with $\mb k_s$ chosen as the $z$-direction. See Fig.\ \ref{fig_4pt_config} for an illustration.

We then consider the nonlocal signal from the single-scalar process, which is denoted by $\mathcal{T}_{\text{NL,1}}^{(h)}$ and is represented by the left diagram in Fig.\ \ref{fig_vector_tree}. This signal can be obtained directly from the two-scalar signal by including momentum permutations:
\begin{align}
\label{eq_TNLh_SingleScalar}
  \mathcal{T}_{\text{NL,1}}^{(h)} =&~\FR{1}{4}\bigg[\mathcal{T}_{\text{NL,2}}^{(h)}+(k_1\leftrightarrow k_2)+(k_3\leftrightarrow k_4)+\binom{k_1\leftrightarrow k_2}{k_3\leftrightarrow k_4}\bigg].
\end{align}

\paragraph{Squeezed limit.} One can see from the above expressions that the amplitudes for the massive spin-1 exchange have complicated dependence on various momenta, and are in general not directly factorized into the product of a single dynamical factor and a single kinematic factor. The factorization $\mathcal{T}\sim\mathcal{JK}$ holds only in the squeezed limit $k_s\to 0$, where these expressions are much simplified. We write the squeezed limit as:
\begin{align}
\label{eq_TNLh}
  \lim_{k_s\to 0}\mathcal{T}_{\text{NL},N}^{(h)} =&~\mathcal{J}_{\wt\mu,\wt\nu,N}^{(h)}(k_I)\mathcal{K}_N^{(h)}(k_I),~~~~~~(N=1,2)
\end{align}
where the label $N=1,2$ is used to distinguish single-scalar signal ($N=1$) from the two-scalar signal ($N=2$). For the two-scalar case, the dynamical factors for all helicities $h=\pm,L$ are given by:
\begin{align}
&\mathcal{J}_{\wt\mu,\wt\nu,2}^{(\pm)}=
  -\FR{\lam^2e^{\mp\pi\wt\mu}}{2^{10} (k_1k_3)^{9/2}}    
    \FR{(1-\ii\sinh\pi\wt\nu )(7+2\ii\wt\nu)^2\Gamma^2(\fr{3}{2}+\ii\wt\nu)\Gamma^2(-2\ii\wt\nu)}{\Gamma(\fr{1}{2}-\ii\wt\mu-\ii\wt\nu)\Gamma(\fr{1}{2}+\ii\wt\mu-\ii\wt\nu)} (4r_1r_2)^{\ii\wt\nu}+\text{c.c.},\\
&\mathcal{J}_{\wt\mu,\wt\nu,2}^{(L)}= -\FR{\lam^2 e^{-\pi\wt\nu} (1-\ii\sinh\pi\wt\nu )}{2^{14}\pi m^2(k_1k_3)^{9/2}}   \Gamma^2(-\ii\wt\nu)\Gamma^2\Big(\FR{3}{2}+\ii\wt\nu\Big)(7-12\ii\wt\nu+4\wt\nu^2)^2\Big(\FR{r_1r_2}{4}\Big)^{\ii\wt\nu} +\text{c.c.}.
\end{align}
Here we have put $k_2=k_1$, $k_4=k_3$, $k_{12}=2k_1$, $k_{34}=2k_3$, $u_2=u_4=1/2$, and  $r_1=r_2=0$ everywhere except in the nonalytical powers $(r_1r_2)^{\pm\ii\wt\nu}$. The kinematic factors are:
\begin{align}
  \mathcal{K}_2^{(\pm)}=&~\FR{1}{2}\sin\theta_1\sin\theta_3e^{\pm\ii\phi_{13}},\\
  \mathcal{K}_2^{(L)}=&~\cos\theta_1\cos\theta_3.
\end{align}
Here again we have used the fact that $-\mb k_2\simeq \mb k_1$, $-\mb k_4\simeq \mb k_3$, and thus $\theta_2\simeq\pi-\theta_1$, $\theta_4\simeq \pi-\theta_3$, and $\phi_{24}\simeq \phi_{13}$.
One can clearly see the ``$P$-wave'' dependence from the kinematic factor, and the parity-violating phase $e^{\mp\ii\phi_{13}}$, showing that a spin-1 state is mediating the signal in a helicity-dependent way.

The results for the single-scalar case are obtained by taking squeezed limit of (\ref{eq_TNLh_SingleScalar}). For the dynamical factors, we have:
\begin{align}
&\mathcal{J}_{\wt\mu,\wt\nu,1}^{(\pm)}=
   \FR{\lam^2e^{\mp\pi\wt\mu}}{2^{13} (k_1k_3)^{9/2}}    
    \FR{(1-\ii\sinh\pi\wt\nu )(7+2\ii\wt\nu)^2\Gamma^2(\fr{3}{2}+\ii\wt\nu)\Gamma^2(-2\ii\wt\nu)}{\Gamma(\fr{1}{2}-\ii\wt\mu-\ii\wt\nu)\Gamma(\fr{1}{2}+\ii\wt\mu-\ii\wt\nu)} (4r_1r_2)^{1+\ii\wt\nu}+\text{c.c.},\\
&\mathcal{J}_{\wt\mu,\wt\nu,1}^{(L)}=  \FR{\lam^2 e^{-\pi\wt\nu} (1-\ii\sinh\pi\wt\nu )}{2^{12}\pi m^2(k_1k_3)^{9/2}}   \Gamma^2(-\ii\wt\nu)\Gamma^2\Big(\FR{3}{2}+\ii\wt\nu\Big)(7-12\ii\wt\nu+4\wt\nu^2)^2\Big(\FR{r_1r_2}{4}\Big)^{1+\ii\wt\nu} +\text{c.c.}.
\end{align}
The kinematic factors are:
\begin{align}
\label{eq_KsingleSq1}
  \mathcal{K}_1^{(\pm)}=& ~\sin2\theta_1\sin2\theta_3e^{\pm\ii\phi_{13}},\\
\label{eq_KsingleSq2}
  \mathcal{K}_1^{(L)}=&  ~\cos 2\theta_1\cos2 \theta_3 . 
\end{align}
Comparing the above results with the two-scalar results, we see two important differences: First, the powers of the momentum ratio $r_1r_2$ are different. For the two-scalar signal it is $(r_1r_2)^{\pm\ii\wt\nu}$ whereas for the single-scalar signal it becomes $(r_1r_2)^{1 \pm\ii\wt\nu}$. That is, the signal is more suppressed for the single-scalar process in the squeezed limit, due to the cancelation of leading-order ``$P$-wave'' contribution. Using the parameters we defined in (\ref{eq_JNLpar2}), we have $\al=0$ for the two-scalar signal and $\al=1$ for the single-scalar signal. Second, the angular dependence is different, again due to the cancelation of $P$-waves. As a result, we see a ``$D$-wave'' behavior from (\ref{eq_KsingleSq1}) and (\ref{eq_KsingleSq2}).

\begin{figure}[t]
\centering
 {\includegraphics[height=0.45\textwidth]{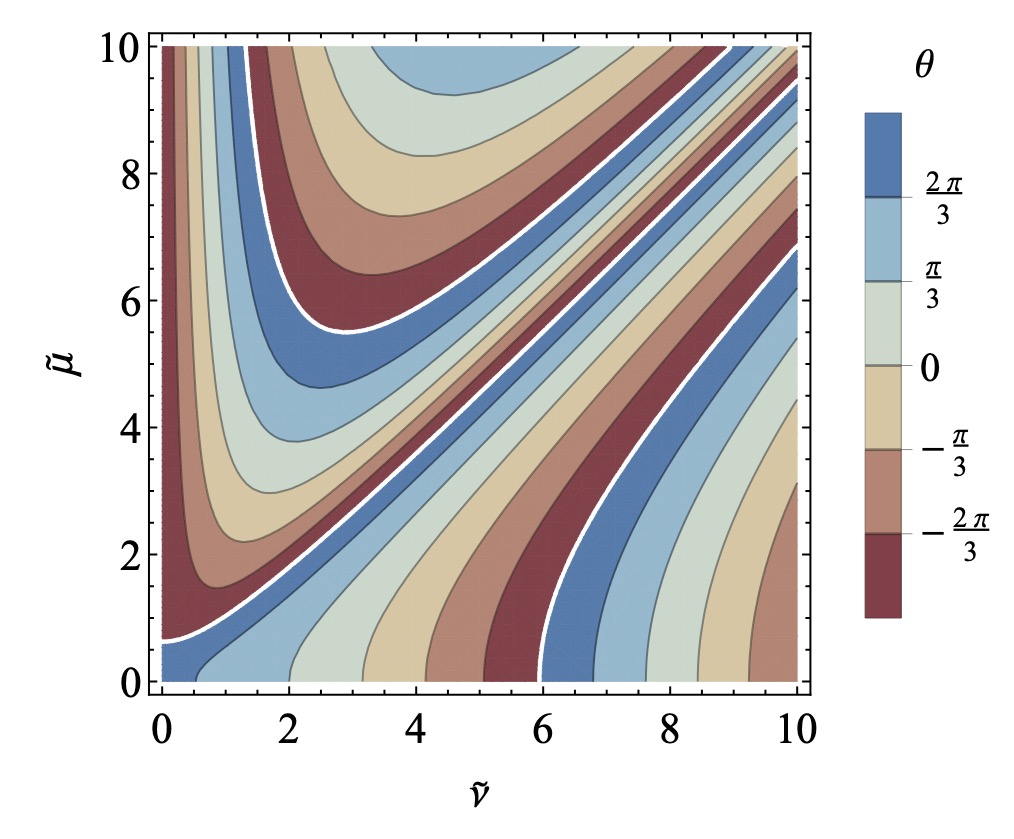}}
 \raisebox{0.03\height}{\includegraphics[height=0.43\textwidth]{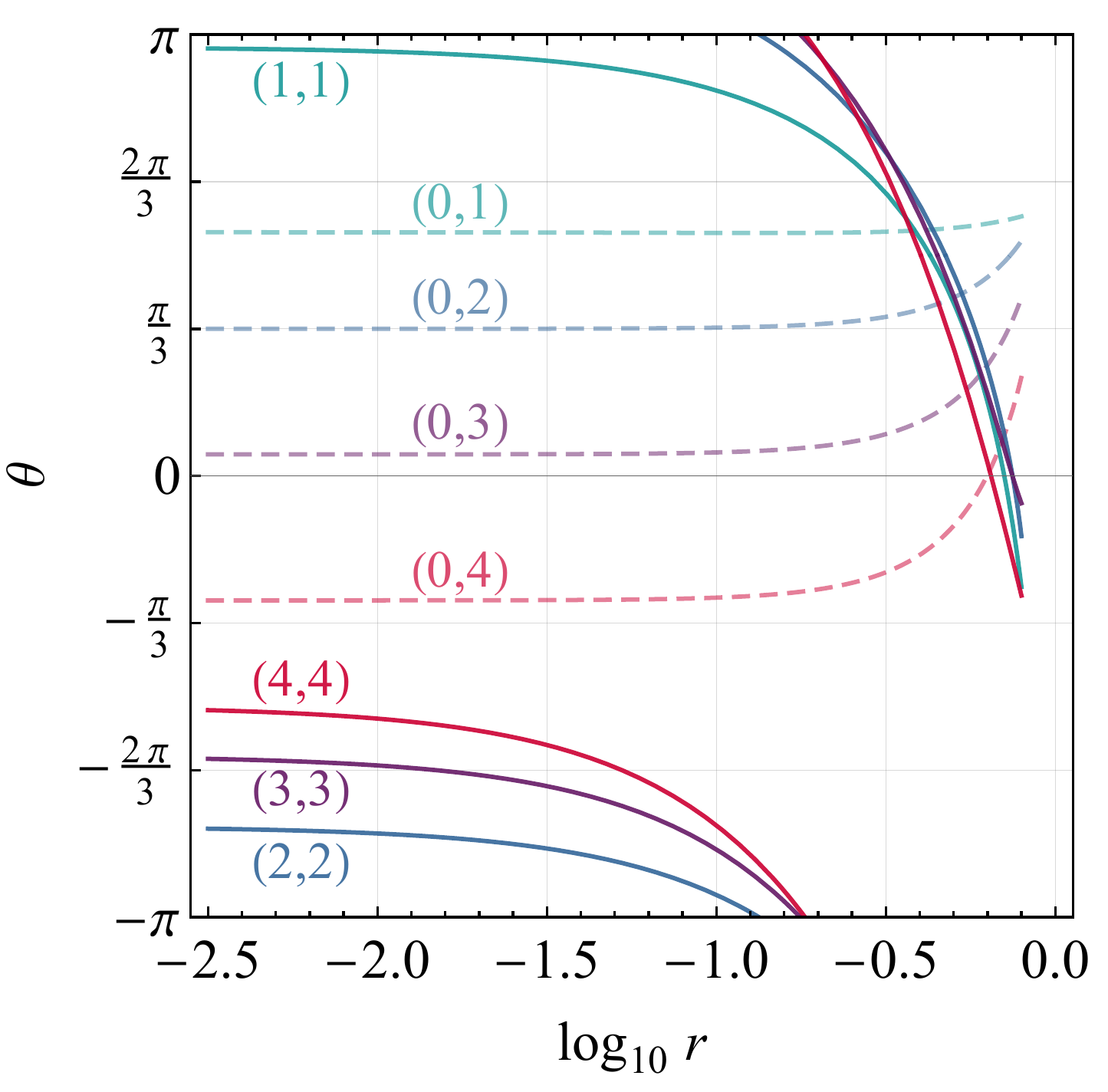}} 
\caption{The phases of nonlocal signal from a massive spin-1 boson. The left panel shows the phase in the squeezed limit $r_1=r_2\to 0$ as a function of mass parameter $\wt\nu$ and the chemical potential $\wt\mu=\mu/H$. The right panel shows the phases as functions of $r$ for several choices of $(\wt\mu,\wt\nu)$ and with $k_1=k_2=k_3=k_4$.}
\label{fig_vectortree}
\end{figure}

\paragraph{Signal phase.} 

Finally, we present the signal phase for the massive spin-1 exchange, for the single-scalar signal. It is clear from the previous result that the signal strength is proportional to $e^{-h\pi\mu}$. For moderately large $\mu$, the signal is dominated by $h=-1$ state. Therefore, we present the phase only for this state in the squeezed limit:
\begin{keyeqn}
\begin{align} 
  \lim_{k_s\to 0}\vartheta_{\varphi'(\pd_i\varphi)A^i} =\,\text{Arg}\,\bigg[\FR{4^{\ii\wt\nu}(1-\ii\sinh\pi\wt\nu )(7+2\ii\wt\nu)^2\Gamma^2(\fr{3}{2}+\ii\wt\nu)\Gamma^2(-2\ii\wt\nu)}{\Gamma(\fr{1}{2}-\ii\wt\mu-\ii\wt\nu)\Gamma(\fr{1}{2}+\ii\wt\mu-\ii\wt\nu)}\bigg]. 
\end{align}
\end{keyeqn}
It is clear that the phase depends on both the mass and the chemical potential. We show the dependence in the left panel of Fig.\ \ref{fig_vectortree}. The phase variation away from the squeezed limit is shown in the right panel of Fig.\ \ref{fig_vectortree}, in which we fix $k_1=k_2=k_3=k_4$, and vary $k_s$.

The phase shift as a function of $r=r_1=r_2$ is more significant than the previous scalar exchange, in part because the function $ {\mathbf{G}}_{\wt\mu,\wt\nu}$ in (\ref{eq_Gmunu}) depends on $r$ rather than on $r^2$ as in the scalar exchange.

\section{Phase Information at the One-Loop Level}
\label{sec_loop}

Now we consider the nonlocal signal from 1-loop processes. The phase information in this case is more useful than it is in tree-level processes. In particular, since the spin information of the intermediate particle is buried in the loop, we can no longer read the spin directly from the angular dependence. The phase information is then very useful for us to identify the species of the intermediate particles.

In general, the loop signal is more suppressed in the squeezed limit than in the tree signal, not only because of the loop factor $1/(4\pi)^2$, but also, more crucially, due to the larger comoving dilution. (See discussions in Sec.\;\ref{sec_dissect}.) Consequently, the loop signal should be searched for from not-too-squeezed configurations, with $r_1$ and $r_2$ smaller than 1 but not too close to 0. Therefore, we must take account of nontrivial dependences of the phase on $r_1$ and $r_2$. For nonlocal signals in the loop, there is no closed-form result in terms of known special functions, as opposed to the tree-level processes. To deal with this problem, we make use of the Mellin-Barnes representation of the intermediate loop propagators, which allows us to perform the loop momentum integral and the time integral completely, and to express the final result in terms of a series expansion in   $r_1$ and $r_2$. In practice, by going to high enough orders in $r_1$ and $r_2$, we can get  reliable results for nonlocal signals away from the squeezed limit. 

For concreteness, in the following, we will first consider the 1-loop process mediated by a massive scalar field $\si$, which is relatively easy to compute. Then we will consider a 1-loop process from massive spin-1 boson enhanced by a dS-boost-breaking and parity-odd chemical potential. This process is more relevant to phenomenological study as the signal can be observably large, although the computation of this process is also considerably more involved.
 
\subsection{Massive Scalar Loop}

\begin{figure}[tbph]
\centering  
  \parbox{0.31\textwidth}{\includegraphics[width=0.31\textwidth]{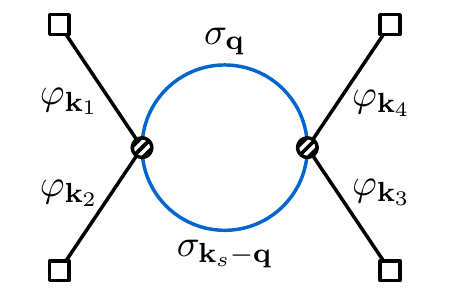}}
\caption{The 4-point function mediated by a pair of massive scalars $\si$ at 1-loop level.}
\label{fig_scalar_loop}
\end{figure}

Now we consider the 1-loop diagram mediated by a massive scalar particle $\si$. For definiteness, we shall consider the following coupling which is the simplest one that preserves the shift symmetry of the inflaton fluctuation $\varphi$:
\bge
\label{eq_scalarLoopCoup}
  \Delta\ld = -\FR{\lam}{4} a^2\varphi'^2\si^2,
\ede
where $\lam$ is the coupling constant of dimension $-2$, and $a$ is again the scale factor. From a model-building perspective, the above coupling appears directly from the Lorentz covariant operator $(\pd_\mu\phi)^2\si^2$ although in this case there should be a corresponding spatial-derivative coupling. One can also find models in which only the time-derivative coupling (\ref{eq_scalarLoopCoup}) contributes to the signal. As an example, we can have the following Lorentz covariant terms in the Lagrangian:
\bge
  \Delta\ld =\sqrt{-g}\bigg[ -\FR{1}{2}m_\chi^2\chi^2-\FR{g}{4}\chi^2\si^2-\FR{1}{2\Lambda}(\pd_\mu\phi)^2\chi\bigg].
\ede
Here we introduced a new heavy scalar $\chi$. When evaluated with the rolling inflaton background, the field $\chi$ acquires a two-point mixing with the inflaton fluctuation in the form of $\varphi'\chi$. Then, by integrating out $\chi$ we get the desired effective coupling in (\ref{eq_scalarLoopCoup}). The resulting 1-loop process is shown in Fig.\;\ref{fig_scalar_loop}.
 
Since the result for this loop process is new, we provide a bit more details. First, following the diagrammatic rule \cite{Chen:2017ryl}, it is straightforward to write down an expression for the diagram: 
\begin{align}
\label{eq_scalarloopInt}
  \mathcal{T}(k_I)=&-\FR{\lam^2}{2}\sum_{\mathsf{a},\mathsf{b}=\pm}\mathsf{ab}\int_{-\infty}^0\FR{\di\tau_1}{(- \tau_1)^2}\FR{\di\tau_1}{(- \tau_2)^2}\,G_\mathsf{a}'(k_1,\tau_1)G_\mathsf{a}'(k_2,\tau_1)G_\mathsf{b}'(k_3,\tau_2)G_\mathsf{b}'(k_4,\tau_2)\n\\
  &\times \int\FR{\di^3\mb q}{(2\pi)^3}\,D_\mathsf{ab}(q;\tau_1,\tau_2)D_\mathsf{ab}(|\mb k_s-\mb q|;\tau_1,\tau_2),
\end{align}
where $G$ and $D$ are propagators for $\varphi$ and $\si$, respectively, and $\mathsf{a},\mathsf{b}=\pm$ are Schwinger-Keldysh indices \cite{Chen:2017ryl}. 
Focusing on the signal part, we can apply the cutting rule, which in this case is equilavent to symmetrizing the two loop propagators $D_{\pm\pm}$ in $\tau_1\leftrightarrow \tau_2$, and thus trivialize the time ordering in the various Schwinger-Keldysh propagators.\footnote{We have checked with explicit calculation that the antisymmetric part of the intermediate propagator does not give rise to nonlocal signals at 1-loop order.} We then use the Mellin-Barnes representations for the massive propagator $D$, on which we elaborate in App.\ \ref{app_MB}. Then, the three integrals in (\ref{eq_scalarloopInt}) can be carried out analytically. The final result is obtained by closing the Mellin-Barnes contours and picking up appropriate poles corresponding to the nonlocal signal, namely, terms proportional to $(r_1r_2)^{\pm2\ii\wt\nu}$. 
 The kinematic factor for this process is trivial, $\mathcal{K}=1$, and the final result for the nonlocal dynamical factor $\mathcal{J}$ is:
\begin{align}
\label{eq_Jscalarloop}
    \mathcal{J}_\text{NL}(k_1,k_2,k_3,k_4,k_s)&~=
     \FR{\lam^2}{256\pi^{7/2}k_1k_2k_3k_4(k_{12}k_{34})^{5/2}}
      \Big(\FR{r_1r_2}{4}\Big)^{3/2+2\ii\wt\nu}\mathbf{H}_{\wt\nu}(r_1,r_2)+\text{c.c.},
\end{align}
where the function $\mathbf{H}_{\wt\nu}(r_1,r_2)$ can be expressed as a power series in $r_1$ and $r_2$:
\begin{align}
\label{eq_Hnur1r2}
    \mathbf{H}_{\wt\nu}(r_1,r_2) =&~(1-\cosh2\pi\wt\nu)\sum_{n_1,n_2,n_3,n_4=0}^\infty
    \FR{(-1)^{n_{1234}}}{n_1!n_2!n_3!n_4!}\Big(\FR{r_1}{2}\Big)^{2n_{13}}
    \Big(\FR{r_2}{2}\Big)^{2n_{24}}
    \n\\
    &\times 
    \Gamma\Big[4+2n_{13}+2\ii\wt\nu,4+2n_{24}+2\ii\wt\nu,-n_1-\ii\wt\nu,-n_2-\ii\wt\nu,-n_3-\ii\wt\nu,-n_4-\ii\wt\nu\Big]\n\\
    &\times\Gamma\bigg[\bgm \fr32+n_{12}+\ii\wt\nu,\fr32+n_{34}+\ii\wt\nu,-n_{1234}-\fr32-2\ii\wt\nu \\ -n_{12}-\ii\wt\nu,-n_{34}-\ii\wt\nu,3+n_{1234}+2\ii\wt\nu \edm\bigg].
\end{align}
Here we use the notation $n_{ij}=n_i+n_j~(i,j=1,\cdots,4)$ and $n_{1234}=n_1+\cdots+n_4$. We also write Euler $\Gamma$ function using a compact notation, whose definition is given in App.\ \ref{app_notation}.
The result in (\ref{eq_Jscalarloop}) does not assume the squeezed limit $k_s\to0$, although the expression in (\ref{eq_Hnur1r2}) is presented as an expansion in $r_1$ and $r_2$. 

\begin{figure}[t]
\centering
\includegraphics[width=0.45\textwidth]{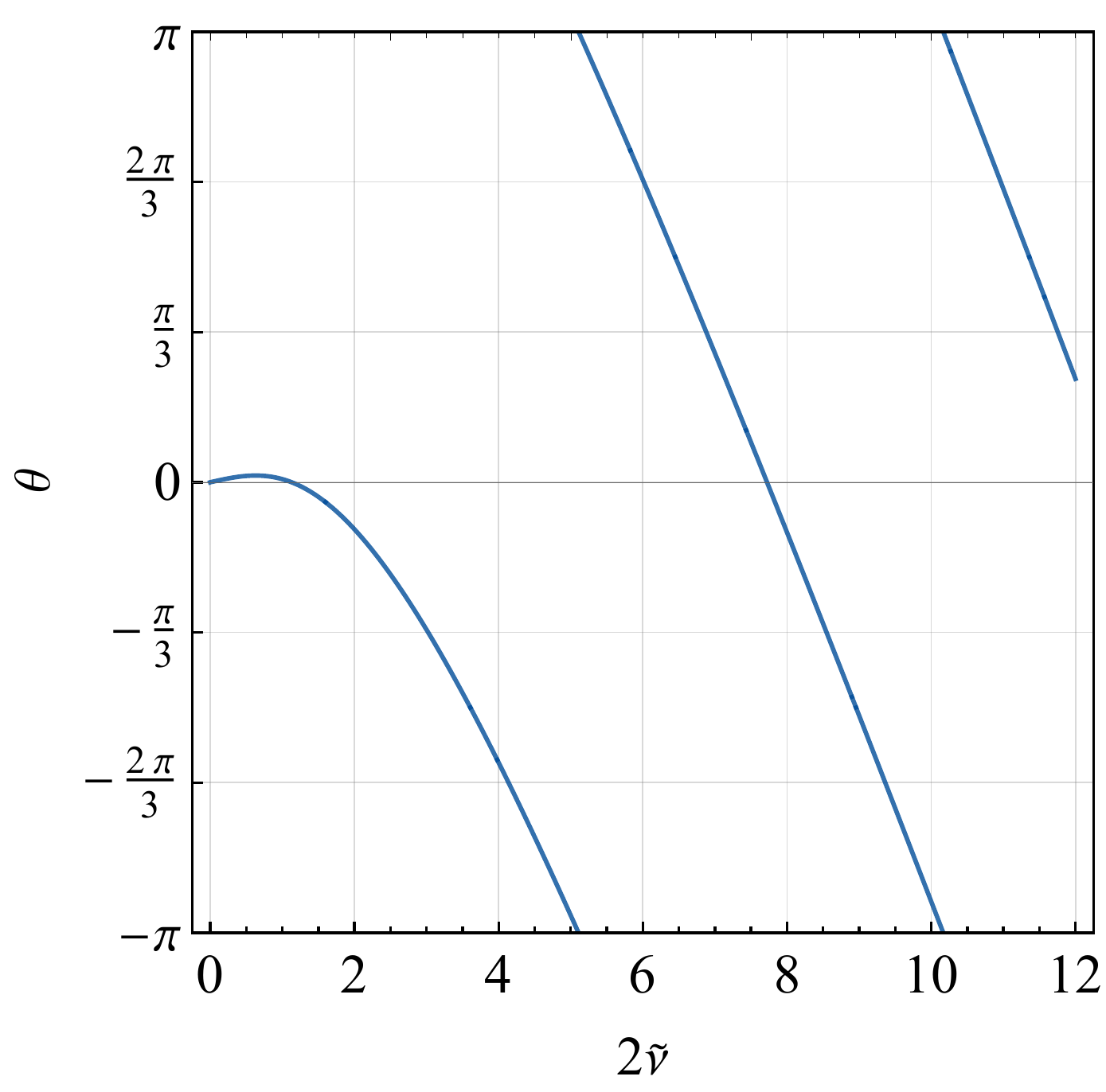} \hspace{5mm}
\includegraphics[width=0.45\textwidth]{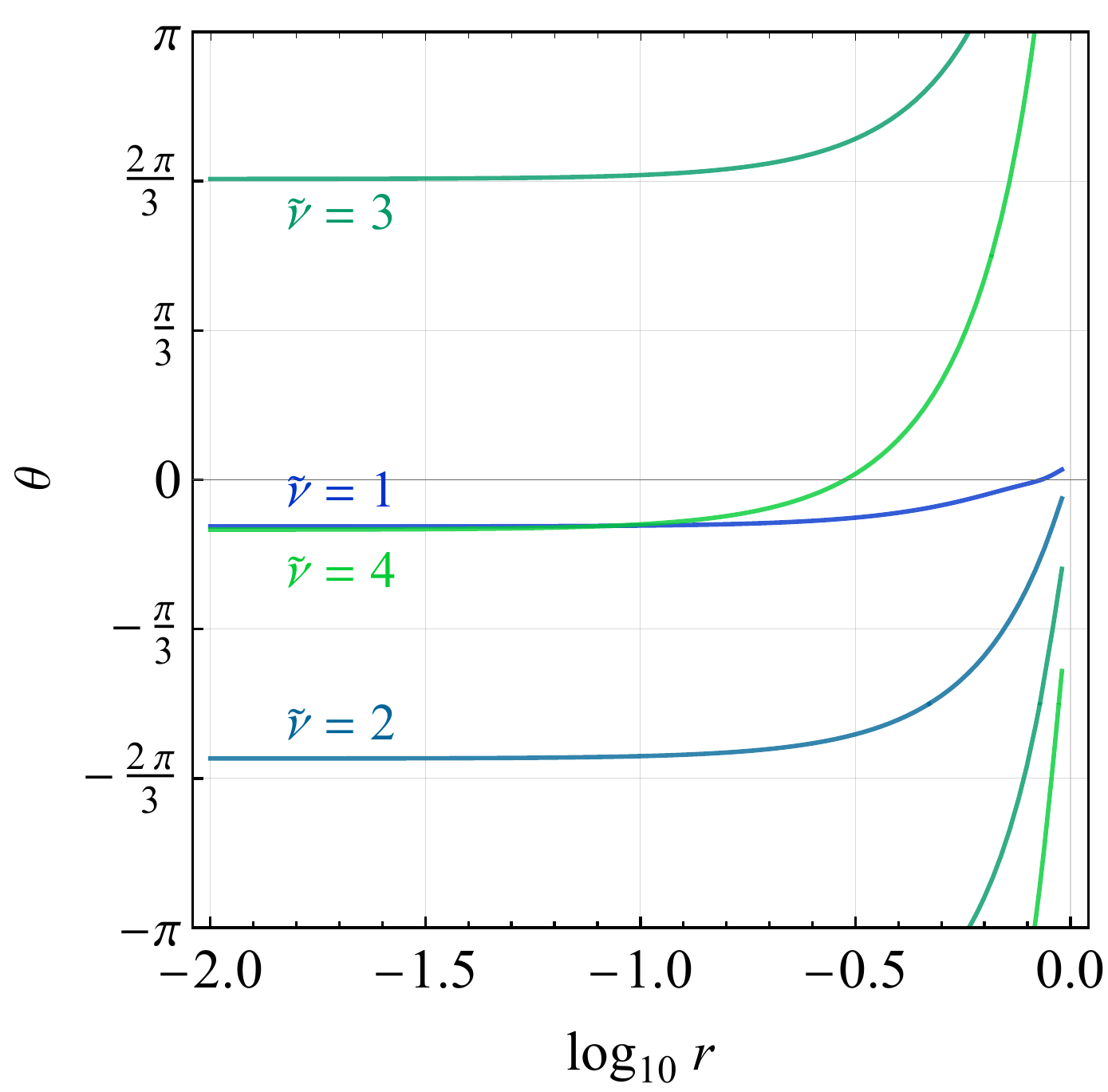} 
\caption{The phases $\theta_T(r,r)$ of the nonlocal signal from the $s$-channel 1-loop exchange of a massive scalar. The left panel shows the phase in the squeezed limit $r\to 0$ as a function of oscillation frequency $\omega=2\wt\nu$, for time-derivative coupling. The right panel shows the phases from time-derivative coupling as functions of momentum ratio $r$, for several values of $\wt\nu$.}
\label{fig_phaseScalarLoop}
\end{figure}

Now let us look at a particular squeezed configuration where $k_1=k_2=k_{12}/2$, $k_3=k_4=k_{34}/2$. We also take $r_1,r_2\to 0$ with the ratio $r_1/r_2$ fixed in order to remove any possible local signal. Then, we can express the nonlocal signal as an expansion in $r_1$ and $r_2$ as below:
\begin{align}
    \mathcal{J}_\text{NL}(k_1,k_3,k_s)&=
     \FR{\lam^2}{2^{13}\pi^{7/2}(k_1k_3)^{9/2}}
      \Big(\FR{r_1r_2}{4}\Big)^{3/2+2\ii\wt\nu}\bigg[\mathbf{H}_{\wt\nu,0} +\sum_{i=1}^\infty \mathbf{H}_{\wt\nu,2i}(r_1,r_2) \bigg]+\text{c.c.},
\end{align}
with the first several orders given by:
\begin{align}
\label{eq_H0}
 \mathbf{H}_{\wt\nu,0}=&~
     (1-\cosh2\pi\wt\nu)(3+2\ii\wt\nu)\Gamma^2(\fr{3}{2}+\ii\wt\nu)\Gamma(-\fr{3}{2}-2\ii\wt\nu)\Gamma(4+2\ii\wt\nu)\Gamma^2(-\ii\wt\nu), \\
 \mathbf{H}_{\wt\nu,2}=&~\mathbf{H}_{\wt\nu,0}
      \fr{(5+2\ii\wt\nu)(2+\ii\wt\nu)}{5+4\ii\wt\nu} (r_1^2+r_2^2),\\ 
\label{eq_H4}
 \mathbf{H}_{\wt\nu,4}=&~\mathbf{H}_{\wt\nu,0}
      \fr{(5+2\ii\wt\nu)(2+\ii\wt\nu)}{5+4\ii\wt\nu}\Big[ \fr{(7+2\ii\wt\nu)(3+\ii\wt\nu)}{14+8\ii\wt\nu}(r_1^4+r_2^4) -\fr{65+8\ii\wt\nu[12+(6+\ii\wt\nu)\ii\wt\nu]}{4(1+\ii\wt\nu)(7+4\ii\wt\nu)}(r_1r_2)^2\Big].
\end{align} 
 
The phase of the signal can be readily read from the full result (\ref{eq_Jscalarloop}) as 
\begin{align}
\label{eq_scalarLoopPhaseFull}
  \vartheta_{\varphi'^2\si^2}=&~\text{Arg}\,\Big[4^{-2\ii\wt\nu}\mathbf{H}_{\wt\nu}(r_1,r_2)\Big].
\end{align}
 
In particular, the leading order result in the squeezed limit $r_1,r_2\to 0$ is obtained by setting $\mathbf{H}_{\wt\nu}(r_1,r_2)=\mathbf{H}_{\wt\nu,0}$, and the result is:
\begin{keyeqn}
\begin{align}
\label{eq_thetasqScaLoop}
   \lim_{k_s\to 0}\vartheta_{\varphi'^2\si^2}=&~\text{Arg}\,\Big[-4^{-2\ii\wt\nu}(3+2\ii\wt\nu)\Gamma^2(\fr{3}{2}+\ii\wt\nu)\Gamma(-\fr{3}{2}-2\ii\wt\nu)\Gamma(4+2\ii\wt\nu)\Gamma^2(-\ii\wt\nu)\Big].
\end{align}
\end{keyeqn}
As we can see, for the given form of interaction, the squeezed-limit phase is uniquely fixed in terms of the mass parameter $\wt\nu$.
Now that the mass parameter $\wt\nu$ can be separately measured from the oscillation frequency $\omega=2\wt\nu$, we see that the phase provides crucial information that can help to identify the nature of the loop particle and its interaction with the external mode.

\begin{figure}[t]
\centering
\includegraphics[width=0.45\textwidth]{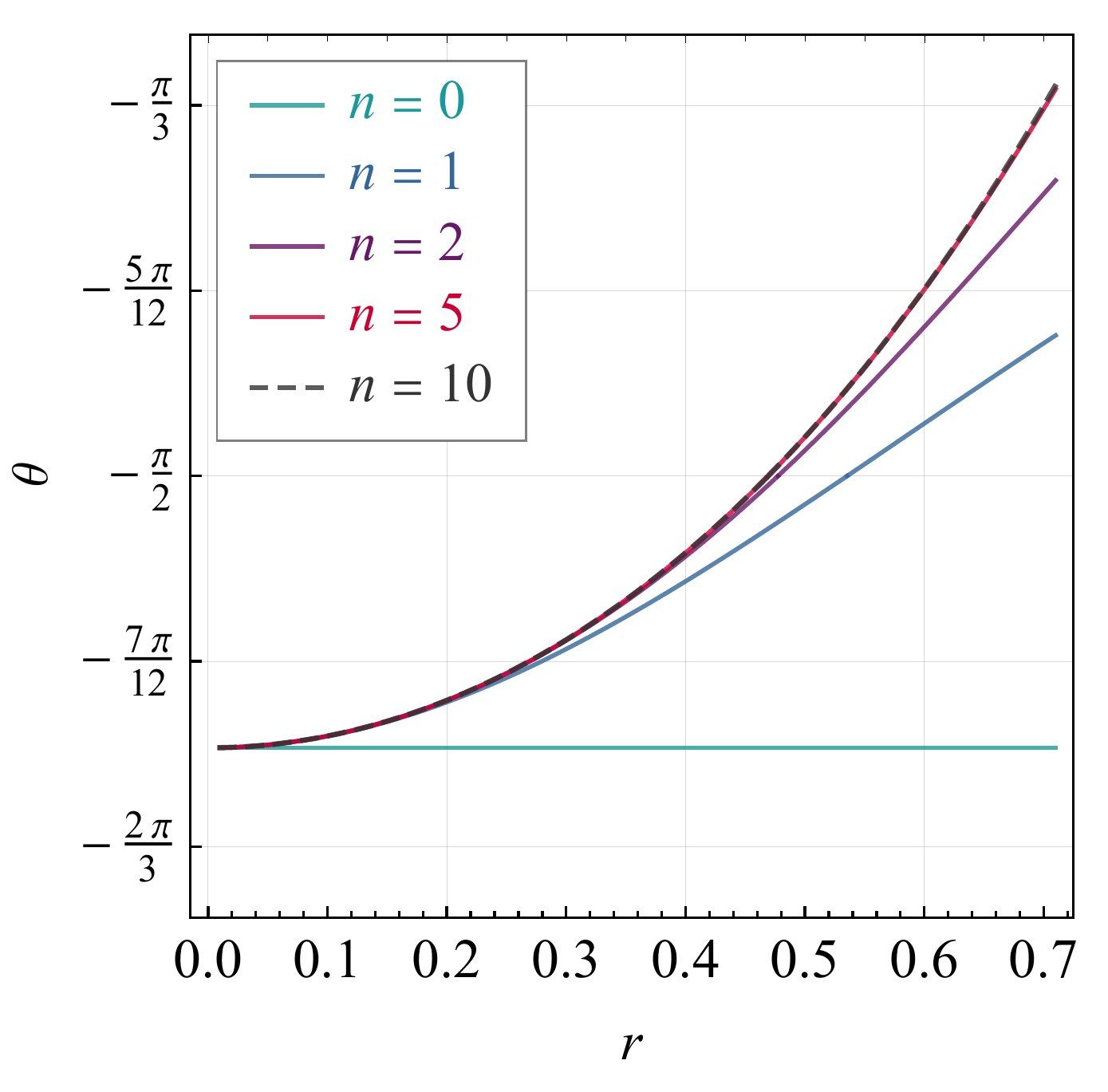}
\includegraphics[width=0.45\textwidth]{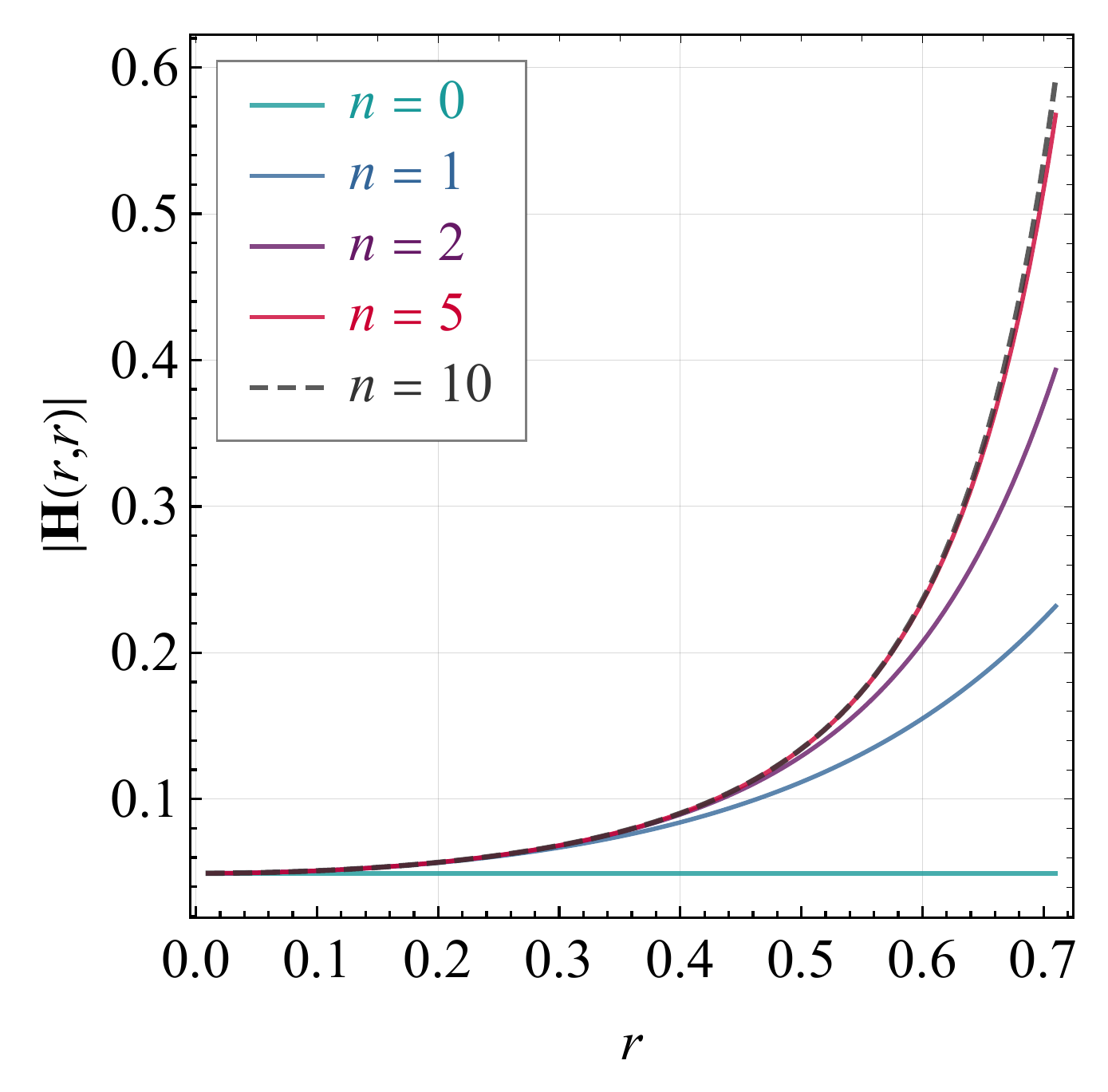}
\caption{The convergence of the series in (\ref{eq_Hnur1r2}). We show the result for both the phase and the magnitude of (\ref{eq_Hnur1r2}) by including terms with $0\leq n_1,\cdots,n_4\leq n$, for $n=0,1,2,5,10$. }
\label{fig_convergence}
\end{figure}

In Fig.\;\ref{fig_phaseScalarLoop} we show the phase of the scalar loop diagram. In the left panel, we plot the phase in the squeezed limit (\ref{eq_thetasqScaLoop}) as a function of oscillation frequency $2\wt\nu$. In the right panel, we set $r_1=r_2=r$ and show the $r$ dependence of the full phase (\ref{eq_scalarLoopPhaseFull}), for several values of $\wt\nu$. It is clear that the $r$ dependence becomes significant when $r\gtrsim 0.1$, and it is thus important to include the higher order terms in (\ref{eq_Hnur1r2}) for a reliable result in this region. In Fig.\;\ref{fig_phaseScalarLoop} we have included the terms with $0\leq n_1,\cdots,n_4\leq 12$. To check the convergence of the summation in (\ref{eq_Hnur1r2}), we show in Fig.\ \ref{fig_convergence} the results for both the phase $\vartheta(r,r)$ and the magnitude $|\mathbf{H}_{\wt\nu}(r,r)|$ from summing over terms with $0\leq n_1,\cdots,n_4\leq n$ for $\wt\nu=2$ and for several choices of $n$. We show the result up to $r=1/\sqrt{2}\simeq 0.7$, since for our choice of the configuration, the $t$-channel oscillation would kick in for $r>0.7$. It is clear from the figure that the series in (\ref{eq_Hnur1r2}) converges quickly for our choices of parameters.

\subsection{Gauge boson loop}
\begin{figure}[t]
\centering  
  \parbox{0.31\textwidth}{\includegraphics[width=0.31\textwidth]{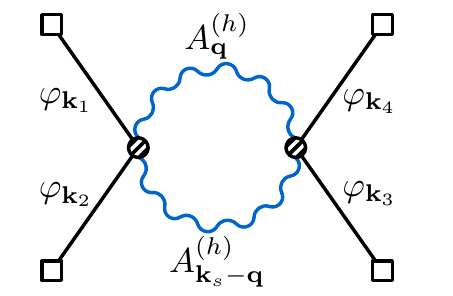}}
\caption{The 4-point function mediated by a pair of massive vectors $A$ at 1-loop level.}
\label{fig_vector_loop}
\end{figure}

Finally, we consider the 1-loop diagram mediated by a massive spin-1 boson with chemical potential enhancement. Similarly, we consider the simplest coupling respecting the shift symmetry of the inflaton fluctuation:
\bge
\label{eq_vectorLoopCoup}
  \Delta\ld = -\FR{\lam}{4} \varphi'^2 A_\mu A^\mu,
\ede
where $\lam$ is the coupling constant of dimension $-2$, and the Lorentz indices here are contracted by the Minkowski metric $\eta_{\mu\nu}$. It is easy to find Lorentz covariant UV completion for this effective coupling. An example was already studied in \cite{Wang:2020ioa}. In short, one can introduce a complex scalar $\chi$ charged under a local U(1) group with $A_\mu$ being the gauge field. Then, with the following terms in the Lagrangian,
\bge
\label{eq_VecLoopLag}
  \Delta\ld =\sqrt{-g}\bigg[-|\D_\mu\chi|^2-\lam_4|\chi|^4-\FR{1}{4\Lambda}(\pd_\mu\phi)^2|\chi|^2 \bigg]-\FR{1}{4\Lambda'}\phi F\wt F,
\ede
the scalar $\chi$ would acquire a nonzero VEV due to the dim-6 operator $(\pd_\mu\phi)^2|\chi|^2$ and the inflaton background $\la(\pd_\mu\phi)^2\ra=-\dot\phi_0^2$. The same dim-6 operator then introduces a two-point mixing $\varphi'\chi$ between the inflaton fluctuation $\varphi$ and the radial mode of $\chi$, while the gauge field $A_\mu$ becomes massive and acquires its P-odd  chemical potential from $\phi F\wt F$ term. Then, from the coupling $\chi^2A^2$ and the two-point mixing $\varphi'\chi$, the above effective coupling (\ref{eq_vectorLoopCoup}) is naturally obtained after integrating out the heavy $\chi$ field. We thus consider the 1-loop process mediated by the massive gauge boson $A$ with the coupling (\ref{eq_vectorLoopCoup}), shown in Fig.\;\ref{fig_vector_loop}.

It is worth noting that we can also couple the massive gauge boson $A_\mu$ to the external inflaton $\varphi$ via chemical-potential coupling $\phi F\wt F$ in (\ref{eq_VecLoopLag}). This coupling will also generate a 1-loop contribution to the 4-point function via a gauge boson box diagram. In the squeezed limit $r_{1,2}\to 0$, this box diagram contains a nonlocal signal, similar to the 3-point process considered in \cite{Wang:2020ioa}. We shall ignore this box diagram in this work as it is more complicated than Fig.\;\ref{fig_vector_loop}. Qualitatively, we expect that the box-diagram signal would have a larger scaling exponent $\al$ than the signal from Fig.\ \ref{fig_vector_loop}, as there are additional derivatives acting on the loop mode. Therefore, a measurement of the scaling exponent can tell the two processes apart. More generally, when there are multiple processes contributing to the signal, we can use various filtering techniques to filter them out. More discussions on this issue will be given in Sec.\ \ref{subsec_discussions}.
 
The expression for the diagram from the diagrammatic rule is: 
\begin{align}
\label{eq_vectorloopInt}
  &\mathcal{T}(k_I)=-\FR{\lam^2}{2}\sum_{\mathsf{a},\mathsf{b}=\pm}\mathsf{ab}\int_{-\infty}^0\di\tau_1\di\tau_2\,G_\mathsf{a}'(k_1,\tau_1)G_\mathsf{a}'(k_2,\tau_1)G_\mathsf{b}'(k_3,\tau_2)G_\mathsf{b}'(k_4,\tau_2)\n\\
  &\times\sum_{h,h'} \int\FR{\di^3\mb q}{(2\pi)^3}\,D^{(h)}_\mathsf{ab}(q;\tau_1,\tau_2)D^{(h')}_\mathsf{ab}(|\mb k_s-\mb q|;\tau_1,\tau_2)  \Big|\eta^{\mu\nu} e^{(h)}_\mu(\mb q)  e^{(h')*}_\nu(\mb k_s - \mb q)\Big|^2.
\end{align}

To find the nonlocal signal, we apply the cutting rule and then extract the nonlocal part of the integral. Notice that, although the loop is mediated by a massive spin-1 boson, the kinematic factor is still trivial: $\mathcal K=1$, which differs from the spin-1 signal at the tree-level. The reason is that the polarization vectors are all contracted with each other, other than the external momenta. In other words, the kinematic factor probes the total angular momentum of the composite operator $A^2$, which is a scalar.

As shown in (\ref{eq_vectorloopInt}), the total signal is the sum of the contributions from all helicity combinations. Here we will make a very mild approximation, that is, we will include only the chemical-potential-enhanced contribution, namely $h=-1$ when $\wt\mu>0$. The contribution from this state is larger than other states by a factor of $e^{2\pi\wt\mu}$. Therefore, the error from this approximation is always under percent level for mildly large $\wt\mu\gtrsim 1$.

 \begin{figure}[t]
\centering
 {\includegraphics[height=0.45\textwidth]{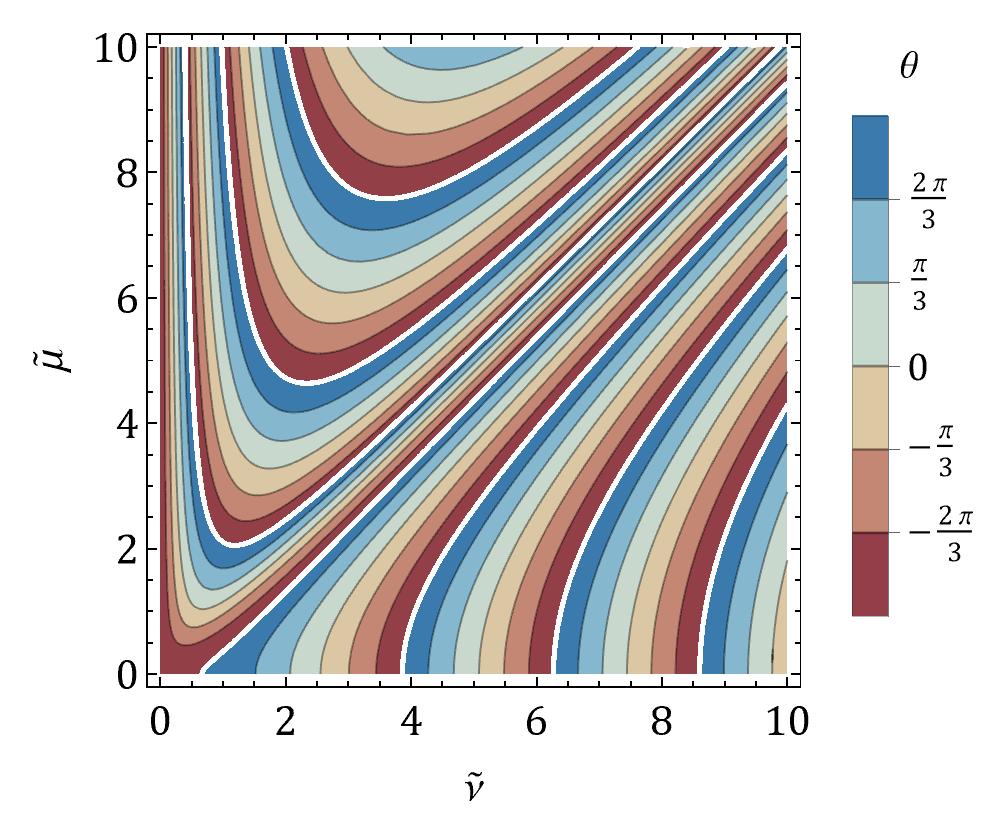}}
 \raisebox{0.03\height}{\includegraphics[height=0.43\textwidth]{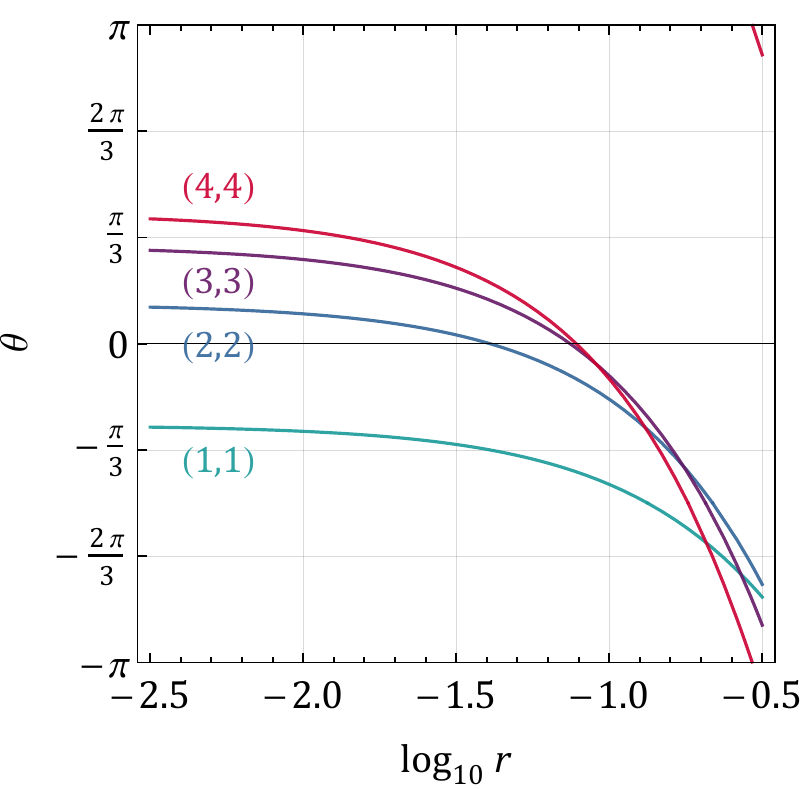}} 
\caption{The phases $\theta_T(r,r)$ of the nonlocal signal from the s-channel 1-loop exchange of a massive spin-1 boson. The left panel shows the phase in the squeezed limit $r\to 0$ as a function of mass parameter $\wt\nu=\sqrt{m^2/H^2-1/4}$ and the chemical potential $\wt\mu=\mu/H$, for time-derivative coupling. The right panel shows the phases from time-derivative coupling as functions of momentum ratio $r$ for several choices of $(\wt\mu,\wt\nu)$.}
\label{fig_phaseVectorLoop}
\end{figure}

With the above approximation, we can keep $h=h'=-1$ in (\ref{eq_vectorloopInt}) and proceed as before. We again put the details in App.\ \ref{app_details}, and the result for the dynamical factor $\mathcal J_{\text{NL}}$ is:
\begin{align}
\label{eq_Jvectorloop}
    \mathcal{J}_\text{NL}(k_1,k_2,k_3,k_4,k_s)=&~
     \FR{\lam^2e^{2\pi\wt\mu}}{2^{14}\pi^6k_1k_2k_3k_4(k_{12}k_{34})^{5/2}}
      (4r_1r_2)^{3/2+2\ii\wt\nu}\n\\
      &~\times\Big[\mathbf{I}_{\wt\mu,\wt\nu}(r_1,r_2)\theta(r_2-r_1)+\mathbf{I}_{\wt\mu,\wt\nu}(r_2,r_1)\theta(r_1-r_2)\Big]+\text{c.c.},
\end{align}
where the function $\mathbf{I}_{\wt\mu,\wt\nu}(r_1,r_2)$ can be also expressed as a power series in $r_1$ and $r_2$:\footnote{The term $(\wt\nu\to-\wt\nu)^*$ denotes that we should first change $\wt\nu$ to $-\wt\nu$ and then take the complex conjugate.}
\begin{align}
\label{eq_Hmunur1r2}
    \mathbf{I}_{\wt\mu,\wt\nu}(r_1,r_2) =&~\big(\cosh2\pi\wt\mu+\cosh2\pi\wt\nu\big)^2\sum_{n_1,n_2,n_3,n_4=0}^\infty\sum_{m_1,m_2=0}^\infty\sum_{\ell=0}^{m_1+m_2}
    \FR{(-2)^{n_{1234}}}{n_1!n_2!n_3!n_4!}
    r_1^{n_{13}+m_1}r_2^{n_{24}+m_2}\n\\
    &\times\bigg\{\Big(1-(-1)^{n_{13}+m_1} e^{2\pi\wt\nu}\Big) 
    \binom{-4-n_{13}-2\ii\wt\nu}{m_1}  \binom{-4-n_{24}-2\ii\wt\nu}{m_2}\binom{m_1+m_2}{\ell}\n\\
    &\times\Gamma\Big[4+n_{13}+2\ii\wt\nu,4+n_{24}+2\ii\wt\nu,-n_1-2\ii\wt\nu,-n_2-2\ii\wt\nu,-n_3-2\ii\wt\nu,-n_4-2\ii\wt\nu\Big]\n\\
    &\times\Gamma\Big[n_1+\FR12+\ii\wt\mu+\ii\wt\nu,n_2+\FR12-\ii\wt\mu+\ii\wt\nu,n_3+\FR12+\ii\wt\mu+\ii\wt\nu,n_4+\FR12-\ii\wt\mu+\ii\wt\nu\Big]\n\\
    &\times \mathbf{L}(n_{12}+\ell+2\ii\wt\nu,n_{34}+m_1+m_2-\ell+2\ii\wt\nu)
    +(\wt\nu\to-\wt\nu)^* \bigg\} ,
\end{align}
where $\binom{a}{b}=\Gamma(a+1)/[\Gamma(b+1)\Gamma(a-b+1)]$ is the binomial coefficient, and the function $\mathbf{L}(a,b)$ is defined by
\begin{align}
  \mathbf{L}(a,b)=&~\FR{2\Gamma(a)\Gamma(b)}{\Gamma(5+a+b)}\Big[
  -(1+a+b)(3+a+b)+\FR{\sin\pi a+\sin\pi b}{\sin \pi(a+b)}\n\\&\times (2a^2b^2+4a^2b+4ab^2+a^2+10ab+b^2+4a+4b+3)\Big].
\end{align}
This complicated expression converges for non-squeezed $k_s$, though slower than the scalar case,
and becomes much simpler when we take the squeezed limit:
\begin{align}
\label{eq_ImunuSqueezed}
\lim_{r_1,r_2\to0}\mathbf{I}_{\wt\mu,\wt\nu}(r_1,r_2)
=&~\FR{2\pi^6\mathrm{sech}^2(\pi\wt\nu) \Gamma^2(4+2\ii\wt\nu)\Gamma^2(-2\ii\wt\nu)}{\wt\nu^2\Gamma^2(\fr12-\ii\wt\mu-\ii\wt\nu)\Gamma^2(\fr12+\ii\wt\mu-\ii\wt\nu)\Gamma(5+4\ii\wt\nu)}\n\\
&\times \Big[(3+16\ii\wt\nu-16\wt\nu^2)-(3+16\ii\wt\nu-48\wt\nu^2-64\ii\wt\nu^3+32\wt\nu^4)\text{sech}(2\pi\wt\nu)\Big].
\end{align}

\begin{figure}[t]
\centering
\includegraphics[width=0.45\textwidth]{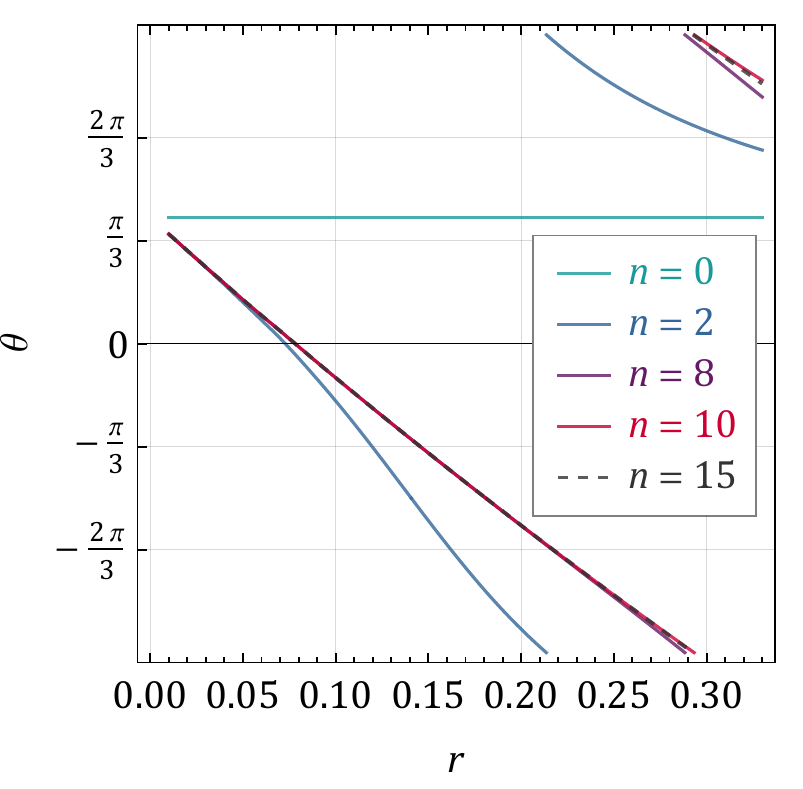}
\includegraphics[width=0.45\textwidth]{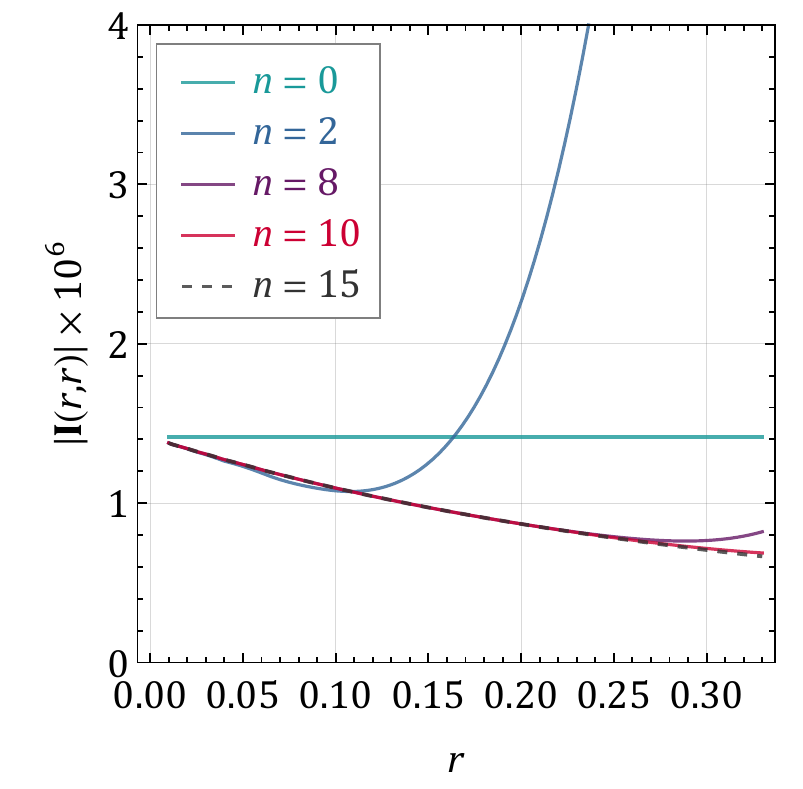}
\caption{The convergence of the series in (\ref{eq_Hmunur1r2}). We show the result for both the phase and the magnitude of (\ref{eq_Hmunur1r2}) by including terms with $0\leq n_{13}+m_1,n_{24}+m_2\leq n$, for $n=0,2,8,10,15$. }
\label{fig_convergence_vector}
\end{figure}

The phase of the signal can be read from the full result as:
\begin{align}
  \label{eq_vectorLoopPhaseFull}
    \vartheta_{\varphi'^2A^2}=&~\text{Arg}\,\Big[4^{2\ii\wt\nu}\mathbf{I}_{\wt\mu,\wt\nu}(r_1,r_2)\Big],
  \end{align}
and in the squeezed limit, it becomes:
\begin{keyeqn}
  \begin{align}
  \label{eq_thetasqVecLoop}
     \lim_{k_s\to 0}\vartheta_{\varphi'^2A^2}=&~\text{Arg}\,\bigg\{\FR{4^{2\ii\wt\nu}\Gamma^2(4+2\ii\wt\nu)\Gamma^2(-2\ii\wt\nu)}{\Gamma^2(\fr12-\ii\wt\mu-\ii\wt\nu)\Gamma^2(\fr12+\ii\wt\mu-\ii\wt\nu)\Gamma(5+4\ii\wt\nu)}\n\\
     &\times\Big[(3+16\ii\wt\nu-16\wt\nu^2)-(3+16\ii\wt\nu-48\wt\nu^2-64\ii\wt\nu^3+32\wt\nu^4)\text{sech}(2\pi\wt\nu)\Big]\bigg\}.
  \end{align}
  \end{keyeqn}
From this expression we observe that, even with the mass parameter $\wt\nu$ separately determined from measuring the oscillation frequency $\omega=2\wt\nu$, the squeezed-limit phase still depends on the chemical potential $\wt\mu$, and can change continuously when we vary $\wt\mu$. If the interaction type is also separately known, then a measurement of the squeezed-limit phase (\ref{eq_thetasqVecLoop}) will be a way to measure the chemical potential. However, in general, the interaction type cannot be determined independently, and therefore we have to go away from the squeezed limit, and try to measure the $r$-dependence of the phase. This $r$-dependence in the phase thus can provide more information to fully break the degeneracy, allowing us to determine both the interaction type and the chemical potential, as is clear from both Fig.\ \ref{fig_phaseVectorLoop} and Fig.\ \ref{fig_signal}.

In Fig.\;\ref{fig_phaseVectorLoop} we show the phase of the vector loop diagram. In the left panel, we plot the phase in the squeezed limit (\ref{eq_thetasqVecLoop}) as a function of mass parameter $\wt\nu$ and chemical potential $\wt\mu$.
In the right panel, we set $r_1=r_2=r$ and show the $r$ dependence of the full phase (\ref{eq_vectorLoopPhaseFull}), for several values of $\wt\mu$ and $\wt\nu$. We find the $r$ dependence becomes significant when $r\gtrsim 0.01$,
so we have to include higher order terms in (\ref{eq_Hmunur1r2}) for a reliable result. In Fig.\;\ref{fig_phaseVectorLoop}, we have included the terms with $0\leq n_{13}+m_1,n_{24}+m_2\leq 10$. To check the convergence of the summation in (\ref{eq_Hmunur1r2}), we show in Fig.\;\ref{fig_convergence_vector} the results for both the phase $\vartheta(r,r)$ and the magnitude $|\mathbf{I}_{\wt\mu,\wt\nu}(r,r)|$ from summing over terms with $0\leq n_{13}+m_1,n_{24}+m_2\leq n$ for $\wt\mu=\wt\nu=4$ and for several choices of $n$. We show the result up to $r=1/3$. It is clear from the figure that the series in (\ref{eq_Hmunur1r2}) converges well for our choices of parameters.

We have selected two representative signals obtained in this section and plot them in Fig.\ \ref{fig_signal}. In both cases we have fixed $k_1=k_2=k_3=k_4$, and we show the signal by varying $k_s$. For the upper panel with the scalar loop, we show the result in (\ref{eq_Jscalarloop}) with $\wt\nu=4$ by the blue curve, and the light gray curve in the background shows the squeezed-limit result, setting $\mathbf{H}_{\wt\nu}(r_1,r_2)=\mathbf{H}_{\wt\nu,0}$ given in (\ref{eq_H0}). The background alternating shadings serve to highlight the phase of the signal. Similarly, we plot in the lower panel the vector-loop signal (\ref{eq_Jvectorloop}) with $\wt\nu=\wt\mu=4$ with the red curve. The background light gray curve shows the squeezed-limit result with (\ref{eq_ImunuSqueezed}). As already discussed in the first section, two signals in Fig.\ \ref{fig_signal} have the identical frequency and the angular dependence. However, they have very different phase structure, not only in the values of the phase in the squeezed limit, but also in the $r$-dependences of the phases away from the squeezed limit. 

\subsection{Discussions}
\label{subsec_discussions}

With the phases of nonlocal signals for several tree and 1-loop processes, we can make a few direct observations, listed as below. Our discussion here only focuses on the nonlocal signals in the trispectrum. Generalizations to other types of signals should be straightforward. Also, we shall ignore possible effect of non-unit sound speed $c_s\neq 1$. Modes with $c_s\neq 1$ can introduce additional phase shift and lead to distinguishable new shapes. We refer readers to \cite{Pimentel:2022fsc,Jazayeri:2022kjy} for details about this case.

\begin{enumerate}
  \item  We first use scaling exponent $\al$ in (\ref{eq_JNLpar2}) to tell tree-level processes and one-loop processes apart. As shown above, the nonlocal signals in tree-level processes typically have $\al=0$ while the nonlocal signal one-loop processes typically have $\al=3/2$. One should also bare in mind that exceptions exist in both cases. At the tree level, the spin-1 exchange with 4 identical external legs would give $\al=1$ due to the cancelation at the leading ($\al=0$) order. Also, the 1-loop process mediated by Dirac fermions gives $\al=5/2$, again, due to a cancelation at the leading $(\al=3/2)$ order. These special cases give us additional information about the underlying physical process which can also be useful.
  
 \item At the tree level, we see that the phase of a nonlocal signal depends on the mass, the spin, and the chemical potential of the intermediate particle. It also depends on the type of couplings. For tree-level processes, the mass can be measured separately from the oscillation frequency, and the spin may be independently measured from the angular dependence. The type of coupling may leave imprints in the angular dependence and in the scaling dimension as well.\footnote{We have seen an example where the coupling types could affect the scaling exponent $\al$ and the angular dependence: In the spin-1 exchanges in  Fig.\ \ref{fig_vector_tree}, the left diagram with four identical external scalars gives $\al=1$ and ``$D$-wave'' angular dependence, while the right diagram with two distinct types of external propagators gives $\al=0$ and ``$P$-wave'' angular dependence. As a second and more generic example, we can couple a spin-$s$ state $\psi_{i_1\cdots i_s}$ to the external scalar $\varphi$ in different ways: $\varphi'(\pd_{i_1}\cdots\pd_{i_n}\varphi)\pd_{i_{n+1}}\cdots \pd_{i_s}\psi_{i_1\cdots i_s}$, where $1\leq n\leq s$ is an arbitrary integer. It is clear that couplings with different $n$ lead to different angular dependences and different scaling exponents.} Then, the information from the signal phase would help us to further constrain the type of couplings. In the lucky situations where the coupling type is separately known, the signal phase in the deep squeezed limit can be used for consistency check.
  
  \item The existence of a chemical potential can be diagnosed in several ways. At the tree-level, a helicity-dependent chemical potential normally leads to parity-breaking shapes that can be measured from angular dependences. Alternatively, we observe that, in the presence of a chemical potential, the phase evolves faster as a function of momentum ratios $r_{1,2}$ for not-so-squeezed configurations. The technical reason for this faster dependence is that the nonlocal signal contain a hypergeometric function with argument $2r/(1\pm r)$ which goes as $\order{r}$ in the small $r$ limit. On the contrary, for dS covariant states without chemical potentials, the corresponding hypergeometric function has the argument $r^2$, which is of higher order than $\order{r}$ when $r\to 0$, implying a weaker dependence of the signal phase on $r$. We suspect this different small $r$ behavior has its origin in the mode equation: The dS covariant scalar mode equation with momentum $k_s$ is invariant under $k_s\to -k_s$, while this symmetry is broken by the chemical potential term in the spin-1 mode equation. Therefore, one can infer the existence of the chemical potential by measuring the evolution of the phase as a function of momentum ratio. We note that this method also applies to 1-loop processes: The phases in the loop signals with chemical potential typically show stronger dependences on the momentum ratio, as is clear from Fig.\ \ref{fig_signal}.

  \item Suppose that we have measured the mass from the oscillation frequency and the chemical potential from the phase evolution. Then, it remains to determine the spin and the coupling types. For tree-level processes, this information is usually known from the angular dependence as mentioned above. For 1-loop processes, the angular dependence is not enough. In such cases, the phase information can again be helpful. The key point is that the spin and the coupling types are discrete variables. So, there are only a discrete number of possible values for the phase with fixed mass and chemical potential, which can be compiled into a table. Then, from a given signal, one can determine the spin and the coupling types by comparing the measured phase with the ``phase table.''  
      
  \item It is well possible that a number of processes can generate nonlocal CC signals at the same time. In such cases one can use filtering techniques to filter out as many individual signals as possible, and then extracts the phase information from each oscillating signal. We note that different processes are added up directly in the correlation function, so there is no interference effect from different processes, as opposed to cross sections measured at the terrestrial collider experiments. 
  
  \item Finally, the phase shift in the momentum ratios $r_{1,2}$ is described by rather complicated functions of both the momentum ratios $r_{1,2}$ and the model parameters such as the mass and the chemical potential. The dependence is particularly complicated in the one-loop processes. In practice, one could use of the phase function in its expanded form, such as in (\ref{eq_H0})-(\ref{eq_H4}). It would also be useful to look for simpler functions to approximately fit the phase function for most values of $r_{1,2}$. We leave the search for such fitting functions in a future work. 

\end{enumerate}

\section{Conclusions and Outlooks}
\label{sec_discussions}

The phase of the cosmological collider signals contains rich physical information about the intermediate heavy particles, including its mass, the spin, the chemical potential, as well as the interaction form. A good understanding of the phase is essential for both digging the signal out from observational data and extracting physical properties of the intermediate heavy particles. 

In this paper, we have initiated a study of phase of oscillating CC signals, and computed the phases of several typical CC signals at both tree and 1-loop levels. In particular, we have allowed for chemical-potential-enhanced processes that breaks both the parity and dS boosts. Such processes can have large signal size and have good prospects for future cosmological observations. 

Extracting the phase of CC signals requires a more precise computation of the signal away from the squeezed limit. In this work, we have computed the nonlocal CC signals analytically, to all orders in the momentum ratios $k_s/k_{12}$ and $k_s/k_{34}$. The tree signals we obtained were known in the literature while our 1-loop results are new.

There are physical effects that can affect the signal phase during the inflation that we did not include in this work. The most important one is the slow-roll correction. The slow-roll correction can in principle be measured from the scale dependence of $n$-point correlators, including the power spectrum. One can then use the measured scale dependence as the input to compute its effect on the phase. Including these effects probably requires a more delicate numerical approach. 

The slow-roll correction can also make the masses of the intermediate states time dependent, and thereby generates signals with scale-dependent frequencies and scale-dependent magnitudes \cite{Wang:2019gok,Reece:2022soh}. The phases in such scenarios can in principle be calculated as well. As a side remark, we note that the scale dependence of the frequency induced by the slow-roll corrections persists in the deeply squeezed limit, while the phase considered in this work reduces to a constant in the same limit. Therefore, the scale-dependent frequencies can be disentangled from the $r_{1,2}$-dependent phases studied in this work, by measuring the signals in the deeply squeezed limit $r_{1,2}\ll 1$.

In this work we only computed the nonlocal signals in the trispectrum as a proof of concept. It would be interesting to extend this analysis to local signals in the trispectrum, and also to the CC signals in the bispectrum (3-point correlators). Finally, one can make use of the phase information to develop better templates for future observations. We leave all these topics for future studies.

\paragraph{Acknowledgments.} We thank Xi Tong and Dong-Gang Wang for helpful discussions. We also thank Xingang Chen and Xi Tong for useful comments on an earlier draft. This work is supported by the National Key R\&D Program of China (2021YFC2203100), an Open Research Fund of the Key Laboratory of Particle Astrophysics and Cosmology, Ministry of Education of China, and a Tsinghua University Initiative Scientific Research Program.

\newpage
\begin{appendix}

\section{Summary of Notations and Conventions}
\label{app_notation}

In this paper we have assumed the background spacetime to be the inflation patch of the dS. We use the inflation coordinates $(\tau,\mb x)$ with $\tau\in(-\infty,0)$ being the conformal time and $\mb x$ being the comoving spatial coordinates. The distance element is $\di s^2=a^2(\tau)(-\di\tau^2+\di\mb x^2)$ with the scale factor $a(\tau)=-1/(H\tau)$ and the Hubble parameter $H$ being constant. We take $H=1$ in this paper with a few exceptions where we spell $H$ out explicitly.  We use mostly plus signature for the metric.

Below we list the frequently used momentum variables and momentum ratios in the text:
\begin{align}
  &k_{s}=|\mb k_1+\mb k_2|, 
  &&k_t=|\mb k_1+\mb k_4|, 
  &&k_{12}=k_1+k_2, &&k_{34}=k_3+k_4,\n\\
  &r_1=k_s/k_{12}, &&r_2=k_s/k_{34},\n\\
  &u_1=k_1/k_{12}, &&u_2=k_2/k_{12},
  &&u_3=k_3/k_{34}, &&u_4=k_4/k_{34}. \n
\end{align}

In addition, we use the following compact notation to denote products of Euler $\Gamma$ functions and their inverse:
\begin{align}
  &\Gamma\left[ z_1,\cdots,z_m \right]\equiv \Gamma(z_1)\cdots \Gamma(z_m), 
  &&\Gamma\left[\bgm z_1,\cdots,z_m \\w_1,\cdots, w_n\edm\right]\equiv\FR{\Gamma(z_1)\cdots \Gamma(z_m)}{\Gamma(w_1)\cdots \Gamma(w_n)}.
\end{align}
We also used the binomial coefficients $\binom{a}{b}$, which can be written in terms of Euler $\Gamma$ functions as:
\bge
  \binom{a}{b}\equiv\Gamma\left[\bgm a+1 \\ b+1,a-b+1\edm\right].
\ede

Finally, in Table \ref{tab_notations} we list some of frequently used symbols in the paper.

\begin{table}[tbph] 
 \centering
  \caption{List of symbols}
  \vspace{2mm}
 \begin{tabular}{lll}
  \toprule[1.5pt]
    Symbol 
   &\multicolumn{1}{c}{} & Equation \\ \hline 
   $\varphi_{\mb k}$ & Fourier mode of inflaton fluctuation & (\ref{eq_varphi4vev})\\
   $\mathcal{T}$ & four-point correlator of inflaton fluctuations  & (\ref{eq_varphi4vev})  \\ 
   $\mathcal{J}$ & dynamical factor of the 4-point correlator & (\ref{eq_Tsqueezed})\\
   $\mathcal{J}_\text{EFT},\mathcal{J}_\text{L},\mathcal{J}_\text{NL}$ 
   & backgroud, local, and nonlocal signal parts of $\mathcal{J}$ & (\ref{eq_TEFTLNL}) \\
   $\mathcal{T}_\text{EFT},\mathcal{T}_\text{L},\mathcal{T}_\text{NL}$ 
   & similar to above & \\
   $\mathcal{K}$ & kinematic factor of the 4-point correlator & (\ref{eq_Tsqueezed})\\
   $\mathcal{A}$ & signal size & (\ref{eq_JNLpar})\\
   $\mathcal{G}$ & complex coefficient of the oscillating signal & (\ref{eq_JNLpar}) \\
   $\omega$ & signal frequency & (\ref{eq_JNLpar})\\
   $\vartheta$ & signal phase & (\ref{eq_JNLpar2})\\
   $\al$ & scaling exponent of the signal & (\ref{eq_JNLpar})\\
   $\wt\nu$ & mass parameter & below (\ref{eq_TNL})\\
   $\wt\mu=\mu/H$ & dimensionless chemical potential for spin-1 fields & below (\ref{eq_TNLh}) \\
   $G_\mathsf{a}(k;\tau_1)$ & bulk-to-boundary propagator of inflaton fluctuation $\varphi$ & (\ref{eq_scalarloopInt}) \\
   $D_\mathsf{ab}(k;\tau_1,\tau_2)$ & bulk propagator of intermediate massive field &  (\ref{eq_scalarloopInt}) \\
   \bottomrule[1.5pt] 
 \end{tabular}
 \label{tab_notations}
 \end{table}

\section{Partial Mellin-Barnes Representation}
\label{app_MB}

The Mellin-Barnes (MB) representation is a way to represent various types of functions in terms of contour integrals on the complex plane, with the integrand typically containing products of Euler $\Gamma$ functions. The MB representation is sometimes convenient for evaluating Schwinger-Keldysh integrals with special functions in the integrand. By representing those special functions, typically Hankel functions or Whittaker functions, as MB integrals, the original time integral or loop momentum integral can often be trivialized and directly finished. Then, one can finish the final contour integral from the MB representation by picking up appropriate poles of the integrand and applying the residue theorem.

The MB representation has already found wide application in computation of flat-space Feynman diagrams \cite{Smirnov:2006ry}. It could be particularly convenient in AdS and dS space, since the MB representation is nothing but the inverse Mellin transformation, which can be viewed as a sort of Fourier transformation in the radial direction (in AdS) or the time direction (in dS). This motivates a complete Mellin-space representation of cosmic correlators in \cite{Sleight:2019mgd,Sleight:2019hfp,Sleight:2020obc,Sleight:2021plv}. However, for our applications in CC physics, we still keep the external modes pinned at the future boundary $\tau=0$, and therefore, we shall only perform the inverse Mellin transformation to the internal modes, a strategy we call the partial MB representation. 

Below, we briefly review the Mellin-Barnes (MB) representation, and explicitly show the MB representations of Hankel functions and Whittaker functions, which appear in the mode functions, and thus the propagators, of scalar fields and gauge bosons with chemical potential, respectively.  

Let $f(z)$ be a locally integrable function on $(0,\infty)$, the Mellin transform of $f(z)$ is defined as:
\begin{equation}
  F(s) = \mathcal M \big[f(z)\big]= \int_0^\infty \di z\, z^{s-1} f(z),
\end{equation}
where $s$ is a complex number. In usual case, $F(s)$ is analytic in a strip $a<\Re s < b$, and the inverse transform is given as a Barnes integral:
\begin{equation}
  f(z) = \FR{1}{2\pi \ii}\int_{c-\ii\infty}^{c+\ii\infty} \di s\,z^{-s} F(s),~~~~ a<c<b.
\end{equation}
This kind of Barnes-integral-like representation is known as the MB representation.

Below, we list out the MB representation of several sepcial functions, namely Hankel functions and Whittaker functions:

\bge
\label{eq_Hmb}
  \text{H}_{\nu}^{(j)}(az)=\int_{-\ii\infty}^{\ii\infty}\FR{\di s}{2\pi\ii}\,\FR{(az/2)^{-s}}{2\pi}e^{(-1)^{j+1}(s-\nu-1)\pi\ii/2}\Gamma\Big[\FR{s-\nu}{2},\FR{s+\nu}{2}\Big],~~~~(j=1,2)
\ede

\bge
\label{eq_Wmb}
  \text{W}_{\mu,\nu}(az)=e^{-az/2}\int_{-\infty}^{\infty} \FR{\di z}{2\pi \ii}\,(az)^{-s}\Gamma\left[\bgm \fr12+\nu+s,\fr12-\nu+s,-\mu-s \\ \fr12+\nu-\mu,\fr12-\nu-\mu\edm\right].
\ede

In particular, we write down the MB representation of these functions with certain arguments ($r,z>0$):
\begin{equation}
  \left\{
  \begin{aligned}  
  &\text{H}_{\ii\wt\nu}^{(1)}(rz)=-\FR{\ii}{\pi}e^{\pi\wt\nu/2}\int_{-\ii\infty}^{\ii\infty} \FR{\di s}{2\pi \ii}\, e^{\ii\pi s}\Big(\FR{rz}2\Big)^{-2s}\Gamma\Big[s-\ii\FR{\wt\nu}2,s+\ii\FR{\wt\nu}2\Big],\\
  &\text{H}_{-\ii\wt\nu}^{(2)}(rz)=\FR{\ii}{\pi}e^{\pi\wt\nu/2}\int_{-\ii\infty}^{\ii\infty} \FR{\di s}{2\pi \ii}\, e^{-\ii\pi s}\Big(\FR{rz}2\Big)^{-2s}\Gamma\Big[s-\ii\FR{\wt\nu}2,s+\ii\FR{\wt\nu}2\Big],
  \end{aligned}
  \right.
\end{equation}
where we have changed the variable $s\to 2s$ in \eqref{eq_Hmb}, and
\begin{equation}
  \left\{
  \begin{aligned}  
  &\text{W}_{-\ii\wt\mu,\ii\wt\nu}(-2\ii rz)=e^{-\ii\pi/4+\ii rz}\int_{-\ii\infty}^{\ii\infty} \FR{\di s}{2\pi \ii}\,e^{\ii\pi s/2}(2rz)^{1/2-s}\Gamma\left[\bgm s+\ii\wt\nu,s-\ii\wt\nu,-s+\fr12+\ii\wt\mu  \\ \fr12+\ii\wt\nu+\ii\wt\mu,\fr12-\ii\wt\nu+\ii\wt\mu\edm\right],\\
  &\text{W}_{\ii\wt\mu,-\ii\wt\nu}(2\ii rz)=e^{\ii\pi/4-\ii rz}\int_{-\ii\infty}^{\ii\infty} \FR{\di s}{2\pi \ii}\,e^{-\ii\pi s/2}(2rz)^{1/2-s}\Gamma\left[\bgm s+\ii\wt\nu,s-\ii\wt\nu,-s+\fr12-\ii\wt\mu  \\ \fr12+\ii\wt\nu-\ii\wt\mu,\fr12-\ii\wt\nu-\ii\wt\mu\edm\right],
  \end{aligned}
  \right.
\end{equation}
where we have changed the variable $s\to s-1/2$ and deformed the integration path in \eqref{eq_Wmb}.

Finally, as we will show in the next section, we always use the MB representation of the bulk propagators. So below we list the SK propagators for both the scalar and the gauge boson with chemical potential.

For the scalar:
\begin{equation}
\left\{
\begin{aligned}
  D_{-+}(k_s;\tau_1,\tau_2) =&~ \FR\pi 4 e^{-\pi\wt\nu} (\tau_1\tau_2)^{3/2} \text{H}_{\ii\wt\nu}^{(1)}(-k_s\tau_1)\text{H}_{-\ii\wt\nu}^{(2)}(-k_s\tau_2)\\
  =&~\FR{1}{4\pi} \int_{-\ii\infty}^{\ii\infty} \FR{\di s_1}{2\pi \ii}\FR{\di s_2}{2\pi \ii}\, e^{\ii\pi(s_1-s_2)}\Big(\FR{k_s}{2}\Big)^{-2s_{12}}(-\tau_1)^{-2s_1+3/2}(-\tau_2)^{-2s_2+3/2}\\
  &\times \Gamma\Big[s_1-\ii\FR{\wt\nu}2,s_1+\ii\FR{\wt\nu}2,s_2-\ii\FR{\wt\nu}2,s_2+\ii\FR{\wt\nu}2\Big],\\
  D_{+-}(k_s;\tau_1,\tau_2) =&~ D_{-+}^*(k_s;\tau_1,\tau_2),\\
  D_{++}(k_s;\tau_1,\tau_2) =&~ D_{-+}(k_s;\tau_1,\tau_2)\theta(\tau_1-\tau_2)+D_{+-}(k_s;\tau_1,\tau_2)\theta(\tau_2-\tau_1),\\
  D_{--}(k_s;\tau_1,\tau_2) =&~ D_{+-}(k_s;\tau_1,\tau_2)\theta(\tau_1-\tau_2)+D_{-+}(k_s;\tau_1,\tau_2)\theta(\tau_2-\tau_1).
\end{aligned}
\right.
\end{equation}

For the gauge boson with chemical potential $\wt \mu = \mu/H$, different helicity modes $h=\pm,0$ have different mode functions.
In general,
\begin{equation}
  D_{\mathsf{ab},\mu\nu}^{(h)}(\mathbf k_s;\tau_1,\tau_2) = D^{(h)}_{\mathsf{ab}}(k_s;\tau_1,\tau_2)e^{(h)}_\mu(\mathbf k_s) e^{(h)*}_\nu(\mathbf k_s).
\end{equation}

For $h=\pm$,
\begin{equation}
    \left\{
\begin{aligned}
  D_{-+}^{(h)}(k_s;\tau_1,\tau_2) =&~ \FR{e^{-h\pi\wt\mu}}{2k_s} \text{W}_{\ii h\wt\mu,\ii\wt\nu}(2\ii k_s \tau_1)\text{W}_{-\ii h\wt\mu,-\ii\wt\nu}(-2\ii k_s \tau_2)\\
  =&~\FR{e^{-h\pi\wt\mu}}{2\pi^2}(\cosh2\pi\wt\mu+\cosh2\pi\wt\nu)
  e^{-\ii k_s\tau_1 + \ii k_s \tau_2}\\
  &\times \int_{-\ii\infty}^{\ii\infty} \FR{\di s_1}{2\pi \ii}\FR{\di s_2}{2\pi \ii}\, e^{\ii\pi(s_1-s_2)/2}(2k_s)^{-s_{12}}
  (-\tau_1)^{1/2-s_1}(-\tau_2)^{1/2-s_2}\\
  &\times \Gamma\Big[-s_1+\FR12-\ii h\wt\mu,-s_2+\FR12+\ii h\wt\mu,s_1-\ii\wt\nu,s_1+\ii\wt\nu,s_2-\ii\wt\nu,s_2+\ii\wt\nu\Big],\\
  D_{+-}^{(h)}(k_s;\tau_1,\tau_2) =&~ D_{-+}^{(h)*}(k_s;\tau_1,\tau_2),\\
  D_{++}^{(h)}(k_s;\tau_1,\tau_2) =&~ D_{-+}^{(h)}(k_s;\tau_1,\tau_2)\theta(\tau_1-\tau_2)+D_{+-}(k_s;\tau_1,\tau_2)\theta(\tau_2-\tau_1),\\
  D_{--}^{(h)}(k_s;\tau_1,\tau_2) =&~ D_{+-}^{(h)}(k_s;\tau_1,\tau_2)\theta(\tau_1-\tau_2)+D_{-+}(k_s;\tau_1,\tau_2)\theta(\tau_2-\tau_1).
\end{aligned}
\right.
\end{equation}

For $h=0$, we explicilty write down each component of the mode function (we have set $H=1$, so $m=m/H$):
\begin{align}
  A^{(L)}_0(\tau,k) =&~ e^{\ii\pi/4}\FR{\sqrt \pi}{2m\sqrt{k}}e^{-\pi \wt\nu/2}(-k\tau)^{3/2}\text{H}_{\ii\wt\nu}^{(1)}(-k\tau),\\
  \mb A^{(L)}(\tau,k) =&~ e^{3\ii\pi/4}\FR{\sqrt \pi}{2m\sqrt{k}}e^{-\pi \wt\nu/2}(-k\tau)^{1/2}\Big(\tau\partial_\tau-\FR12\Big)\text{H}_{\ii\wt\nu}^{(1)}(-k\tau) \wh{\mb k}.
\end{align}
In particular, the spatial component of the propagators can be expressed as
\begin{equation}
  D^{(L)}_{\mathsf{ab},ij}(\mathbf k_s;\tau_1,\tau_2) = D^{(L)}_{\mathsf{ab}}(k_s;\tau_1,\tau_2)e^{(0)}_i(\mathbf k_s)e^{(0)*}_j(\mathbf k_s),
\end{equation}
where
\begin{equation}
    \left\{
\begin{aligned}
  D^{(L)}_{-+}(k_s;\tau_1,\tau_2) =&~ \FR{\pi}{4m^2} e^{-\pi\wt\nu} (\tau_1\tau_2)^{1/2} \Big(\tau_1\partial_{\tau_1}-\FR12\Big)\Big(\tau_2\partial_{\tau_2}-\FR12\Big)\text{H}_{\ii\wt\nu}^{(1)}(-k_s\tau_1)\text{H}_{-\ii\wt\nu}^{(2)}(-k_s\tau_2)\\
  =&~\FR{1}{4\pi m^2} \int_{-\ii\infty}^{\ii\infty} \FR{\di s_1}{2\pi \ii}\FR{\di s_2}{2\pi \ii}\, e^{\ii\pi(s_1-s_2)}\Big(\FR{k_s}{2}\Big)^{-2s_{12}}(-\tau_1)^{-2s_1+1/2}(-\tau_2)^{-2s_2+1/2}\\
  &\times \Big(2s_1+\FR12\Big)\Big(2s_2+\FR12\Big)\Gamma\Big[s_1-\ii\FR{\wt\nu}2,s_1+\ii\FR{\wt\nu}2,s_2-\ii\FR{\wt\nu}2,s_2+\ii\FR{\wt\nu}2\Big],\\
  D^{(L)}_{+-}(k_s;\tau_1,\tau_2) =&~ D^{(L)*}_{-+}(k_s;\tau_1,\tau_2),\\
  D^{(L)}_{++}(k_s;\tau_1,\tau_2) =&~ D^{(L)}_{-+}(k_s;\tau_1,\tau_2)\theta(\tau_1-\tau_2)+D^{(L)}_{+-}(k_s;\tau_1,\tau_2)\theta(\tau_2-\tau_1),\\
  D^{(L)}_{--}(k_s;\tau_1,\tau_2) =&~ D^{(L)}_{+-}(k_s;\tau_1,\tau_2)\theta(\tau_1-\tau_2)+D^{(L)}_{-+}(k_s;\tau_1,\tau_2)\theta(\tau_2-\tau_1).
\end{aligned}
\right.
\end{equation}
\section{Intermediate Steps}
\label{app_details}

The MB representation is very useful when we calculate the correlation functions.
At the tree level, both local and nonlocal signals can be explicitly calculated;
while at 1-loop level, the nonlocal signal can be expressed as a power series in momentum ratio.
Here we summarize the main procedures:
\begin{enumerate}
  \item Write down the expression of the correlation function following the diagrammatic rule \cite{Chen:2017ryl}. Apply the cutting rule to the SK propagators. This will extract the signal out of the background \cite{Tong:2021wai}.
  \item Express only the bulk propagators via the MB representation. That is,  we introduce two Mellin variables for each bulk propagator with the two endings at $\tau_1$ and $\tau_2$. Then the dependence of the Schwinger-Keldysh integral on $\tau_1$ and $\tau_2$ becomes some simple power functions and exponential functions so long as the external lines are conformal or massless scalars. We call this a partial MB representation.
  \item Analytically calculate the time integrals, and also the momentum loop integral for 1-loop processes.
  \item Complete the Barnes integrals using the residue theorem, and pick up the desired CC signals by selecting appropriate poles of the integrand.
\end{enumerate}

In this appendix, we present the intermediate calculations in details, using the partial MB representation shown above. 
\subsection{Scalar tree diagrams}
\paragraph{\bm{$\lam_T^2$} term}
First we consider the following interaction as a simple example:
\begin{equation}
  \Delta \ld = \FR12 \lam_T a^2\varphi'^2\sigma,
\end{equation}
where $\varphi$ is a massless scalar, and the bulk-to-boundary propagator is given by:
\begin{equation}
  G_{\mp}(k,\tau) = \FR{1}{2k^3}(1\pm \ii k\tau)e^{\mp \ii k\tau}.
\end{equation}
This is the case if we take $\lam_S \to 0$ in \eqref{eq_lTlS}. It is straightforward to write down the 4-point function in the following expression:
\begin{align}
  \mathcal{T}_{\varphi'^2\si}=&-\lam_T^2\sum_{\mathsf{a},\mathsf{b}=\pm}\mathsf{ab}\int_{-\infty}^0\FR{\di\tau_1}{(-\tau_1)^2}\FR{\di\tau_2}{(-\tau_2)^2}G_\mathsf{a}'(k_1,\tau_1)G_\mathsf{a}'(k_2,\tau_1)G_\mathsf{b}'(k_3,\tau_2)G_\mathsf{b}'(k_4,\tau_2)D_\mathsf{ab}(k_s;\tau_1,\tau_2)\n\\
  =&-\FR{\lam_T^2}{16k_1\cdots k_4}\sum_{\mathsf{a},\mathsf{b}}\mathsf{ab}\int_{-\infty}^0 \di\tau_1  \di\tau_2\,e^{\ii\mathsf{a}k_{12}\tau_1+\ii\mathsf{b}k_{34}\tau_2} D_\mathsf{ab}(k_s;\tau_1,\tau_2).
\end{align}
Since we only focus on the signal, we can apply the cutting rule, namely:
\begin{align}
  D_{\pm\pm}(k_s;\tau_1,\tau_2) &= D_{\mp\pm}(k_s;\tau_1,\tau_2)\theta(\tau_1-\tau_2)+D_{\pm\mp}(k_s;\tau_1,\tau_2)\theta(\tau_2-\tau_1)\n\\
  &\To 
  D_{\mp\pm}(k_s;\tau_1,\tau_2)\theta(k_{12}-k_{34})+D_{\pm\mp}(k_s;\tau_1,\tau_2)\theta(k_{34}-k_{12}). 
\end{align}
If we focus on the case $k_{12}>k_{34}$, then we obtain:
\begin{align}
  \mathcal{T}_{\text{signal},\varphi'^2\si}
  =&~\FR{\lam_T^2}{16k_1\cdots k_4}
  \int_{-\infty}^0 \di\tau_1  \di\tau_2\,\big(e^{-\ii k_{12}\tau_1+\ii k_{34}\tau_2}
  -e^{\ii k_{12}\tau_1+\ii k_{34}\tau_2}\big)\times D_{-+}(k_s;\tau_1,\tau_2)+\text{c.c.}.
\end{align}

Next, we express the propagator $D_{-+}(k_s;\tau_1,\tau_2)$ by MB representation, and simplify the integral by introducing dimensionless variables $z_1 = -k_{12} \tau_1$, $z_2 = -k_{34}\tau_2$ and momentum ratios $r_1 = k_s/k_{12}$, $r_2=k_s/k_{34}$:
\begin{align}
  \mathcal{T}_{\text{signal},\varphi'^2\si}
  =&~\FR{\lam_T^2}{64\pi k_1\cdots k_4(k_{12}k_{34})^{5/2}}
  \int_{-\ii\infty}^{\ii\infty} \FR{\di s_1}{2\pi \ii}\FR{\di s_2}{2\pi \ii}\,
  e^{\ii\pi(s_1-s_2)}\Big(\FR{r_1}{2}\Big)^{-2s_1}\Big(\FR{r_2}{2}\Big)^{-2s_2}\n\\
  &\times
  \int_0^\infty \di z_1  \di z_2\,z_1^{-2s_1+3/2}z_2^{-2s_2+3/2}
  \big(e^{\ii z_1-\ii z_2}
  -e^{-\ii z_1-\ii z_2}\big)\n\\
  &\times \Gamma\Big[s_1-\ii\FR{\wt\nu}2,s_1+\ii\FR{\wt\nu}2,s_2-\ii\FR{\wt\nu}2,s_2+\ii\FR{\wt\nu}2\Big]
  +\text{c.c.}.
\end{align}
The integral in the second line can be calculated analytically:
\begin{align}
  &\int_0^\infty \di z_1  \di z_2\,z_1^{-2s_1+3/2}z_2^{-2s_2+3/2}
  \big(e^{\ii z_1-\ii z_2}
  -e^{-\ii z_1-\ii z_2}\big)\n\\
  =&~\big(e^{-\ii\pi(s_1-s_2)}+\ii e^{\ii\pi s_{12}}\big)\times \Gamma\Big[-2s_1+\FR52,-2s_2+\FR52\Big],
\end{align}
so the signal can be expressed as:
\begin{align}
  \label{eq_scaTtree}
  \mathcal{T}_{\text{signal},\varphi'^2\si}
  =&~\FR{\lam_T^2}{64\pi k_1\cdots k_4(k_{12}k_{34})^{5/2}}
  \int_{-\ii\infty}^{\ii\infty} \FR{\di s_1}{2\pi \ii}\FR{\di s_2}{2\pi \ii}\,
  \Big(\FR{r_1}{2}\Big)^{-2s_1}\Big(\FR{r_2}{2}\Big)^{-2s_2}\n\\
  &\times
  \big(1+\ii e^{2\ii\pi s_1}\big)\times \Gamma\Big[-2s_1+\FR52,-2s_2+\FR52\Big]\n\\
  &\times \Gamma\Big[s_1-\ii\FR{\wt\nu}2,s_1+\ii\FR{\wt\nu}2,s_2-\ii\FR{\wt\nu}2,s_2+\ii\FR{\wt\nu}2\Big]
  +\text{c.c.}.
\end{align}

Finally, we calculate the complex integral using the residue theorem. We close the contour from left,
then all the poles are from the gamma functions in the last line of \eqref{eq_scaTtree}. There are four sets of poles:
\begin{equation}
  \label{eq_scapoletree}
  \left\{
    \begin{aligned}
    s_1 = -n_1 - \ii\FR{\wt\nu}2,\quad s_2 = -n_2 - \ii\FR{\wt\nu}2,\\
    s_1 = -n_1 + \ii\FR{\wt\nu}2,\quad s_2 = -n_2 + \ii\FR{\wt\nu}2,\\
    s_1 = -n_1 - \ii\FR{\wt\nu}2,\quad s_2 = -n_2 + \ii\FR{\wt\nu}2,\\
    s_1 = -n_1 + \ii\FR{\wt\nu}2,\quad s_2 = -n_2 - \ii\FR{\wt\nu}2.
    \end{aligned}
  \right.
\end{equation}
We can observe that the first two sets of poles in \eqref{eq_scapoletree}
contribute to the nonlocal signal, namely proportional to $(r_1r_2)^{\pm \ii\wt\nu}$, and the last two sets contribute to the local signal,
proportional to $(r_1/r_2)^{\pm \ii\wt\nu}$. 

If we focus on the nonlocal signal, we sum over the residues of the first two sets of poles.
The factor $1+\ii e^{2\ii\pi s_1}=1+\ii e^{\pm \pi\wt\nu}$ at each pole, and we obtain:
\begin{align}
  \mathcal{T}_{\text{NL},\varphi'^2\si}
  =&~\FR{\lam_T^2}{64\pi k_1\cdots k_4(k_{12}k_{34})^{5/2}}
  \bigg\{\Big(\FR{r_1r_2}4\Big)^{\ii\wt\nu}\big(1+\ii e^{\pi\wt\nu}\big)\n\\
  &\times \sum_{n_1,n_2=0}^\infty \FR{(-1)^{n_{12}}}{n_1!n_2!}
  \Big(\FR{r_1}2\Big)^{2n_1}\Big(\FR{r_2}2\Big)^{2n_2}\n\\
  &\times\Gamma\Big[2n_1+\FR52+\ii\wt\nu,2n_2+\FR52+\ii\wt\nu,-n_1-\ii\wt\nu,-n_2-\ii\wt\nu\Big]+(\wt\nu\to-\wt\nu)\bigg\}
  +\text{c.c.}\n\\
  =&~\FR{\lam_T^2}{32\pi k_1\cdots k_4(k_{12}k_{34})^{5/2}}
  \Big(\FR{r_1r_2}4\Big)^{\ii\wt\nu}(1+\ii\sinh\pi\wt\nu)\n\\
  &\times \Gamma^2(-\ii\wt\nu)\Gamma^2\Big(\FR52+\ii\wt\nu\Big)
  \mathbf{F}_{\wt\nu}(r_1)\mathbf{F}_{\wt\nu}(r_2)+\text{c.c.},
\end{align}
where we have defined:
\bge
\mathbf{F}_{\wt\nu}(r)\equiv {}_2\mathrm{F}_1\left[\bgm\fr{5}{4}+\fr{\ii\wt\nu}{2},\fr{7}{4}+\fr{\ii\wt\nu}{2}\\1+\ii\wt\nu\edm\middle|\,r^2\right].
\ede
Notice that the above nonlocal signal is symmetric with $r_1\leftrightarrow r_2$. Therefore, the same result also applies to the case of $k_{12}<k_{34}$.

\paragraph{\bm{$\lam_S^2$} term}

Now we calculate the nonlocal signal from the interaction:
\begin{equation}
  \Delta\ld = \FR12 \lam_S a^2 (\partial_i\varphi)^2 \sigma.
\end{equation}
In the momentum space, $\partial_i \to \ii k_i$. Applying the cutting rule and taking $k_{12}>k_{34}$, the signal can be expressed as:
\begin{align}
  \mathcal{T}_{\text{signal},(\partial_i \varphi)^2\si}
  =&~\FR{\lam_S^2(\mb k_1\cdot \mb k_2)(\mb k_3\cdot \mb k_4)}{16(k_1\cdots k_4)^3}
  \Big(1-k_{12}\partial_{k_{12}}+k_1k_2\partial^2_{k_{12}}\Big)\Big(1-k_{34}\partial_{k_{34}}+k_3k_4\partial^2_{k_{34}}\Big)\n\\
  &\times\int_{-\infty}^0 \FR{\di\tau_1}{\tau_1^2}  \FR{\di\tau_2}{\tau_2^2}\,
  \big(e^{-\ii k_{12}\tau_1+\ii k_{34}\tau_2}
  -e^{\ii k_{12}\tau_1+\ii k_{34}\tau_2}\big)
  \times D_{-+}(k_s;\tau_1,\tau_2)+\text{c.c.}.
\end{align}
Here the operator in the second line is introduced by observation:
\begin{equation}
  \label{eq_difoperator}
  (1-k_{12}\partial_{k_{12}}+k_1k_2\partial_{k_{12}}^2)e^{\mp \ii k_{12}\tau_1}=(1\pm \ii k_1\tau_1)(1\pm \ii k_2\tau_1)e^{\mp \ii k_{12}\tau_1}.
\end{equation}
Introduce $u_1=k_1/k_{12}$, $u_2=k_2/k_{12}$, $u_3=k_3/k_{34}$, $u_4=k_4/k_{34}$, and $z_i$ and $r_i$ as before, we express the signal in the MB representation:
\begin{align}
  \mathcal{T}_{\text{signal},(\partial_i \varphi)^2\si}
  =&~\FR{\lam_S^2(\mb k_1\cdot \mb k_2)(\mb k_3\cdot \mb k_4)}{64\pi(k_1\cdots k_4)^3(k_{12}k_{34})^{1/2}}
  \mathcal D_1\mathcal D_2
  \int_{-\ii\infty}^{\ii\infty} \FR{\di s_1}{2\pi \ii}\FR{\di s_2}{2\pi \ii}\,
  e^{\ii\pi(s_1-s_2)}\Big(\FR{r_1}{2}\Big)^{-2s_1}\Big(\FR{r_2}{2}\Big)^{-2s_2}\n\\
  &\times
  \int_0^\infty \di z_1  \di z_2\,z_1^{-2s_1-1/2}z_2^{-2s_2-1/2}
  \big(e^{\ii z_1-\ii z_2}
  -e^{-\ii z_1-\ii z_2}\big)\n\\
  &\times \Gamma\Big[s_1-\ii\FR{\wt\nu}2,s_1+\ii\FR{\wt\nu}2,s_2-\ii\FR{\wt\nu}2,s_2+\ii\FR{\wt\nu}2\Big]
  +\text{c.c.}.
\end{align}
where $\mathcal D_i$ is the differential operator obtained by changing $\partial_{k_{12}}$, $\partial_{k_{34}}$ in \eqref{eq_difoperator} to $\partial_{r_1}$, $\partial_{r_2}$, respectively:
\begin{equation}
  \mathcal D_1 \equiv \FR32 +r_1\partial_{r_1}+u_1u_2\big(r_1^2\partial_{r_1}^2+3r_1\partial_{r_1}+\FR34\big),\quad \mathcal D_2 \equiv \FR32+r_2\partial_{r_2}+u_3u_4\big(r_2^2\partial_{r_2}^2+3r_2\partial_{r_2}+\FR34\big).
\end{equation}
Complete the time integral and we obtain:
\begin{align}
  \label{eq_scaStree}
  \mathcal{T}_{\text{signal},(\partial_i \varphi)^2\si}
  =&~\FR{\lam_S^2(\mb k_1\cdot \mb k_2)(\mb k_3\cdot \mb k_4)}{64\pi (k_1\cdots k_4)^3(k_{12}k_{34})^{1/2}}
  \mathcal D_1\mathcal D_2
  \int_{-\ii\infty}^{\ii\infty} \FR{\di s_1}{2\pi \ii}\FR{\di s_2}{2\pi \ii}\,
  \Big(\FR{r_1}{2}\Big)^{-2s_1}\Big(\FR{r_2}{2}\Big)^{-2s_2}\n\\
  &\times
  \big(1+\ii e^{2\ii\pi s_1}\big)\times \Gamma\Big[-2s_1+\FR12,-2s_2+\FR12\Big]\n\\
  &\times \Gamma\Big[s_1-\ii\FR{\wt\nu}2,s_1+\ii\FR{\wt\nu}2,s_2-\ii\FR{\wt\nu}2,s_2+\ii\FR{\wt\nu}2\Big]
  +\text{c.c.}.
\end{align}
To do the complex integral, we also close the contour from left, and thus the poles are still given by \eqref{eq_scapoletree}.
We focus on the nonlocal signal:
\begin{align}
  \mathcal{T}_{\text{NL},(\partial_i \varphi)^2\si}
  =&~\FR{\lam_S^2(\mb k_1\cdot \mb k_2)(\mb k_3\cdot \mb k_4)}{64\pi (k_1\cdots k_4)^3(k_{12}k_{34})^{1/2}}
  \mathcal D_1\mathcal D_2 \bigg\{ \Big(\FR{r_1r_2}4\Big)^{\ii\wt\nu}\big(1+\ii e^{\pi\wt\nu}\big)\n\\
  &\times \sum_{n_1,n_2=0}^\infty \FR{(-1)^{n_{12}}}{n_1!n_2!}
  \Big(\FR{r_1}2\Big)^{2n_1}\Big(\FR{r_2}2\Big)^{2n_2}\n\\
  &\times \Gamma\Big[2n_1+\FR52+\ii\wt\nu,2n_2+\FR52+\ii\wt\nu,-n_1-\ii\wt\nu,-n_2-\ii\wt\nu\Big]+(\wt\nu\to-\wt\nu)\bigg\}
  +\text{c.c.}\n\\
  =&~\FR{\lam_S^2(\mb k_1\cdot \mb k_2)(\mb k_3\cdot \mb k_4)}{32\pi (k_1\cdots k_4)^3(k_{12}k_{34})^{1/2}}\n\\
  &\times \Big(\FR{r_1r_2}4\Big)^{\ii\wt\nu}(1+\ii\sinh\pi\wt\nu)\Gamma^2(-\ii\wt\nu)\Gamma^2\Big(\FR12+\ii\wt\nu\Big)
  \mathbf{F}_{\wt\nu}(u_1u_2,r_1)\mathbf{F}_{\wt\nu}(u_3u_4,r_2)+\text{c.c.},
\end{align}
where we have defined:
\begin{align}
  \mathbf{F}_{\wt\nu}(\alpha,r)\equiv&~
  \big(1-\FR{\alpha}{1-r^2}(\FR14+\wt\nu^2)\big)
  {}_2\mathrm{F}_1\left[\bgm\fr{1}{4}+\fr{\ii\wt\nu}{2},\fr{3}{4}+\fr{\ii\wt\nu}{2}\\1+\ii\wt\nu\edm\middle|\,r^2\right]
  \n\\
  &+\big(1+\FR{2\alpha}{1-r^2}\big)\big(\FR12+\ii\wt\nu\big){}_2\mathrm{F}_1\left[\bgm\fr{3}{4}+\fr{\ii\wt\nu}{2},\fr{5}{4}+\fr{\ii\wt\nu}{2}\\1+\ii\wt\nu\edm\middle|\,r^2\right].
  \end{align}
Notice again that the above nonlocal signal is symmetric with $r_1\leftrightarrow r_2$. Therefore, the same result also applies to the case of $k_{12}<k_{34}$.

\paragraph{General case}
Finally we consider the general coupling \eqref{eq_lTlS}.
In principle, we should also calculate the $\lam_T\lam_S$ term and do the summation. However, at the tree level,
the signal is totally factorized, we can therefore directly write down the result:
\begin{align}
 \mathcal{T}_{\text{NL}}=\FR{\big(1+\ii\sinh\pi\wt\nu\big)\Gamma^2(-\ii\wt\nu)}{32\pi k_1\cdots k_4(k_{12}k_{34})^{5/2}}\mathcal{B}(k_1,k_2,k_s)\mathcal{B}(k_3,k_4,k_s)\Big(\FR{r_1r_2}{4}\Big)^{\ii\wt\nu}+\text{c.c.},
\end{align}
where
\begin{align}
  \mathcal{B}(k_1,k_2,k_s)= \lam_T \Gamma\Big(\FR{5}{2}+\ii\wt\nu\Big) \mathbf{F}_{\wt\nu}(\FR{k_s}{k_{12}})
  -\lam_S\FR{ k_s^2- k_1^2- k_2^2}{2 k_1 k_2}\FR{ k_{12}^2}{ k_1 k_2}\Gamma\Big(\FR{1}{2}+\ii\wt\nu\Big) \mathbf{F}_{\wt\nu}(\FR{k_1k_2}{k_{12}^2},\FR{k_s}{k_{12}}).
\end{align}
The minus sign before $\lam_S$ comes from $\partial_i^2\to - k_i^2$.

\subsection{vector tree diagrams}
Now we come to the vector case. We consider the interaction \eqref{eq_phiphiA},
and the 4-point function is:
\begin{equation}
\mathcal{T}_{\text{1}}^{(h)} =\FR{1}{4}\bigg[\mathcal{T}_{\text{2}}^{(h)}+(k_1\leftrightarrow k_2)+(k_3\leftrightarrow k_4)+\binom{k_1\leftrightarrow k_2}{k_3\leftrightarrow k_4}\bigg],
\end{equation}
where
\begin{align}
  \mathcal{T}_{\text{2}}=&-\lam^2\sum_{\mathsf{a},\mathsf{b}=\pm}\mathsf{ab}\int_{-\infty}^0\FR{\di\tau_1}{(-\tau_1)}\FR{\di\tau_2}{(-\tau_2)}\,
   G_\mathsf{a}'(k_1,\tau_1) \partial_i G_\mathsf{a}(k_2,\tau_1)
   G_\mathsf{b}'(k_3,\tau_2) \partial_j G_\mathsf{b}(k_4,\tau_2)
   \n\\
  &\times D_{\mathsf{ab},ij}(\mathbf k_s;\tau_1,\tau_2)\n\\
  =&~\FR{\lam^2}{16k_1\cdots k_4}\FR{(\mb k_2)_i(\mb k_4)_j}{k_2^2k_4^2}
  \sum_{h=\pm,0}\big[\mb e_{h}(\mathbf k_s)\big]_i \big[\mb e^*_{h}(\mathbf k_s)\big]_j\big(1 - \ii \mathsf{a}k_2\tau_1\big)\big(1 - \ii \mathsf{b}k_4\tau_2\big)\n\\
  &\times \sum_{\mathsf{a},\mathsf{b}}\mathsf{ab}\int_{-\infty}^0 \di\tau_1\di\tau_2\,
  e^{\ii\mathsf{a}k_{12}\tau_1+\ii\mathsf{b}k_{34}\tau_2}
  D_{\mathsf{ab}}^{(h)}(k_s;\tau_1,\tau_2)\n\\
  =&~\FR{\lam^2}{16k_1\cdots k_4}
  \sum_{h=\pm,0}\bigg[\FR{\big(\wh{\mb k}_2\cdot\mb{e}_h\big)\big(\wh{\mb k}_4\cdot\mb{e}_h^*\big)}{k_2k_4}
 \big(1-k_2\partial_{k_{12}}\big)\big(1-k_4\partial_{k_{34}}\big)\bigg]\n\\
  &\times \sum_{\mathsf{a},\mathsf{b}}\mathsf{ab}\int_{-\infty}^0 \di\tau_1\di\tau_2\,
  e^{\ii\mathsf{a}k_{12}\tau_1+\ii\mathsf{b}k_{34}\tau_2}
  D_{\mathsf{ab}}^{(h)}(k_s;\tau_1,\tau_2).
\end{align}
where we write $\mb e_h \equiv \mb e_h(\mb k_s)$ for abbreviation.

Since the propagators of the transverse modes and the longitudinal mode are different, we calculate their contributions separately.

\paragraph{Transverse states}

First we consider the transverse states $h=\pm$.
Apply the cutting rule with the assumption $k_{12}>k_{34}$, and then apply the partial MB representation, we obtain:
\begin{align}
  &\sum_{\mathsf{a},\mathsf{b}}\mathsf{ab}\int_{-\infty}^0 \di\tau_1\di\tau_2\,
  e^{\ii\mathsf{a}k_{12}\tau_1+\ii\mathsf{b}k_{34}\tau_2}
  D_{\mathsf{ab}}^{(\pm)}(k_s;\tau_1,\tau_2)\n\\
  \to&-\FR{e^{\mp \pi\wt\mu}}{2\pi^2(k_{12}k_{34})^{3/2}}(\cosh2\pi\wt\mu+\cosh2\pi\wt\nu)
  \int_{-\ii\infty}^{\ii\infty}
  \FR{\di s_1}{2\pi\ii}\FR{\di s_2}{2\pi\ii}\, e^{\ii\pi(s_1-s_2)/2}(2r_1)^{-s_1}(2r_2)^{-s_2}\n\\
  &\times \int_0^\infty \di z_1\di z_2\,
  z_1^{-s_1+1/2}z_2^{-s_2+1/2}e^{\ii r_1z_1 - \ii r_2z_2}
  \big(e^{\ii z_1-\ii z_2}
  -e^{-\ii z_1-\ii z_2}\big)\n\\
  &\times \Gamma\Big[-s_1+\FR12\mp \ii\wt\mu,-s_2+\FR12\pm \ii\wt\mu\Big]\times \Gamma\Big[s_1-\ii\wt\nu,s_1+\ii\wt\nu,s_2-\ii\wt\nu,s_2+\ii\wt\nu\Big]+\text{c.c.}\n\\
  =&-\FR{e^{\mp \pi\wt\mu}}{2\pi^2(k_{12}k_{34})^{3/2}}(\cosh2\pi\wt\mu+\cosh2\pi\wt\nu)
  \int_{-\ii\infty}^{\ii\infty}
  \FR{\di s_1}{2\pi\ii}\FR{\di s_2}{2\pi\ii}\, (2r_1)^{-s_1}(2r_2)^{-s_2}\n\\
  &\times \Big[(1+r_1)^{s_1-3/2}(1+r_2)^{s_2-3/2}-\ii e^{\ii\pi s_1}(1-r_1)^{s_1-3/2}(1+r_2)^{s_2-3/2}\Big]\n\\
  &\times\Gamma\Big[-s_1+\FR32,-s_2+\FR32\Big]\times \Gamma\Big[-s_1+\FR12\mp \ii\wt\mu,-s_2+\FR12\pm \ii\wt\mu\Big]\n\\
  &\times \Gamma\Big[s_1-\ii\wt\nu,s_1+\ii\wt\nu,s_2-\ii\wt\nu,s_2+\ii\wt\nu\Big]+\text{c.c.}.
\end{align}
The signal is also factorized, and there are four sets of poles:
\begin{equation}
  \label{eq_vecpoletree}
  \left\{
    \begin{aligned}
      &s_1 = -n_1 - \ii\wt\nu,\quad s_2 = -n_2 - \ii\wt\nu,\\
      &s_1 = -n_1 + \ii\wt\nu,\quad s_2 = -n_2 + \ii\wt\nu,\\
      &s_1 = -n_1 - \ii\wt\nu,\quad s_2 = -n_2 + \ii\wt\nu,\\
      &s_1 = -n_1 + \ii\wt\nu,\quad s_2 = -n_2 -\ii\wt\nu,
    \end{aligned}
  \right.
\end{equation}
and the first two sets in \eqref{eq_vecpoletree} contribute to the nonlocal signal.
To express the result in a more compact form, we write
\begin{equation}
  \int_{-\ii\infty}^{\ii\infty}\FR{\di s}{2\pi\ii}\,\FR{1}{(1+r)^{3/2}}
  \Big(\FR{2r}{1+r}\Big)^{-s}\Gamma\Big[-s+\FR32,-s+\FR12-\ii\wt\mu,s-\ii\wt\nu,s+\ii\wt\nu\Big]
  =\wt{\mathbf{G}}_{\wt\mu,\wt\nu}(r)+\wt{\mathbf{G}}_{\wt\mu,-\wt\nu}(r),
\end{equation}
where $\wt{\mb G}_{\wt\mu,\wt\nu}(r)$ is the contribution of the first set in \eqref{eq_vecpoletree}:
\begin{align}
  \wt{\mathbf{G}}_{\wt\mu,\wt\nu}(r)
  =&~\FR{1}{(1+r)^{3/2}}\sum_{n=0}^\infty \FR{(-1)^n}{n!}\Big(\FR{2r}{1+r}\Big)^{n+\ii\wt\nu}\Gamma\Big[n+\FR32+\ii\wt\nu,n+\FR12-\ii\wt\mu+\ii\wt\nu,-n-2\ii\wt\nu\Big]\n\\
  =&~\FR{1}{(1+r)^{3/2}}\Big(\FR{2r}{1+r}\Big)^{\ii\wt\nu}\Gamma\Big[-2\ii\wt\nu,\FR32+\ii\wt\nu,\FR12-\ii\wt\mu+\ii\wt\nu\Big]{}_2\mathrm{F}_1\left[\bgm\fr{3}{2}+\ii\wt\nu,\fr12-\ii\wt\mu+\ii\wt\nu\\1+2\ii\wt\nu\edm\middle|\,\FR{2r}{1+r}\right].
\end{align}
Then the nonlocal signal for $k_{12}>k_{34}$ is given by:
\begin{align}
  \mathcal{T}^{(\pm)}_{\text{NL,2},>} =& -\FR{\lam^2e^{\mp\pi\wt\mu}(\cosh 2\pi\wt\mu+\cosh2\pi\wt\nu)}{32\pi^2 k_1\cdots k_4(k_{12}k_{34})^{3/2}}\n\\
  &\times \FR{\big(\wh{\mb k}_2\cdot\mb{e}_\pm\big)\big(\wh{\mb k}_4\cdot\mb{e}_\pm^*\big)}{k_2k_4}
 \big(1+\FR32 u_2+ u_2 r_1\partial_{r_1}\big)\big(1+\FR32u_4 + u_4r_2\partial_{r_2}\big)\n\\
 &\times \bigg\{\bigg[\Big(\wt{\mathbf{G}}_{\pm \wt\mu,\wt\nu}(r_1)\wt{\mathbf{G}}_{\mp \wt\mu,\wt\nu}(r_2)
 -\ii \wt{\mathbf{G}}_{\pm \wt\mu,\wt\nu}(e^{-\ii\pi}r_1)\wt{\mathbf{G}}_{\mp \wt\mu,\wt\nu}(r_2)
 \Big)
 +(\wt\nu\to-\wt\nu)\bigg]+\text{c.c.}\bigg\}.
\end{align}
For convenience we define:
\begin{align}
 {\mb G}_{\wt\mu,\wt\nu}(u,r) \equiv&~  \big(1+\FR32 u+ u r\partial_r \big) \wt{\mb G}_{\wt\mu,\wt\nu}(r)\n\\
  =&~\FR{1}{(1+r)^{3/2}}\Big(\FR{2r}{1+r}\Big)^{\ii\wt\nu}\Gamma\Big[-2\ii\wt\nu,\FR32+\ii\wt\nu,\FR12-\ii\wt\mu+\ii\wt\nu\Big] \n\\
  &\times\bigg\{{}_2\mathrm{F}_1\left[\bgm\fr{3}{2}+\ii\wt\nu,\fr12-\ii\wt\mu+\ii\wt\nu\\1+2\ii\wt\nu\edm\middle|\,\FR{2r}{1+r}\right]
 +\FR{u(\fr32+\ii\wt\nu)}{1+r}{}_2\mathrm{F}_1\left[\bgm\fr{5}{2}+\ii\wt\nu,\fr12-\ii\wt\mu+\ii\wt\nu\\1+2\ii\wt\nu\edm\middle|\,\FR{2r}{1+r}\right]\bigg\},
\end{align}
then
\begin{align}
  &\mathcal{T}^{(\pm)}_{\text{NL},2,>}(k_1,k_2,k_3,k_4,k_s) = -\FR{\lam^2e^{\mp\pi\wt\mu}(\cosh 2\pi\wt\mu+\cosh2\pi\wt\nu)}{32\pi^2 k_1\cdots k_4(k_{12}k_{34})^{3/2}}
  \FR{\big(\wh{\mb k}_2\cdot\mb{e}_\pm\big)\big(\wh{\mb k}_4\cdot\mb{e}_\pm^*\big)}{k_2k_4}\n\\
  &\times \Big\{\Big[\Big({\mathbf{G}}_{\pm \wt\mu,\wt\nu}(u_2,r_1){\mathbf{G}}_{\mp \wt\mu,\wt\nu}(u_4,r_2)-\ii {\mathbf{G}}_{\pm \wt\mu,\wt\nu}(u_2,e^{-\ii\pi}r_1){\mathbf{G}}_{\mp \wt\mu,\wt\nu}(u_4,r_2)
 \Big)
 +(\wt\nu\to-\wt\nu)\Big]+\text{c.c.}\Big\}.
\end{align}
We note that the above result applies only to the case of $k_{12}>k_{34}$, namely when $r_1<r_2$. Although not quite obvious, one can check that this result is not symmetric with $(k_1,k_2)\leftrightarrow (k_3,k_4)$. On the other hand, the result for $k_{12}<k_{34}$ can be obatined by the direct replacement $(k_1,k_2)\leftrightarrow (k_3,k_4)$ in the above expression. Therefore, the general result for arbitrary $(r_1,r_2)$ should be written as:
\begin{align}
  \mathcal{T}^{(\pm)}_{\text{NL},2}(k_1,k_2,k_3,k_4,k_s) =&~\mathcal{T}^{(\pm)}_{\text{NL},2,>}(k_1,k_2,k_3,k_4,k_s)\theta(k_{12}-k_{34})\n\\
  &~+\mathcal{T}^{(\pm)}_{\text{NL},2,>}(k_3,k_4,k_1,k_2,k_s)\theta(k_{34}-k_{12}).
\end{align}
To make sense of this expression for $k_{12}=k_{34}$, we use the prescription that $\theta(0)=1/2$.

This complicated expression becomes much simpler in the squeezed limit $r_1,r_2\to 0$, $u_i\to 1/2$, where
\begin{equation}
   {\mb G}_{\wt\mu,\wt\nu}(u,r)\to \FR14 (2r)^{\ii\wt\nu}(7+2\ii\wt\nu)
  \Gamma\Big[-2\ii\wt\nu,\FR32+\ii\wt\nu,\FR12-\ii\wt\mu+\ii\wt\nu\Big],
\end{equation}
and
\begin{equation}
   {\mb G}_{\wt\mu,\wt\nu}(u,e^{\mp \ii\pi}r) \to e^{\pm \pi\wt\nu} {\mb G}_{\wt\mu,\wt\nu}(u,r).
\end{equation}
In particular, this squeezed-limit result is symmetric with respect to $(k_1,k_2)\leftrightarrow (k_3,k_4)$.

\paragraph{Longitudinal state}
Now we calculate the contribution of the longitudinal state. Following the similar procedure, we apply the cutting rule assuming $k_{12}>k_{34}$, and write down the partial MB representation:
\begin{align}
  &\sum_{\mathsf{a},\mathsf{b}}\mathsf{ab}\int_{-\infty}^0 \di\tau_1\di\tau_2\,
  e^{\ii\mathsf{a}k_{12}\tau_1+\ii\mathsf{b}k_{34}\tau_2}
  D_{\mathsf{ab}}^{(L)}(k_s;\tau_1,\tau_2)\n\\
  \to&-\FR{1}{4\pi m^2(k_{12}k_{34})^{3/2}}
  \int_{-\ii\infty}^{\ii\infty}
  \FR{\di s_1}{2\pi\ii}\FR{\di s_2}{2\pi\ii}\, e^{\ii\pi(s_1-s_2)}
  \Big(\FR{r_1}{2}\Big)^{-2s_1}\Big(\FR{r_2}{2}\Big)^{-2s_2}\n\\
  &\times \int_0^\infty \di z_1\di z_2\,
  z_1^{-2s_1+1/2}z_2^{-2s_2+1/2}
  \Big[e^{\ii z_1-\ii z_2}
  -\FR12\big(e^{-\ii z_1-\ii z_2}+e^{\ii z_1+\ii z_2}\big) \Big]\n\\
  &\times \Big(2s_1+\FR12\Big)\Big(2s_2+\FR12\Big)\Gamma\Big[s_1-\ii\wt\nu,s_1+\ii\wt\nu,s_2-\ii\wt\nu,s_2+\ii\wt\nu\Big]+\text{c.c.}\n\\
  =&-\FR{1}{4\pi m^2(k_{12}k_{34})^{3/2}}
  \int_{-\ii\infty}^{\ii\infty} \FR{\di s_1}{2\pi \ii}\FR{\di s_2}{2\pi \ii}\,
  \Big(\FR{r_1}{2}\Big)^{-2s_1}\Big(\FR{r_2}{2}\Big)^{-2s_2}
  \big(1-\ii e^{2\ii\pi s_1}\big)\n\\
  &\times \Big(2s_1+\FR12\Big)\Big(2s_2+\FR12\Big)\Gamma\Big[-2s_1+\FR32,-2s_2+\FR32\Big]\n\\
  &\times \Gamma\Big[s_1-\ii\FR{\wt\nu}2,s_1+\ii\FR{\wt\nu}2,s_2-\ii\FR{\wt\nu}2,s_2+\ii\FR{\wt\nu}2\Big]
  +\text{c.c.}.
\end{align}
We still focus on the nonlocal signal, and taking poles in the first two sets in \eqref{eq_vecpoletree}:
\begin{align}
  \mathcal{T}^{(L)}_{\text{NL},2} =& -\FR{\lam^2}{64\pi m^2 k_1\cdots k_4(k_{12}k_{34})^{3/2}}\n\\
  &\times \FR{\big(\wh{\mb k}_2\cdot\mb{e}_L\big)\big(\wh{\mb k}_4\cdot\mb{e}_L\big)}{k_2k_4}
  \big(1+\FR32 u_2+ u_2 r_1\partial_{r_1}\big)\big(1+\FR32u_4 + u_4r_2\partial_{r_2}\big)
  \times \bigg\{\bigg[\Big(\FR{r_1r_2}4\Big)^{\ii\wt\nu}\big(1-\ii e^{\pi\wt\nu}\big)\n\\
  &\times \sum_{n_1,n_2=0}^\infty \FR{(-1)^{n_{12}}}{n_1!n_2!}
  \Big(\FR{r_1}2\Big)^{2n_1}\Big(\FR{r_2}2\Big)^{2n_2}\Big(\FR12-2n_1-\ii\wt\nu\Big)\Big(\FR12-2n_2-\ii\wt\nu\Big)\n\\
  &\times \Gamma\Big[2n_1+\FR32+\ii\wt\nu,2n_2+\FR32+\ii\wt\nu,-n_1-\ii\wt\nu,-n_2-\ii\wt\nu\Big]+(\wt\nu\to-\wt\nu)\bigg]
  +\text{c.c.}\bigg\}\n\\
  =& -\FR{\lam^2}{32\pi m^2 k_1\cdots k_4(k_{12}k_{34})^{3/2}}
 \FR{\big(\wh{\mb k}_2\cdot\mb{e}_L\big)\big(\wh{\mb k}_4\cdot\mb{e}_L\big)}{k_2k_4}
  \bigg[\Big(\FR{r_1r_2}{4}\Big)^{\ii\wt\nu}\big(1-\ii\sinh\pi\wt\nu\big)\n\\
  &\times \Gamma^2(-\ii\wt\nu)\Gamma^2\Big(\FR{3}{2}+\ii\wt\nu\Big){\mathbf G}^L_{\wt\nu}(u_2,r_1){\mathbf G}^L_{\wt\nu}(u_4,r_2)+\text{c.c.}\bigg],
\end{align}
where
\begin{align}
  {\mathbf G}^L_{\wt\nu}(u,r)\equiv&                 
    \left(3+\FR{u}{1-r^2}(\FR{13}{4}+3r^2+\wt\nu^2)\right)
  {}_2\mathrm{F}_1\left[\bgm\fr{3}{4}+\fr{\ii\wt\nu}{2},\fr{5}{4}+\fr{\ii\wt\nu}{2}\\ 1+\ii\wt\nu\edm \middle|\,r^2\right]\n\\
  &-\Big(1+\FR{u}{1-r^2}(1+3r^2)\Big)\Big(\FR52+\ii\wt\nu\Big)
  {}_2\mathrm{F}_1\left[\bgm\fr{3}{4}+\fr{\ii\wt\nu}{2},\fr{9}{4}+\fr{\ii\wt\nu}{2}\\ 1+\ii\wt\nu\edm \middle|\,r^2\right].
\end{align}

Unlike the signal from the transverse states, the above result for the longitudinal state is symmetric with $r_1\leftrightarrow r_2$. Therefore, the same result also applies to the case of $k_{12}<k_{34}$.

Though we only focus on the nonlocal signal, it is worth noting that
the local signal can also be extracted straightforwardly,
namely considering the contribution of the last two sets of poles in \eqref{eq_scapoletree}
and \eqref{eq_vecpoletree}.

\subsection{Scalar loop diagrams}
Now we consider the loop diagrams, with the interaction \eqref{eq_scalarLoopCoup}.
 The cutting rule for the 4-point function \eqref{eq_scalarloopInt} is similar to the tree level case. Applying the cutting rule with $k_{12}> k_{34}$, we obtain the signal: 
\begin{align}
  \mathcal{T}_{\text{signal},\varphi'^2\si^2}
  =&~\FR{\lam^2}{32k_1\cdots k_4}
  \int_{-\infty}^0 \di\tau_1  \di\tau_2\,\big(e^{-\ii k_{12}\tau_1+\ii k_{34}\tau_2}
  -e^{\ii k_{12}\tau_1+\ii k_{34}\tau_2}\big)\n\\
  &\times \int \FR{\di^3 \mb q}{(2\pi)^3}\,D_{-+}(q;\tau_1,\tau_2)D_{-+}(|\mb{k}_s-\mb q|;\tau_1,\tau_2)+\text{c.c.}.
\end{align}
Now by applying the partial MB representation, both the time integrals and the loop momentum integral can be calculated analytically:
\begin{align}
  \mathcal{T}_{\text{signal},\varphi'^2\si^2}
  =&~\FR{\lam^2k_s^3}{512\pi^2k_1\cdots k_4(k_{12}k_{34})^4}
  \int_{-\ii\infty}^{\ii\infty}
  \FR{\di s_1}{2\pi \ii}\FR{\di s_2}{2\pi \ii}\FR{\di s_3}{2\pi \ii}\FR{\di s_4}{2\pi \ii}\,
  e^{\ii\pi(s_{13}-s_{24})}
  \Big(\FR{r_1}2\Big)^{-2s_{13}}\Big(\FR{r_2}2\Big)^{-2s_{24}}\n\\
  &\times \int \FR{\di^3 \mb s}{(2\pi)^3}\,s^{-2s_{12}}t^{-2s_{34}}
  \int_0^\infty \di z_1  \di z_2\,
  z_1^{-2s_{13}+3}z_2^{-2s_{24}+3}\big(e^{\ii z_1-\ii z_2}
  -e^{-\ii z_1-\ii z_2} \big)\n\\
  &\times \Gamma\Big[s_1-\ii\FR{\wt\nu}2,s_1+\ii\FR{\wt\nu}2,s_2-\ii\FR{\wt\nu}2,s_2+\ii\FR{\wt\nu}2,s_3-\ii\FR{\wt\nu}2,s_3+\ii\FR{\wt\nu}2,s_4-\ii\FR{\wt\nu}2,s_4+\ii\FR{\wt\nu}2\Big]\n\\
  &+\text{c.c.},\n\\
  =&~\FR{\lam^2k_s^3}{512\pi^2k_1\cdots k_4(k_{12}k_{34})^4}
  \int_{-\ii\infty}^{\ii\infty}
  \FR{\di s_1}{2\pi \ii}\FR{\di s_2}{2\pi \ii}\FR{\di s_3}{2\pi \ii}\FR{\di s_4}{2\pi \ii}\,\n\\
  &\times 
  \big(1-e^{2\ii\pi s_{13}}\big)
  \Big(\FR{r_1}2\Big)^{-2s_{13}}\Big(\FR{r_2}2\Big)^{-2s_{24}}
  \Gamma\Big[-2s_{13}+4,-2s_{24}+4\Big]\n\\
  &\times \FR{\pi^{3/2}}{(2\pi)^3}
  \Gamma\bigg[\bgm s_{1234}-\fr32,\fr32-s_{12},\fr32-s_{34}\\
s_{12},s_{34},3-s_{1234}\edm\bigg]\n\\
  &\times \Gamma\Big[s_1-\ii\FR{\wt\nu}2,s_1+\ii\FR{\wt\nu}2,s_2-\ii\FR{\wt\nu}2,s_2+\ii\FR{\wt\nu}2,s_3-\ii\FR{\wt\nu}2,s_3+\ii\FR{\wt\nu}2,s_4-\ii\FR{\wt\nu}2,s_4+\ii\FR{\wt\nu}2\Big]\n\\
  &+\text{c.c.},
\end{align}
where in the loop momentum integral, we have introduced two dimensionless variables:
\begin{equation}
  \mb s \equiv \mb q/k_s,\quad \mb t \equiv (\mb k_s - \mb q)/k_s,
\end{equation}
and the loop integral is given by:
\begin{equation}
\int \FR{\di^3 \mb s}{(2\pi)^3}\, s^at^b = \FR{\pi^{3/2}}{(2\pi)^3}\Gamma\bigg[\bgm \fr a2+\fr32,\fr b2+\fr32,-\fr a2-\fr b2-\fr32\\
-\fr a2,-\fr b2,\fr a2+\fr b2+3 \edm\bigg].
\end{equation}

The final step is using the residue theorem: closing the contour and calculating the residues at approprate poles.
The crucial point is that only the following two sets of poles contribute to the nonlocal signal:
\begin{equation}
  \left\{
  \begin{aligned}
    &s_i = -n_i - \ii\FR{\wt\nu}2,\\
    &s_i = -n_i + \ii\FR{\wt\nu}2.\\  
  \end{aligned}  
  \right.
\end{equation}
The full nonlocal signal can thus be extracted:
\begin{align}
   \mathcal{T}_{\text{NL},\varphi'^2\si^2}
  =&~\FR{\lam^2}{256\pi^{7/2}k_1\cdots k_4(k_{12}k_{34})^{5/2}}(1-\cosh2\pi\wt\nu)
  \Big(\FR{r_1r_2}{4}\Big)^{3/2+2\ii\wt\nu}\n\\
  &\times \sum_{n_1,n_2,n_3,n_4=0}^\infty
    \FR{(-1)^{n_{1234}}}{n_1!n_2!n_3!n_4!}\Big(\FR{r_1}{2}\Big)^{2n_{13}}
    \Big(\FR{r_2}{2}\Big)^{2n_{24}}
    \n\\
    &\times 
    \Gamma\Big[4+2n_{13}+2\ii\wt\nu,4+2n_{24}+2\ii\wt\nu,-n_1-\ii\wt\nu,-n_2-\ii\wt\nu,-n_3-\ii\wt\nu,-n_4-\ii\wt\nu\Big]\n\\
    &\times\Gamma\bigg[\bgm \fr32+n_{12}+\ii\wt\nu,\fr32+n_{34}+\ii\wt\nu,-n_{1234}-\fr32-2\ii\wt\nu \\ -n_{12}-\ii\wt\nu,-n_{34}-\ii\wt\nu,3+n_{1234}+2\ii\wt\nu \edm\bigg]\n\\
    &+\text{c.c.}.
\end{align}
We note that this result is symmetric with $r_1\leftrightarrow r_2$, and thus applies equally to the case of $k_{34}>k_{12}$. 
Furthermore, taking each $n_i = 0$, we automatically obtain the result in the squeezed limit:
\begin{align}
  \lim_{k_s\to 0} \mathcal{T}_{\text{NL},\varphi'^2\si^2}
  =&~\FR{\lam^2}{256\pi^{7/2}k_1\cdots k_4(k_{12}k_{34})^{5/2}}(1-\cosh2\pi\wt\nu)
  \Big(\FR{r_1r_2}{4}\Big)^{3/2+2\ii\wt\nu}\n\\
  &\times(3+2\ii\wt\nu)\Gamma(4+2\ii\wt\nu)\Gamma^2(-\ii\wt\nu)\Gamma^2\Big(\FR32+\ii\wt\nu\Big)\Gamma\Big(-\FR32-2\ii\wt\nu\Big)
  +\text{c.c.}.
\end{align}
It is worth noting that, if the two internal lines have different masses, there will be two nonlocal signals with different frequencies $\omega=|\nu_1 \pm \nu_2|$, each of which can be calculated separately following the same procedures shown above.

\subsection{Vector loop diagrams}
Finally, we consider the vector loop with the interaction \eqref{eq_vectorLoopCoup}.
In principal, we should sum over all the possible helicity combinations in the loop.
Here we only calculate the case that $h=h'=-1$, and other terms can be calculated in a similar procedure.

 Applying the cutting rule and assuming $k_{12}>k_{34}$, the 4-point function \eqref{eq_vectorloopInt} becomes:
\begin{align}
  \mathcal{T}_{\text{signal},\varphi'^2 A^2,>}
  =&~\FR{\lam^2}{32k_1\cdots k_4}
  \int_{-\infty}^0 \di\tau_1  \di\tau_2\, (\tau_1\tau_2)^2\big(e^{-\ii k_{12}\tau_1+\ii k_{34}\tau_2}
  -e^{\ii k_{12}\tau_1+\ii k_{34}\tau_2}\big)\n\\
  &\times\int\FR{\di^3\mb q}{(2\pi)^3}\,D^{(-)}_{-+}(q;\tau_1,\tau_2)D^{(-)}_{-+}(|\mb k_s-\mb q|;\tau_1,\tau_2)  \Big|\eta^{\mu\nu} e^{(-)}_\mu(\mb q)  e^{(-)}_\nu(\mb k_s - \mb q)\Big|^2.
\end{align}

Let $\theta$ be the angle between loop momentum $\mb q$ and $\mb k_s$, then
\begin{align}
  \Big|\eta^{\mu\nu} e^{(-)}_\mu(\mb q)  e^{(-)}_\nu(\mb k_s - \mb q)\Big|^2=&~\Big| \mb e_-(\mb q) \cdot \mb e_-(\mb k_s-\mb q)\Big|^2\n\\
  =&~\FR14\Big(1+\FR{q-k_s\cos\theta}{|\mb k_s-\mb q|}\Big)^2.
\end{align}
Now we apply the partial MB representation:
\begin{align}
  \label{eq_veclooptemp1}
  \mathcal{T}_{\text{signal},\varphi'^2 A^2,>}
  =&~\FR{\lam^2e^{2\pi\wt\mu}(\cosh2\pi\wt\mu+\cosh2\pi\wt\nu)^2k_s^3}{128\pi^4k_1\cdots k_4(k_{12}k_{34})^4}\n\\
  &\times\int_{-\ii\infty}^{\ii\infty}
  \FR{\di s_1}{2\pi \ii}\FR{\di s_2}{2\pi \ii}\FR{\di s_3}{2\pi \ii}\FR{\di s_4}{2\pi \ii}\,
  e^{\ii\pi(s_{13}-s_{24})/2}(2r_1)^{-s_{13}}(2r_2)^{-s_{24}}
  \n\\
  &\times \int \FR{\di^3 \mb s}{(2\pi)^3}\,s^{-s_{12}}t^{-s_{34}}\times \FR14 \Big(1+\FR{s-\cos\theta}{t}\Big)^2\n\\
  &\times \int_0^\infty \di z_1  \di z_2\,
  z_1^{-s_{13}+3}z_2^{-s_{24}+3}e^{\ii (s+t)r_1 z_1 - \ii (s+t)r_2 z_2}\big(e^{\ii z_1-\ii z_2}
  -e^{-\ii z_1-\ii z_2}\big)\n\\
  &\times \Gamma\Big[-s_1+\FR12 + \ii\wt\mu,-s_2+\FR12-\ii\wt\mu,-s_3+\FR12 + \ii\wt\mu,-s_4+\FR12-\ii\wt\mu\Big]\n\\
  &\times \Gamma\Big[s_1-\ii\wt\nu,s_1+\ii\wt\nu,s_2-\ii\wt\nu,s_2+\ii\wt\nu,s_3-\ii\wt\nu,s_3+\ii\wt\nu,s_4-\ii\wt\nu,s_4+\ii\wt\nu\Big]\n\\
  &+\text{c.c.},
\end{align}
then the time integral can be calculated, and we can expand the result: 
\begin{align}
  \label{eq_veclooptemp2}
  &\int_0^\infty \di z_1  \di z_2\,
  z_1^{-s_{13}+3}z_2^{-s_{24}+3}e^{\ii (s+t)r_1 z_1 - \ii (s+t)r_2 z_2}\big(e^{\ii z_1-\ii z_2}
  -e^{-\ii z_1-\ii z_2}\big)\n\\
  =&~\Big[e^{-\ii\pi(s_{13}-s_{24})/2}\big(1+(s+t)r_1\big)^{s_{13}-4}\big(1+(s+t)r_2\big)^{s_{24}-4}\n\\
  &-e^{\ii\pi(s_{13}+s_{24})/2}\big(1-(s+t)r_1\big)^{s_{13}-4}\big(1+(s+t)r_2\big)^{s_{24}-4}\Big]
  \times \Gamma\Big[-s_{13}+4,-s_{24}+4\Big]\n\\
=&~e^{-\ii\pi(s_{13}-s_{24})/2}\Gamma\Big[-s_{13}+4,-s_{24}+4\Big]
\sum_{m_1,m_2=0}^\infty \binom{s_{13}-4}{m_1}\binom{s_{24}-4}{m_2}\n\\
&\times (s+t)^{m_1+m_2}\Big(1-(-1)^{m_1} e^{\ii\pi s_{13}}\Big)r_1^{m_1}r_2^{m_2}\n\\
=&~e^{-\ii\pi(s_{13}-s_{24})/2}\Gamma\Big[-s_{13}+4,-s_{24}+4\Big]
\sum_{m_1,m_2=0}^\infty \sum_{\ell=0}^{m_1+m_2}\binom{s_{13}-4}{m_1}\binom{s_{24}-4}{m_2}\binom{m_1+m_2}{\ell}\n\\
&\times \Big(1-(-1)^{m_1} e^{\ii\pi s_{13}}\Big) s^{\ell}t^{m_1+m_2-\ell}r_1^{m_1}r_2^{m_2}.
\end{align}
Each term in the above expansion is proportional to $s^\ell t^{m_1+m_2-\ell}$, and thus will enter the loop integral:
\begin{align}
  \label{eq_veclooptemp3}
  &\int \FR{\di^3 \mb s}{(2\pi)^3}\,s^{-s_{12}}t^{-s_{34}}\times \FR14 \Big(1+\FR{s-\cos\theta}{t}\Big)^2\times s^\ell t^{m_1+m_2-\ell}\n\\
  =&~ \FR{1}{16\pi^2}\int_0^\infty s^2\di s\int_0^\pi \sin\theta \di \theta\,
  s^{\ell-s_{12}}t^{m_1+m_2-\ell}\Big(1-\FR{s-\cos\theta}{t}\Big)^2\n\\
  =&~\FR{1}{16\pi^2}\mathbf{L}(-s_{12}+\ell,-s_{34}+m_1+m_2-\ell),
\end{align}
where
\begin{align}
  \mathbf{L}(a,b)\equiv &\int_0^\infty s^2\di s\int_0^\pi \sin\theta \di\theta \,s^{a}t^{b}\cdot (1+\FR{s-\cos\theta}t)^2\n\\
  =&~\FR{2\Gamma(a)\Gamma(b)}{\Gamma(5+a+b)}\Big[
  -(1+a+b)(3+a+b)+\FR{\sin\pi a+\sin\pi b}{\sin \pi(a+b)}\n\\&\times (2a^2b^2+4a^2b+4ab^2+a^2+10ab+b^2+4a+4b+3)\Big].
\end{align}

Insert \eqref{eq_veclooptemp2} and \eqref{eq_veclooptemp3} into \eqref{eq_veclooptemp1}, we find:
\begin{align}
  \mathcal{T}_{\text{signal},\varphi'^2 A^2,>}
  =&~\FR{\lam^2e^{2\pi\wt\mu}(\cosh2\pi\wt\mu+\cosh2\pi\wt\nu)^2k_s^3}{2^{11}\pi^6k_1\cdots k_4(k_{12}k_{34})^4}\n\\
  &\times\int_{-\ii\infty}^{\ii\infty}
  \FR{\di s_1}{2\pi \ii}\FR{\di s_2}{2\pi \ii}\FR{\di s_3}{2\pi \ii}\FR{\di s_4}{2\pi \ii}\,
  \Big(1-(-1)^{m_1} e^{\ii\pi s_{13}}\Big) 
  (2r_1)^{-s_{13}}(2r_2)^{-s_{24}}
  \n\\
  &\times\sum_{m_1,m_2=0}^\infty \sum_{\ell=0}^{m_1+m_2}\binom{s_{13}-4}{m_1}\binom{s_{24}-4}{m_2}\binom{m_1+m_2}{\ell}\n\\
&\times r_1^{m_1}r_2^{m_2}\mathbf{L}(-s_{12}+\ell,-s_{34}+m_1+m_2-\ell)
\Gamma\Big[-s_{13}+4,-s_{24}+4\Big]\n\\
  &\times \Gamma\Big[-s_1+\FR12 + \ii\wt\mu,-s_2+\FR12-\ii\wt\mu,-s_3+\FR12 + \ii\wt\mu,-s_4+\FR12-\ii\wt\mu\Big]\n\\
  &\times \Gamma\Big[s_1-\ii\wt\nu,s_1+\ii\wt\nu,s_2-\ii\wt\nu,s_2+\ii\wt\nu,s_3-\ii\wt\nu,s_3+\ii\wt\nu,s_4-\ii\wt\nu,s_4+\ii\wt\nu\Big]\n\\
  &+\text{c.c.}.
\end{align}
Also, the only poles contributing to the nonlocal signal are the following:
\begin{equation}
  \left\{
    \begin{aligned}
      &s_i = -n_i - \ii\wt\nu,\\
      &s_i = -n_i + \ii\wt\nu,
    \end{aligned}
  \right.
\end{equation}
and the final result for $\mathcal{T}_{\text{NL},\varphi'^2 A^2,>}$ is:
\begin{align}
  \mathcal{T}_{\text{NL},\varphi'^2 A^2,>}
  =&~\FR{\lam^2e^{2\pi\wt\mu}(\cosh2\pi\wt\mu+\cosh2\pi\wt\nu)^2}{2^{14}\pi^6k_1\cdots k_4(k_{12}k_{34})^{5/2}}
  (4r_1r_2)^{3/2+2\ii\wt\nu}
  \sum_{n_1,n_2,n_3,n_4=0}^\infty \FR{(-2)^{n_{1234}}}{n_1!n_2!n_3!n_4!}\n\\
  &\times\bigg\{
  \Big(1-(-1)^{n_{13}+m_1} e^{2\pi\wt\nu}\Big) r_1^{n_{13}+m_1}r_2^{n_{24}+m_2}\n\\
  &\times \sum_{m_1,m_2=0}^\infty \sum_{\ell=0}^{m_1+m_2}\binom{-n_{13}-4-2\ii\wt\nu}{m_1}\binom{-n_{24}-4-2\ii\wt\nu}{m_2}\binom{m_1+m_2}{\ell}\n\\
  &\times\mathbf{L}(n_{12}+\ell+2\ii\wt\nu,n_{34}+m_1+m_2-\ell+2\ii\wt\nu)
\Gamma\Big[n_{13}+4+2\ii\wt\nu,n_{24}+4+2\ii\wt\nu\Big]\n\\
  &\times \Gamma\Big[n_1+\FR12 + \ii\wt\mu + \ii\wt\nu,n_2+\FR12-\ii\wt\mu+\ii\wt\nu,n_3+\FR12 + \ii\wt\mu+\ii\wt\nu,n_4+\FR12-\ii\wt\mu+\ii\wt\nu\Big]\n\\
  &\times \Gamma\Big[-n_1-2\ii\wt\nu,-n_2-2\ii\wt\nu, -n_3-2\ii\wt\nu,-n_4-2\ii\wt\nu\Big]\n\\
  &+(\wt\nu\to-\wt\nu)\bigg\}+\text{c.c.}.
\end{align}
Again, the result for $k_{34}>k_{12}$ is obtained by the exchange $r_1\leftrightarrow r_2$, and the full non-local signal for this process is:
\begin{align}
  \mathcal{T}_{\text{NL},\varphi'^2 A^2}(k_1,k_2,k_3,k_4,k_s)=&~\mathcal{T}_{\text{NL},\varphi'^2 A^2,>}(k_1,k_2,k_3,k_4,k_s)\theta(k_{12}-k_{34})\n\\
  &~+\mathcal{T}_{\text{NL},\varphi'^2 A^2,>}(k_3,k_4,k_1,k_2,k_s)\theta(k_{34}-k_{12}).
\end{align}

\end{appendix}

\newpage
\bibliography{CosmoCollider} 
\bibliographystyle{utphys}

\end{document}